\documentclass[journal]{IEEEtran}

\usepackage{mathrsfs,amssymb}
\usepackage{times}
\usepackage{epsfig}
\usepackage{graphicx}
\usepackage{amsmath}
\usepackage{amssymb}
\usepackage{hyperref}
\usepackage{changes}
\usepackage{algorithmic}
\usepackage{multirow} 
\usepackage{amsmath}
\usepackage{xcolor}
\usepackage[nocompress,space]{cite}
\usepackage[normalem]{ulem}
\usepackage{array}
\newcolumntype{C}[1]{>{\centering\let\newline\\\arraybackslash\hspace{0pt}}m{#1}}
\usepackage[linesnumbered,ruled]{algorithm2e}
\usepackage{indentfirst}

\ifCLASSINFOpdf
\else
\fi

\def\refcolor#1#2{\expandafter\xdef\csname#1color\endcsname{#2}}
\refcolor{add13}{red}

\begin{document}
\title{The bilateral solver for quality estimation based multi-focus image fusion}
%
%
%

 \author{
{Jingwei GUAN, Yibo CHEN and Wai-kuen~CHAM
}}



\markboth{Journal of IEEE Transactions on Image Processing,~Vol.~XX, No.~XX, August~2015}%
{Shell \MakeLowercase{\textit{et al.}}: Bare Demo of IEEEtran.cls for IEEE Journals}
%



\maketitle

\begin{abstract}
In this work, a fast Bilateral Solver for Quality Estimation Based multi-focus Image Fusion method (BS-QEBIF) is proposed. 
The all-in-focus image is generated by pixel-wise summing up the multi-focus source images with their focus-levels maps as weights. 
Since the visual quality of an image patch is highly correlated with its focus level,
the focus-level maps are preliminarily obtained based on visual quality scores, as pre-estimations.
However, the pre-estimations are not ideal.
Thus the fast bilateral solver is then adopted to smooth the pre-estimations, and edges in the multi-focus source images can be preserved simultaneously.
The edge-preserving smoothed results are utilized as final focus-level maps.
Moreover, this work provides a confidence-map solution for the unstable fusion in the focus-level-changed boundary regions.
Experiments were conducted on $25$ pairs of source images.
The proposed BS-QEBIF outperforms the other $13$ fusion methods objectively and subjectively.
The all-in-focus image produced by the proposed method can well maintain the details in the multi-focus source images and does not suffer from any residual errors.
Experimental results show that BS-QEBIF can handle the focus-level-changed boundary regions without any blocking, ringing and blurring artifacts.
\end{abstract}

\IEEEpeerreviewmaketitle

\section{Introduction}
\label{sec:intro}

\begin{figure}[h!]
\begin{tabular}
{c@{\hspace{-5mm}}C{30mm}@{\hspace{+1mm}}C{30mm}@{\hspace{+1mm}}C{30mm}}
&\includegraphics[width =1\linewidth]{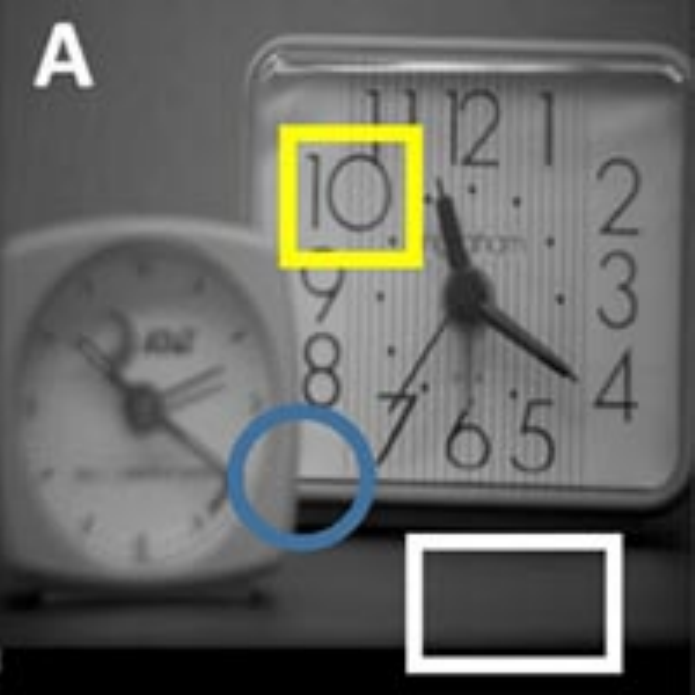}
&\includegraphics[width =1\linewidth]{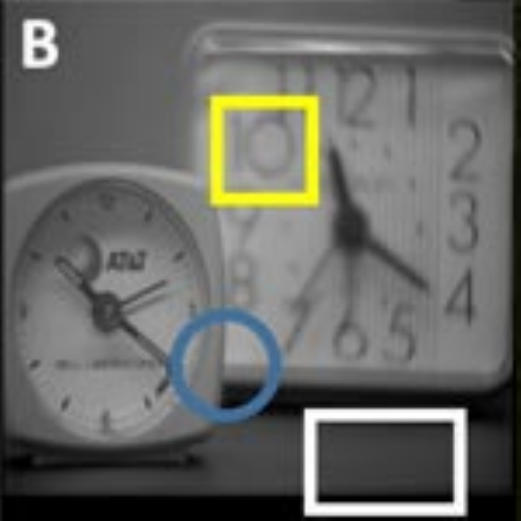}
&\includegraphics[width =1\linewidth]{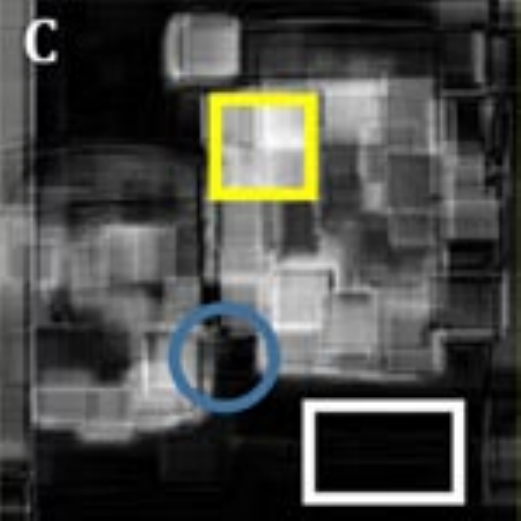}
\\
&
&(a)
&\\
&\includegraphics[width =1\linewidth]{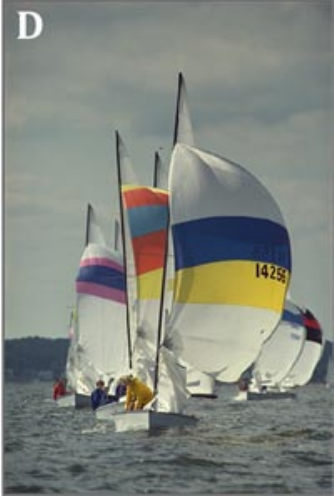}
&\includegraphics[width =1\linewidth]{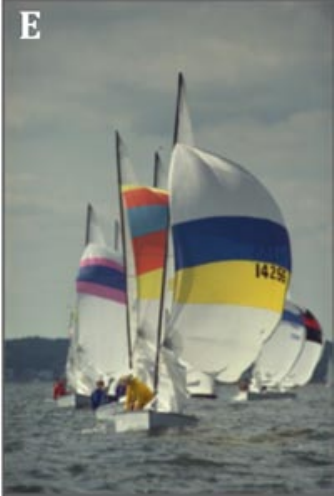}
&\includegraphics[width =1\linewidth]{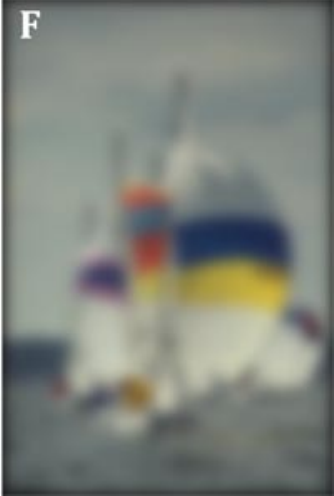}
\\
&\vspace{-6mm}
&(b)\vspace{+6mm}\vspace{-6mm}
&\vspace{-6mm}
 \\
&\includegraphics[width =1\linewidth]{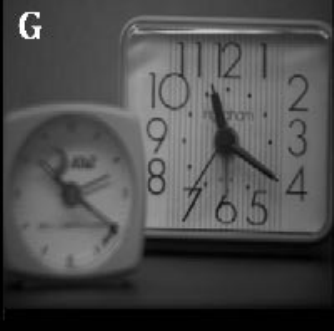}
&\includegraphics[width =1\linewidth]{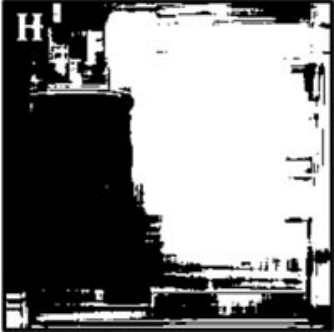}
&\includegraphics[width =1\linewidth]{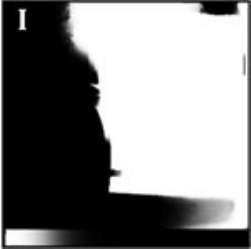}
\\
&
&(c)&
\end{tabular}
\caption{{\small
(a) Examples of two multi-focus source images A, B and their absolute quality difference C.
%
%
(b) Examples of blur images in the image quality assessment (IQA) dataset. 
(c) Examples of pre-estimation, image H, and edge-preserving smoothed result, image I, of source image G.
(Best viewed in color.)
}}
\label{fig:motivation}
\vspace{-5mm}
\end{figure}

\IEEEPARstart{V}arious photographs are taken every day. 
If the light from an object point is not well converged in the focal plane, the object would be out-of-focus which leads to image blurring and loss of information.
Take multi-focus source image A and B in Fig. \ref{fig:motivation} (a) as an example.
These images are taken from the same scene.
The out-of-focus effect leads to information loss in the left clock in A and the right clock in B.
The multi-focus image fusion task \cite{goshtasby2007image,li2017pixel} aims at generating an all-in-focus image based on the source images, where both clocks can be in-focus in this example.

The accurate estimation of focus-level of source images is crucial in generating the all-in-focus image.
In this work, we advocate utilizing image quality assessment (IQA) task to do focus-level estimation. 
The IQA task targets on accurately estimating image visual quality, where the increase in image blurring level makes its visual quality worse.
As an example shown in Fig. \ref{fig:motivation} (b), the focus levels of the images decrease from D to F. 
Their visual quality also decrease from left to right.
Hence, visual quality can be utilized to coarsely measure image focus levels, and generate preliminary focus-level estimation maps, pre-estimation, by comparing the visual quality scores among all source images.
For example in Fig. \ref{fig:motivation} (c), the pre-estimation of image G is shown as image H.

As shown in image H, most estimation results within the object regions are correct.
However, the estimation results in the focus-level-changed boundary regions and background out-of-focus regions are not accurate.
Thus pre-estimations are not appropriate to be directly used as focus-level maps.
In this work, the state-of-the-art edge-preserving smoothing filter, the fast bilateral solver \cite{solver}, is first introduced to the image fusion task.
With the fast bilateral solver, the focus-level map I is generated by smoothing H and preserving edges in G simultaneously.
The focus-level map, image I, is used to generate the all-in-focus image.

To better estimate the focus-levels, we further analyze the source images.
These source images are supposed to be focused at different regions, and one region cannot be focused in all multi-focus images.
However, it is possible for some regions, especially the background, to be out of focus in all multi-focus source images.
For example in Fig. \ref{fig:motivation} (a), image {C} illustrates the absolute visual-quality difference between source image A and B, where the brighter color corresponds to the larger difference.
In the yellow square patches, the difference between A and B is big representing their big difference in visual quality scores and focus levels. 
The focus-level estimations in these regions are more likely to be accurate, and the square regions can be regarded as reliable regions.
While in the rectangular and circular patches, the differences are small.
The rectangular patches in both source images are out-of-focus, and the circular patches locate at the focus-level-changed boundary regions which cover both in-focus and out-of-focus regions.
It is more challenging to measure the slightly focus-level difference for the rectangular and circular patches, and they should be regarded as less reliable regions.
In this work, a confidence map is proposed to measure the reliability of different regions.
With the confidence map, different regions can be treated differently and adaptively to generate more accurate focus-level maps.

To sum up, a novel multi-focus image fusion method, the fast Bilateral Solver for Quality Estimation Based Image Fusion (BS-QEBIF), is proposed. 
We follow the focus-level weighted summation pipeline where the all-in-focus image is generated by pixel-wise summing up all multi-focus images with their focus-levels as weights.
The contributions are mainly three-fold.
First, the visual quality is first introduced to image fusion task to help estimate image focus levels.
With rich images with subjective quality scores provided in IQA datasets, various supervised training methods can be explored in the fusion procedure. 
Second, the fast bilateral solver \cite{solver} is first adopted to smooth the pre-estimation results and generate the focus-level maps as weights.
{Instead of directly applying the fast bilateral solver to do the fusion, we explore different settings and design the confidence map accordingly.}
Benefiting from the bilateral solver, good and robust fusion results can be obtained.
Finally, the confidence map is proposed to help improve focus-level estimation.
The higher confidence scores correspond to the more reliable regions.
The proposed method employs this concept in the focus-level estimation and effectively improves the fusion performance, especially in the focus-level-changed boundary regions.



\section{Related work}
\label{sec:relatedWork}\vspace{-1mm}
Various multi-focus image fusion methods were proposed in the last decades.
{Generally speaking, the image fusion methods can be divided into three categories, defocus-modeling methods, transform-based methods,  and spatial-frequency methods.}
The \emph{defocus-modelling} methods \cite{subbarao1995accurate,aguet2008model} defocus the input images by a designed filter to reserve the blur effect.
The \emph{transform-based} methods \cite{GF,pajares2004wavelet, MST-SR, chen2016multi,omar2011region,wang2005image,lewis2004region,25,29,311,312,321,33,35,44,45,46,47,22,26} employed various transformations to do the fusion.
The\emph{ spatial-frequency} methods concentrate on accurately measuring the focus or sharpness levels of the multi-focus images \cite{yingjie2007region,tian2010multi, li2008multifocus,13,xiao2008image,9,11,36,39,42,43,12}.
For example, the simplest method is averaging (AVG) where the multi-focus source images are averaged to generate the all-in-focus image.
Besides, MWGF \cite{28} and IM \cite{37} presented multi-scale structure-based and morphological filtering-based focus estimations respectively.
The proposed method also concentrates on exploring the focus level estimations and belongs to spatial-frequency methods.

In the last decades, neural network and learning based methods were widely employed in image processing tasks and achieved impressive success.
Some image fusion methods also try to utilize machine learning to help focus level estimation.
For example, some methods \cite{wang2010multi,li2006region,agrawal2010multifocus,xiao2008image} adopt pulse coupled neural networks (PCNN) to do image fusion where PCNN does not require the training procedure.
Some other methods \cite{li2002multifocus,gao2016multi,yang2014effective} train neural networks or dictionaries where focus level estimation was regarded as a classification problem to divide a local region into an in-focus or out-of-focus one.
To do the training, \cite{gao2016multi,yang2014effective} collect some all-in-focus images and manually blur them using the Gaussian filters, while \cite{wang2010multi} manually labels in-focus/out-of-focus regions from source images as training sets.

However, all above methods have not made fully use of machine learning and trained a deep neural network.
The main reason is the lack of large amount of training data with labels.
\emph{In this work, we first introduce a highly-related task, Image Quality Assessment (IQA), to help effectively estimate focus levels.}
With the large amount of training data with labels provided in the IQA dataset, various deep neural networks could be explored. 

Besides, different edge-preserving smoothing filters were explored in earlier works to make full use of spatial consistency during the fusion process.
CBF \cite{38} adopts the cross bilateral filter (CBF) to compute the weights to measure the strength of details in multi-focus source images.
Li \emph{et al.} \cite{GF}, Nejati \emph{et al.} \cite{Lytro} and Guan \emph{et al.} \cite{QEBIF} adopted the guided filter to do the fusion.
In this work, a recent proposed state-of-the-art edge-preserving smoothing filter, the fast bilateral solver, is first adopted in image fusion task.
Moreover, the confidence map is proposed to help the focus-level estimation with the fast bilateral solver in BS-QEBIF.

\begin{figure*}[t!]
\begin{tabular}
{@{\hspace{-10mm}}c@{\hspace{-10mm}}c}
\includegraphics[width=1.1\linewidth]{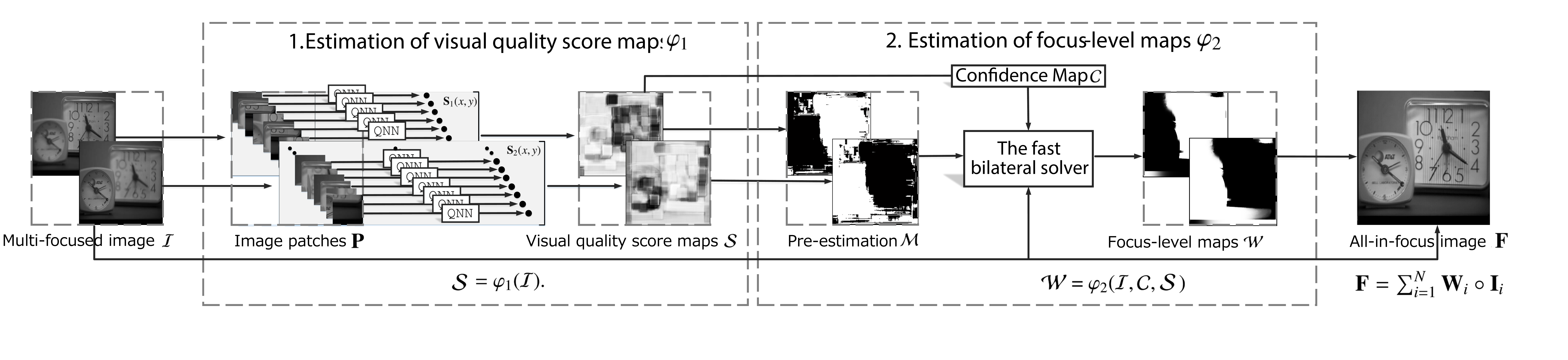}\\
\end{tabular}
\caption{
{\small {Pipeline of the proposed BS-QEBIF method.
}
}}
\label{fig:pipeline}
\end{figure*}

\section{The proposed BS-QEBIF method}
\label{sec:method}\vspace{-1mm}
The pipeline of the proposed BS-QEBIF method is summarized in Fig. \ref{fig:pipeline}. 
The all-in-focus image $\textbf{F}$ is generated by {pixel-wise summarizing} the Hadamard products \cite{HadamardProduct} of multi-focus source images $\cal{I}$$= \{ \textbf{I}_1,\dots,\textbf{I}_N\}$ with their corresponding focus-level maps $\cal{W}$$= \{ \textbf{W}_1,\dots,\textbf{W}_N\}$ as weights, \emph{i.e.}
\begin{align}
\textbf{F} = \sum_{i = 1}^{N}{{\textbf{W}_i}\circ{\textbf{I}_i}}, i\in \{1,2,\dots, N\}. 
\end{align}
$N$ is the number of multi-focus source images.
All notations used in this work are summarized in TABLE \ref{table:notation}.
The focus-level weights $\cal{W}$ are mainly obtained via steps $\varphi_1$ and $\varphi_2$.

In $\varphi_1$, the visual quality score for each pixel is estimated to generate the visual quality score maps $\cal{S}$$= \{ \textbf{S}_1,\dots, \textbf{S}_N\}$.
\begin{align}
{\textbf{S}_i}= \varphi _{1}(\textbf{I}_i).
\label{equ:varphi1}
\end{align}
 


Next, the focus-level maps of all images $\cal{W}$ are estimated based on visual quality score maps $\cal{S}$ and the source images $\cal{I}$.
During this process, the confidence maps $\cal{C}$ are employed to improve the estimation, \emph{i.e.}
\begin{align}
{\cal{W}} = \varphi_2({\cal{I},\cal{C},\cal{S}}).
 \label{equ:varphi2}
\end{align}
In the following sections, $\varphi_1$ (Sec. \ref{sec:varphi_1}) and $\varphi_2$ (Sec. \ref{sec:varphi_2}) would be introduced in details. 
%
\vspace{-2mm}
\subsection{{Estimation of visual quality score maps $\varphi_1$}}
\label{sec:varphi_1}
As discussed earlier, the focus level of a region can be approximately estimated by its visual quality score. 
Thus a Quality deep Neural Network (\texttt{QNN}) is proposed to estimate visual quality.
In $\texttt{QNN}$, the large amount of training data with labels are from IQA datasets.

\begin{table}[!t]
\caption{Notations used in this work}\label{table:notation}
\begin{small}
    \begin{tabular}{C{10mm}|C{70mm}}\hline
    IQA & Image Quality Assessment.\\
    \textbf{F}& Generated all-in-focus image \\
    $\cal{I}$  &  Multi-focus source images.\\
    $\cal{W}$ & Focus-level maps utilized as weights to do the fusion.\\
    $\cal{M}$ & Preliminary focus-level estimation. \\
     $\cal{S}$ & Estimated visual quality score maps of $\cal{I}$. \\
$\cal{C}$ & Confidence map to measure the reliability of different local regions.\\
{Conv} & Convolutional neural layer.\\
\small{DMOS} & Difference mean opinion score: subjective visual quality scores provided in the IQA datasets.\\
FC & Fully connected layer. \\
$\textbf{I}_i$ & $i^{th}$ multi-focus source image in $\cal{I}$.\\
$N$ & Number of multi-focus images. \\
$L_{\texttt{QNN}}$& Loss function for training $\texttt{QNN}$.\\
$n_{ \texttt{QNN}}$ & Mini-batch size when training $ \texttt{QNN}$.\\
$\textbf{P}$ & Image patch utilized to estimate the visual quality of its centering pixel.\\ 
Pool & Pooling layer. \\
\texttt{QNN} & Quality deep neural network to estimate visual quality.\\
 $\cal{W}'$ & Edge-preserving smoothing results using the fast bilateral solver.\\
$(x,y)$ & Spatial coordinates. \\
$\theta_j$ & Parameters in $ \texttt{QNN}$.\\
$\varphi_1$ & Learning based visual quality estimation process.\\
$\varphi_2$ & Edge preserving smoothing process.\\
      \hline
    \end{tabular}
\end{small}
\end{table}

\begin{figure}[t]
\begin{center}
\includegraphics[width=\linewidth]{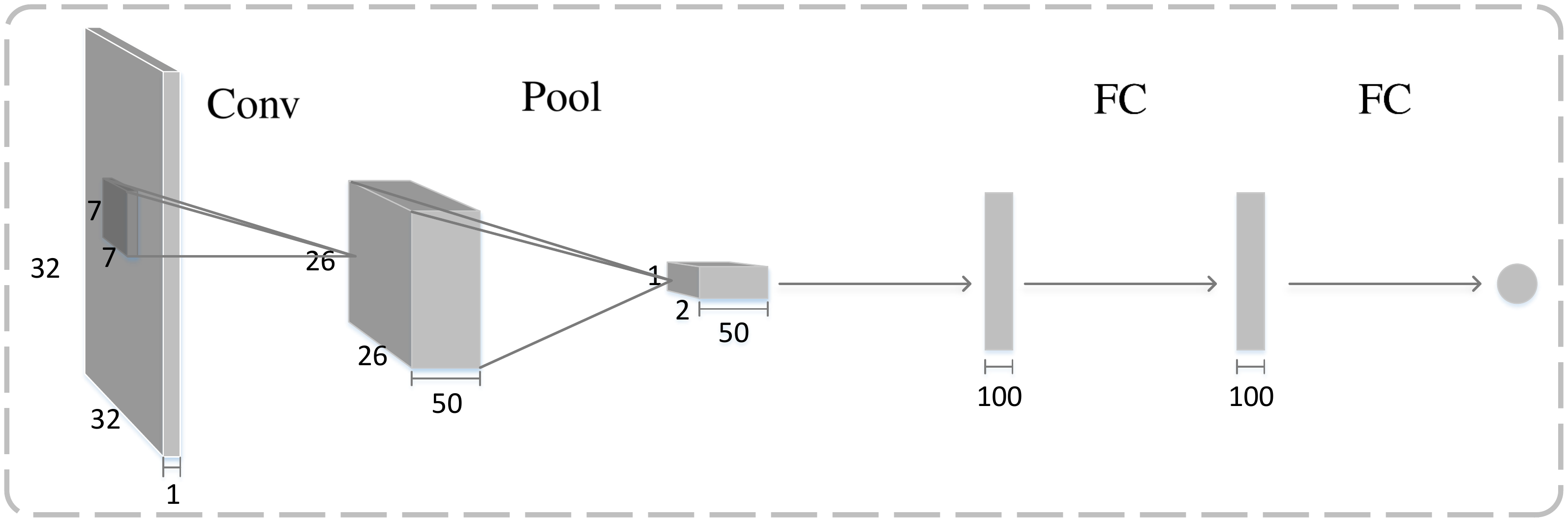}
\end{center}
\caption{{\small
The architecture of  \texttt{QNN}.}}
\label{fig:NN_architecture}
\end{figure}

\textbf{Architecture of \texttt{QNN}} is illustrated in Fig. \ref{fig:NN_architecture}. 
To avoid the blocking artifacts in quality score map $\textbf{S}_i$, every pixel in $\textbf{I}_i$ is evaluated by \texttt{QNN}.
As the input of $\texttt{QNN}$, each pixel in $\textbf{I}_i$ is first normalized by removing the local mean and dividing by the standard deviation.
The local mean and standard deviation are computed within a neighboring area of size $7\times7$ which is popularly utilized in state-of-the-art IQA models \cite{CNN, BRISQUE, TMM, APSIPA}.
After normalization, an image patch $\textbf{P}_i(x,y)$ of size $32\times 32$ centering at $(x,y)$ is utilized as the input of \texttt{QNN} to estimate the visual quality of $\textbf{I}_i(x,y)$, \emph{i.e.} $\textbf{S}_i(x,y) = \texttt{QNN}(\textbf{P}_i(x,y))$. 
The centering position of a $32\times 32$ patch is set as $(17,17)$ in this work.
Details of the three layer types used in \texttt{QNN}, \emph{i.e.} convolutional neural layer
(Conv), pooling layer (Pool) and fully connected layer (FC), are introduced below.
\begin{enumerate}
\item 
The convolutional neural layer (Conv) convolves the image patch $\textbf{P}$ with the $50$ learned filters and obtain $50$ response maps.
The filters are of size $7\times 7$. The stride size is $1$.
\item 
The pooling layer (Pool) is a way of sub-sampling and has been widely used in deep learning. 
In this work, both max-pooling and min-pooling are adopted simultaneously for each response map. 
Both the kernel size and stride size are $26$.

\item
The fully connected layer (FC) takes the flattened results of previous layer as input and connects it with all neurons in this FC layer.
The non-linear activation function is applied to each filter response. In \texttt{QNN}, Rectified Linear Unit (ReLU \cite{ReLU}) and Sigmoid are utilized for $\mathrm{FC_1}(100)$ and $\mathrm{FC_2}(1)$ respectively. The activation function of ReLU and Sigmoid are illustrated as follows.
\begin{align}
ReLU: &~~~ b = max(0,a)\\
Sigmoid: & ~~~ b= \frac{1}{1+e^{-a}}
\end{align}
where $a$ and $b$ represent the input and output of the activation function respectively.
\end{enumerate}




\textbf{Training procedure of \texttt{QNN}:} The parameters in \texttt{QNN} are learned through back-propagation.
The key to success for such supervised learning task is the use of large amount of labeled training images.
Since the principle of out-of-focus blurring is similar to the Gaussian blurring effect, images blurred by Gaussian filter in the IQA dataset, the LIVE dataset \cite{LIVE}, are utilized as training samples.
The subjective visual quality scores provided in the dataset \cite{LIVE}, difference mean opinion scores (DMOS), were utilized as training labels, where the smaller DMOS represents better visual quality.
Since these Gaussian images were generated with a circular-symmetric 2D Gaussian kernel of a certain standard deviation sigma $\sigma$, all pixels in one image can be regarded as of the same focus levels and share the same label.
In the implementation, these blurry images are non-overlappingly divided into image patches $\textbf{P}$ of size $32\times32$ to serve as the training input of \texttt{QNN}, while the corresponding DMOS of the whole image are utilized as the training labels.
The loss function for training \texttt{QNN} is 
\begin{align}
L_{ \texttt{QNN}} = \frac{1}{n_ \texttt{QNN}}\sum_{l = 1}^{n_  \texttt{QNN}}|\texttt{QNN}(\textbf{P}_l)-DMOS|,
\end{align}
where $n_ \texttt{QNN}$ is the mini-batch size \cite{mini-batch} and $l$ is the index of $\textbf{P}$. For any parameter $\theta_j$ in the \texttt{QNN}, the updated process is illustrated as 
\begin{align}
\bigtriangleup \theta_j= \eta \frac{\partial L_{ \texttt{QNN}}}{\partial \theta_j}, \theta_j^{new} = \theta_j^{old}-\bigtriangleup \theta_j
\end{align}
where $\eta$ represents the learning rate, $\frac{\partial L_{ \texttt{QNN}}}{\partial \theta_j}$ is the corresponding derivative.
{It should be noted that the current \texttt{QNN} model was trained on gray images. 
It is easy to extend the model to process color images.
}




\begin{figure}
\centering
\begin{tabular}{@{\hspace{0mm}}c@{\hspace{1mm}}c@{\hspace{4mm}}c}
&\includegraphics[width =0.4\linewidth]{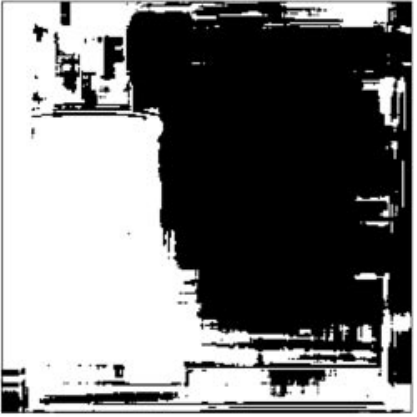}
&\includegraphics[width =0.4\linewidth]{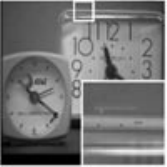}
\\
&(a) pre-estimation $\textbf{M}_1$.
&(b) Fusion result using $\cal{M}$
\\
\end{tabular}
\caption{
Examples of (a) the pre-estimation $\textbf{M}_1$ and (b) the fusion result using $\cal{M}$ as focus level weights.
}
\label{fig:NEW}
\end{figure}

\begin{figure}[t!]
\begin{center}
\includegraphics[width=0.8\linewidth]{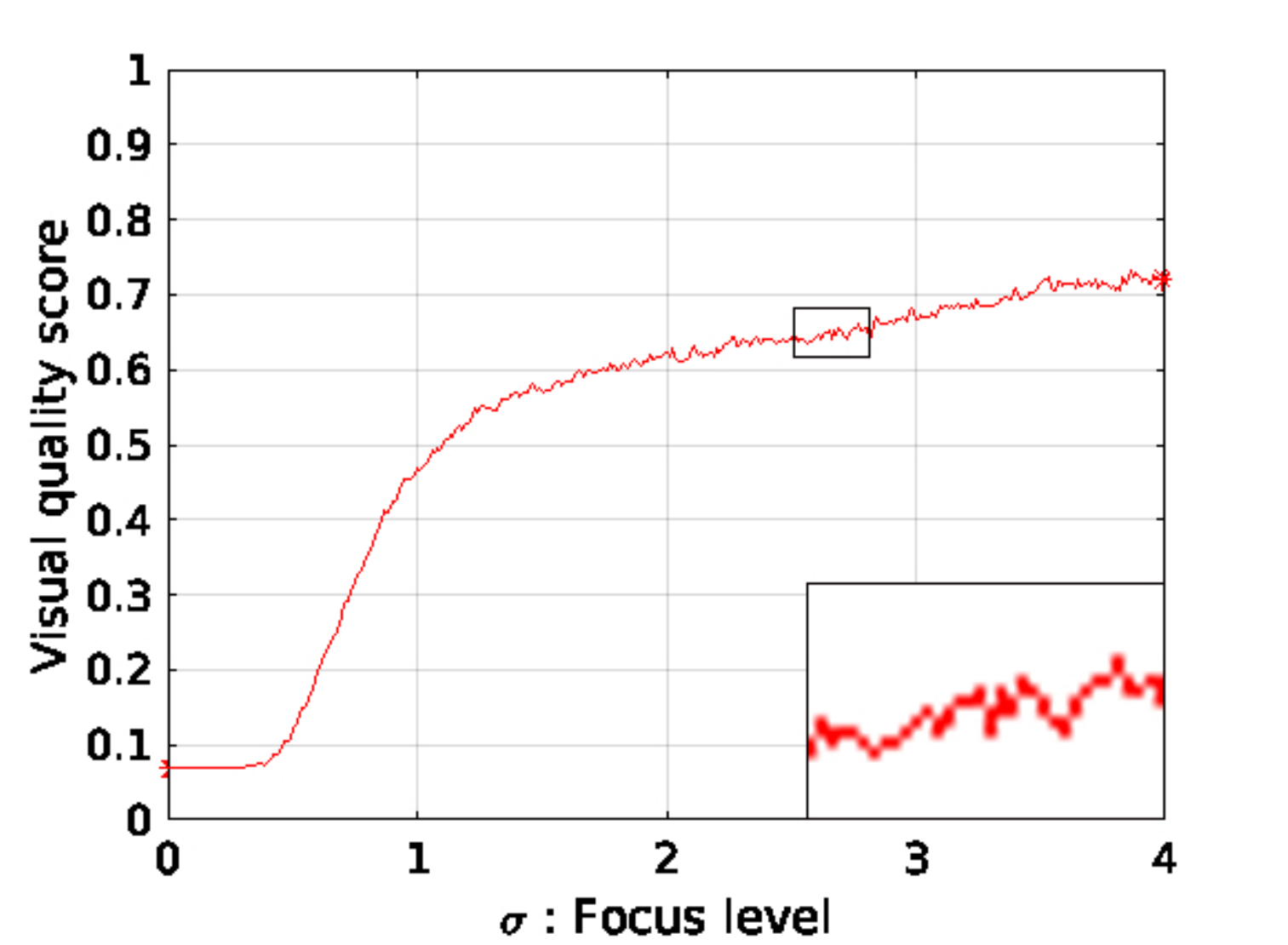}
\end{center} \vspace{-4mm}
\caption{{\small
The change of estimated visual quality scores using $\texttt{QNN}$ (y-axis) with different focus levels (x-axis). 
The standard deviation $\sigma$ of Gaussian blurred images are utilized to measure focus levels.
}}
\label{fig:plot}
\end{figure}

\subsection{Estimation of focus-level maps $\varphi_2$}
\label{sec:varphi_2}
The estimation of focus-level maps can be divided into two steps.
First, the preliminary focus-level estimation results, pre-estimation $\cal{M}$, can be estimated by regarding the image with best visual quality as the in-focused one. 
However, the pre-estimations contain much noise and are not appropriate to be directly used as the weights in the fusion.
Therefore, the fast bilateral solver \cite{solver} is employed in the second step to edge-preserved smooth $\cal{M}$ and generate focus-level maps $\cal{W}$.
During the second stage, the confidence map $\cal{C}$ and the normalization are employed.

\begin{algorithm}[t!]
\KwIn{~~~Visual quality score maps $\cal{S}$ = \{$\textbf{S}_1,\dots, \textbf{S}_N$\}}
\KwOut{~Confidence Map $\cal{C}$ = \{$\textbf{C}_1,\dots, \textbf{C}_N$\}}
\Begin{
{
\textbf{for} every pixel $(x,y)$\\
~~~~1. Calculate the absolute difference $\textbf{S}'_{d}(x,y)$ \\
~~~~~~between the maximum score map $\textbf{S}_{max}(x,y)$ and \\~~~ minimum score map $\textbf{S}_{min}(x,y)$ among all $\cal{I}$$(x,y)$.
\\}
~~~~2. Calculate $\textbf{S}_{d}(x,y)$ by linearly normalizing \\
~~~~ ${\textbf{S}}'_{d}(x,y)$ to $[0,1]$;\\ 
~~~~3. Compute $\textbf{C}_i$ by a  step function with threshold \\
~~~~Thr.\\
~~~~$\textbf{C}_i(\textbf{S}_{d}<Thr)  = 0.1$;\\
~~~~$\textbf{C}_i(\textbf{S}_{d}>Thr)  = 1$;\\
\textbf{end}\\
$\textbf{C}_1 = \textbf{C}_2 = \dots = \textbf{C}_N$\\
}
\caption{Computation of confidence map $\cal{C}$.}
\label{alg:C}
\end{algorithm}

\subsubsection{Pre-estimation $\cal{M}$}

The pre-estimation of focus levels $\cal{M}$ are estimated based on the visual quality score maps $\cal{S}$.
The in-focus image is estimated to be the one with best visual quality. 
%
\emph{i.e.}
\begin{align}
\textbf{M}_i(x,y) = 
\left\{
\begin{aligned}
1, &~\textrm{if min($\textbf{S}_1(x,y),\dots,\textbf{S}_N(x,y))=\textbf{S}_i(x,y)$},\\
0, &~\textrm{otherwise},
\label{equ:M}
\end{aligned}
\right.
\end{align}
where the smaller value in $\textbf{S}_i$ indicates better visual quality. 
Thus $\textbf{M}_i(x,y) = 1$ indicates that $\textbf{I}_i$ is approximated as the in-focused source image at position $(x,y)$.

An example of $\cal{M}$ and the fusion result directly using $\cal{M}$ are shown in Fig. \ref{fig:NEW}.
The fusion result shown in image (b) is not satisfactory since some unwanted sudden changes exist.
{In the object region, $\cal{M}$ is expected to be smooth, such as the right clock in image (a) since they are of the same focus level.}
While in some focus-level-changed boundary region, the lines in $\cal{I}$ are expected to be well preserved in $\cal{M}$, such as the lines between the two clocks. 
These unwanted sudden changes in $\cal{M}$ result in the unwanted sudden changes in the fusion results as shown in (b).

\textbf{Reason of inaccuracy
:} These unwanted sudden changes in $\cal{M}$ are mainly result from {the instability of} \texttt{QNN} when the focus-level changes of inputs are small.
Fig. \ref{fig:plot} illustrates the change of estimated visual quality scores using \texttt{QNN} (y-axis) with different focus levels (x-axis).  
In this experiment, $100$ image patches distorted by Gaussian blur are utilized for evaluation, where the averaging visual quality estimation scores of these $100$ patches are reported in y-axis.
It can be seen that the overall trend
is monotonous. 
However, the estimation results of small focus-level changing images are not stable as shown in the enlarged detail block in the lower right corner.
Therefore, when using $\texttt{QNN}$ to evaluate several regions with similar quality scores, the results may not be stable.

\subsubsection{Edge-preserving smoothing $\cal{M}$ to generate $\cal{W}$}

In order to smooth within object regions and preserve boundary edges, an edge-preserving smoothing filter, the fast bilateral solver \cite{solver}, is utilized to generate the final estimation of focus-level maps $\cal{W}$. 

Besides, a lot of information is lost when calculating $\cal{M}$ in $\varphi_1$.
Because for any same-sign quality difference, $\cal{M}$ would always be given the same decision. 
To make full use of information in quality score maps, the confidence map is proposed to pixel-wise measure the reliability of $\cal{M}$, and help improve focus level estimation accuracy.
In addition, the normalization is employed.
In this section, the confidence map, fast bilateral solver and normalization are introduced.


\begin{figure}[t!]
\begin{center}
\hspace{-3mm}\includegraphics[width=1.05\linewidth]{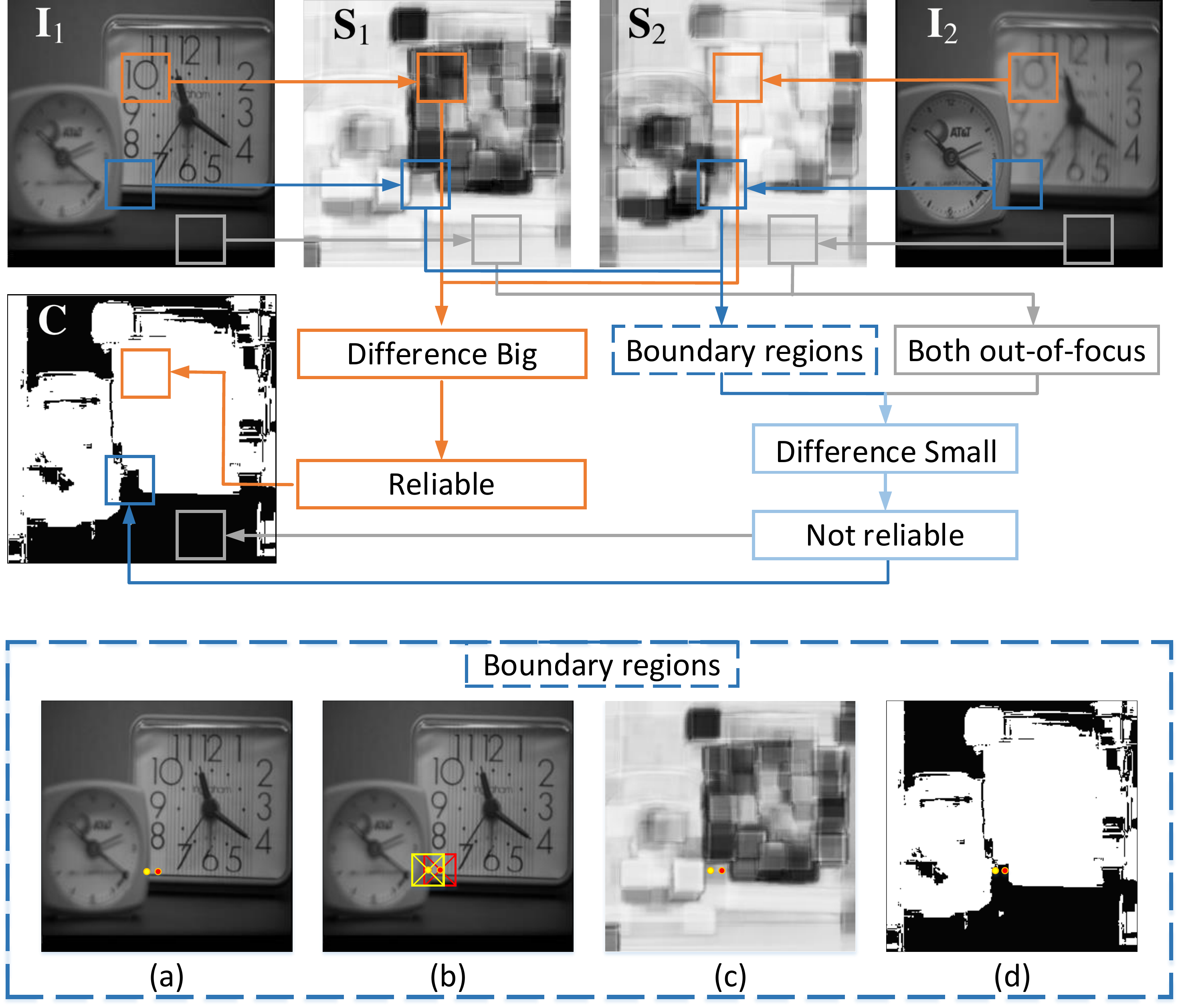}
\end{center} \vspace{-4mm}
\caption{{\small
Demonstration of the way to calculate confidence maps $\cal{C}$.
Darker color in \textbf{S} represents better quality, while brighter color in $\textbf{C}$ represents bigger confidence.
The `Boundary regions' block demonstrates that the unreliability of visual quality estimation results in the focus-level-changed boundary regions.
(Best viewed in color.)
}}
\label{fig:CM}
\end{figure}

\textbf{Confidence map:}
As shown in Fig. \ref{fig:plot}, the quality estimation results tend to be more accurate in regions with larger focus-level difference than those with smaller difference.
Based on such observation, $\cal{C}$ are generated based on the quality difference as illustrated in the upper part of Fig. \ref{fig:CM}.
For the orange patches, the difference in their quality maps is big.
There is a big chance that the estimation result in this region is accurate. 
On the contrary, things become uncertain when the difference is small, which mainly results from two reasons.
%
First, both patches may be out of focus such as the gray ones and result in unreliable results in $\cal{M}$.
Second, the patches may locate at the focus-level-changed boundary region such as the blue ones.
To be specific, the estimation results in the boundary patches are not reliable as illustrated in the `Boundary regions' block of Fig. \ref{fig:CM}.
In (a), the yellow and red point is out-of-focus and in-focus respectively.
However, the patches utilized for estimating the quality scores of these two points cover both the in-focus and out-of-focus regions as shown in (b).
Since the two patches are of similar focus levels, the estimated visual quality scores for these two points are similar as shown in (c), which are not reliable. 
Therefore, the corresponding points in $\cal{M}$ should be regarded as unreliable regions.

Considering the above conditions, the confidence map divides the whole image as reliable and unreliable regions according to their visual-quality-score difference.
The computation of $\cal{C}$ is summarized in Algorithm \ref{alg:C}. Thr = $0.1$ in the implementation.
The confidence maps of all source images are the same.

\begin{figure}
\centering
\begin{tabular}{@{\hspace{-1mm}}c}
\includegraphics[width=1\linewidth]{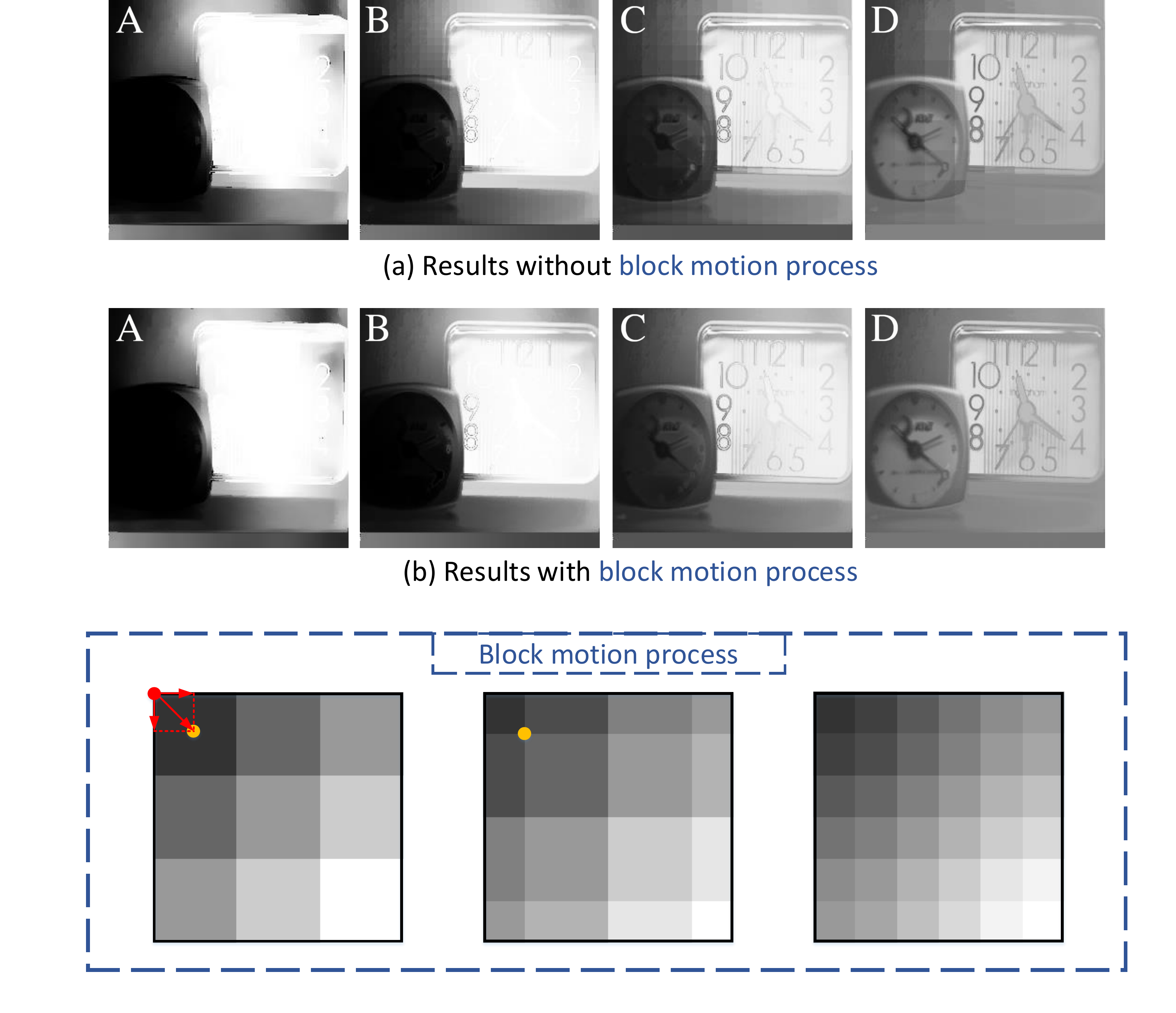}
\\
\end{tabular}
\caption{{\small
Demonstration of the results $\textbf{W}'_i$ with different settings.
(a) and (b) are results without and with the block motion process respectively.
Four columns, A ,B, C and D, are results with $\sigma_{xy} = 3, 8, 16$ and $ 32 $ respectively
The `Block motion process' block illustrates the process to eliminate the blocking artifacts.
}}
\label{fig:shift}
\end{figure}

\textbf{The fast bilateral solver:}
In the fast bilateral solver \cite{solver}, the pre-estimations $\cal{M}$ serves as the filter input which need to be smoothed.
The multi-focus images $\cal{I}$ are utilized as the reference images that guide which lines to be preserved.
Meanwhile, the confidence maps $\cal{C}$ are adopted to measure the {reliability of filter input $\cal{M}$.}

In order to preserve the lines in $\cal{I}$ and smooth $\cal{M}$, an optimization problem is constructed \cite{solver} which includes two terms, a data-fidelity term and a reference image based smoothness term, \emph{i.e.}
\begin{equation}
\begin{split}
\small{
\underset{\textbf{W}'_i} {\arg \min}~\sum_{\textbf{r}} \textbf{C}_i(\textbf{r})(\textbf{W}'_i(\textbf{r})-\textbf{M}_i(\textbf{r}))^2+
\frac{\lambda}{2} \sum_{\textbf{r},\textbf{k}} \hat{B}_i({\textbf{r},\textbf{k}})(\textbf{W}'_i(\textbf{r})-\textbf{W}'_i(\textbf{k}))^2}
\label{equ:solver_min}
\end{split}
\end{equation}
where $\textbf{W}'_i$ is the $i^{th}$ edge-preserving smoothed result. $\textbf{r}$ and $\textbf{k}$ are utilized to represent pixel $(x_r,y_r)$ and $(x_k,y_k)$ respectively in this subsection for simplify.
$\textbf{W}'_i(\textbf{r})$ is a scalar representing the value at pixel $\textbf{r}$ in map $\textbf{W}'_i$.

\emph{The data-fidelity term} $\textbf{C}_i(\textbf{r})(\textbf{W}'_i(\textbf{r})-\textbf{M}_i(\textbf{r}))^2$ help minimize the squared residual between filter input $\cal{M}$ and output $\cal{W'}$ while weighted by confidence map $\cal{C}$.
A bigger weight is given to a reliable region of the input $\cal{M}$ compared with that given to an unreliable region.

\emph{The reference image based smoothness term}
 $\frac{\lambda}{2} \sum_{\textbf{r},\textbf{k}}{}$
$ \hat{B}_i({\textbf{r},\textbf{k}})(\textbf{W}'_i(\textbf{r})-\textbf{W}'_i(\textbf{k}))^2$ tried to minimize the difference between any two pixels, $\textbf{r}$ and $\textbf{k}$, in $\textbf{W}'_i$ weighted by $\hat{B}_i({\textbf{r},\textbf{k}})$. 
Specifically, $\hat{B}_i({\textbf{r},\textbf{k}})$ is a bistochastized version of a bilateral affinity matrix which measures the similarity between $\textbf{r}$ and $\textbf{k}$ in $\textbf{I}_i$ in terms of spatial position and intensity, \emph{i.e.}
\begin{align}
\tiny{\hspace{-3mm}
\hat{B}_{\textbf{r},\textbf{k}} = exp(
-\frac{(\textbf{I}_i\textbf{(r)}-\textbf{I}_i\textbf{(k)})^2}{2\sigma_{in}^2}
-\frac{||[~\textbf{I}_i^x { \textbf{(r)}},\textbf{I}_i^y \textbf{(r)}]-[\textbf{I}_i^x \textbf{(k)},\textbf{I}_i^y \textbf{(k)}]||^2}{2\sigma_{xy}^2}
)
\label{equ:solver_B}
}
\end{align}
where $\textbf{I}_i(\textbf{r})$ and $[\textbf{I}_i^x { \textbf{(r)}},\textbf{I}_i^y \textbf{(r)}]$ represents intensity and the spatial position of pixel $\textbf{r}$ in $\textbf{I}_i$. 
$\sigma_{xy}$ and $\sigma_{in}$ are the parameters control the extent of the spatial and intensity support of the filter respectively. 

Specifically, the weight $\hat{B}_i({\textbf{r},\textbf{k}})$ is big if $\textbf{r}$ and $\textbf{k}$ have similar intensities or are spatially close in $\textbf{I}_i$.
Thus the big weights given to the similar intensity help minimize the difference in \textbf{r} and \textbf{k} in $\textbf{W}_i$, and results in edge-preserving in lines and smoothness in plane regions. 
Besides, the closer pixels may have higher correlation compared with not related ones in the minimization.
To speed up, the fast bilateral solver \cite{solver} treats $\hat{B}$ as a ``splat / blur / slice" procedure so that the optimization process can be solved in the bilateral-space and be reduced into a simple least-square problem.
It is proved that the fast bilateral solver is fast, robust and has been successfully generalized to many new domains, such as stereo, depth super resolution, colorization and semantic segmentation \cite{solver}.



The fusion results are highly correlated with different settings of the bilateral solver.
First, directly applying the fast bilateral solver may lead to obvious block artifact as shown in the first row of Fig. \ref{fig:shift} (a).
Specifically, the block size is equal to $\sigma_{xy}$. 
This is because $\sigma_{xy}$ helps measure the spatial similarity of the filter inputs.
In the implementation, the input is non-overlappingly divided into several blocks for measuring the spatial difference.
%
%
Pixels belonging to the same block and two neighboring block would have a big difference in $\hat{B}$ and result in the block artifact of the filter output.

In order to eliminate the blocking artifact, the block motion process is proposed as illustrated in the `Block motion process' of Fig. \ref{fig:shift}.
In this example, the input image, image (a), is of size $6\times 6$.
In order to measure the spatial similarity in (\ref{equ:solver_B}), the filter input is non-overlappingly divided into blocks of size $\sigma_{xy} \times \sigma_{xy} = 2 \times2$ beginning from the left-up corner. Thus the result shown in (a) suffers from the block artifact from left-up corner.
To eliminate the blocking artifact using block motion process, the fast bilateral solver is processed again and get another filter output as (b). 
This bilateral solver process is repeated for $\sigma_{xy} = 2$ times according to the block sizes. 
During each process, the beginning point of dividing blocks shifted one-pixel in each direction from the red point in (a) to the yellow point in (b).
The final fusion result is obtained by averaging over all the filter outputs as shown in Fig. \ref{fig:shift} (c).
The filter output of `clock' with block motion process is shown in (b) of Fig. \ref{fig:shift}.
 Benefiting from the block motion process, the block artifacts have been effectively removed.

Besides, the choice of $\sigma_{xy}$ has influence on the filter outputs. 
As shown in Fig. \ref{fig:shift}, $\sigma_{xy}$ corresponds to the respective filed.
%
%
In the image fusion task, the estimation results may be not accurate in some boundary regions if $\sigma_{xy}$ is too small.
Besides, the speed of the bilateral solver is highly depended on $\sigma_{xy}$. 
Thus $\sigma_{xy}$ cannot be too big.
To increase the operation speed and maintain the performance, $\sigma_{xy} = 8$ in the implementation. Besides, $\lambda = 64$.

\textbf{Normalization:}
Via the fast bilateral solver \cite{solver}, the edge-preserving smoothed result $\cal{W'}$ are calculated.
Next, the normalization is proceed to get the final focus-level maps $\cal{W}$ as weights.
 
At each position $(x,y)$,
all values in  $N$ $\textbf{W}_i'(x,y), i\in\{1,2,\dots, N\}$ are normalized to sum as 1 first.
Next, a non-linear transformation, sigmoid,  with mean = 0.5,  $\sigma$ = 40, is employed.
In this way, most values in $\cal{W}$ would concentrated around the maximum value $1$ and minimum value $0$, which would help improve the fusion results.

\section{Experiment}
\label{sec:experiment}
Several experiments were conducted to evaluate the performance of the proposed method on $25$ pairs of source images subjectively and objectively.
First, experimental configurations are introduced in Sec. \ref{sec:Config}, including multi-focus source images and objective evaluation metrics.
Moreover, component analysis results are reported in Sec. \ref{Sec:EX_CM}.

\begin{figure}[t!]
\begin{tabular}
{c@{\hspace{-5mm}}C{30mm}@{\hspace{+1mm}}C{30mm}@{\hspace{+1mm}}C{30mm}}
&\includegraphics[width =1\linewidth]{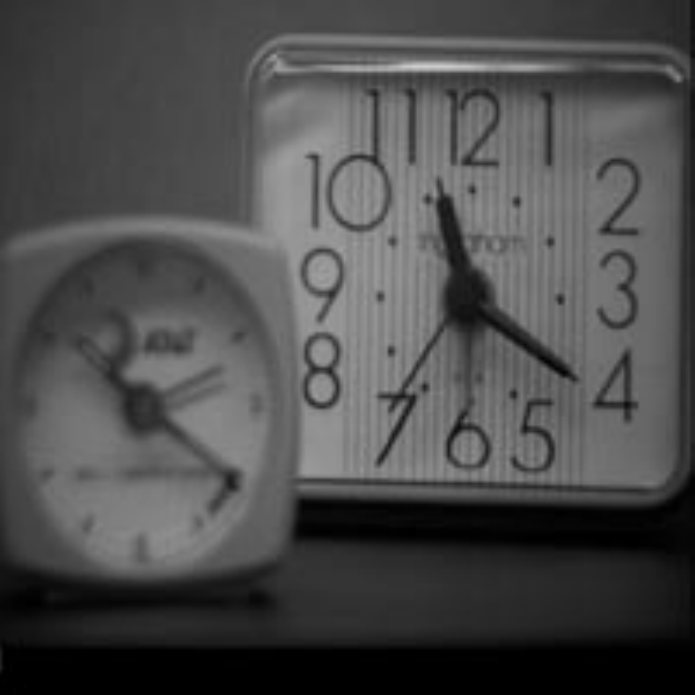}
&\includegraphics[width =1\linewidth]{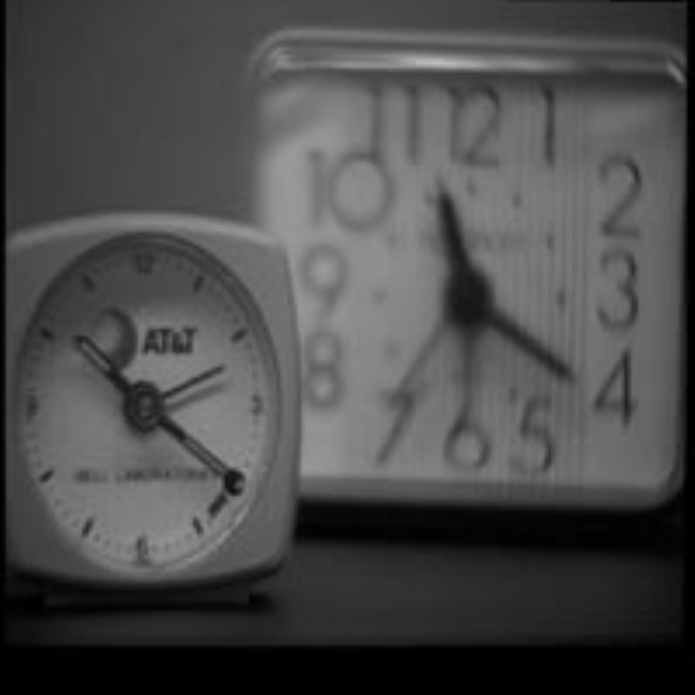}
&\includegraphics[width =1\linewidth]{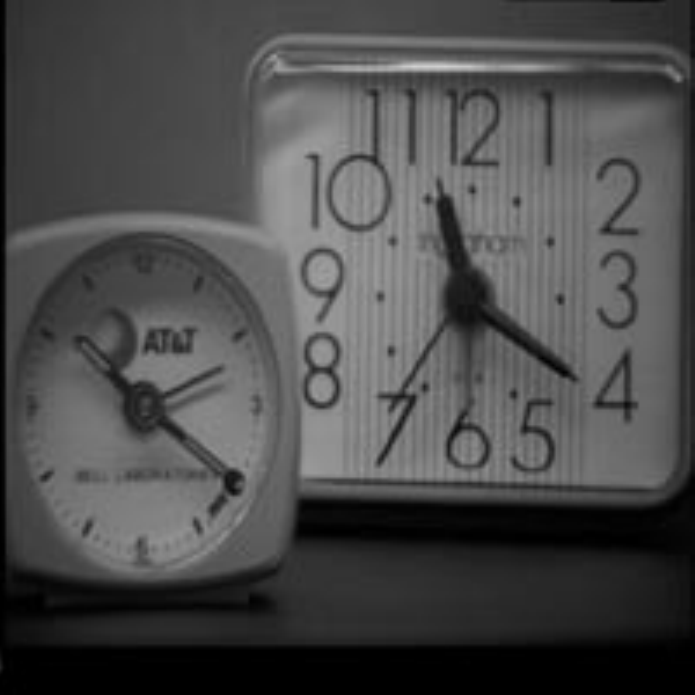}
\\
&\includegraphics[width =1\linewidth]{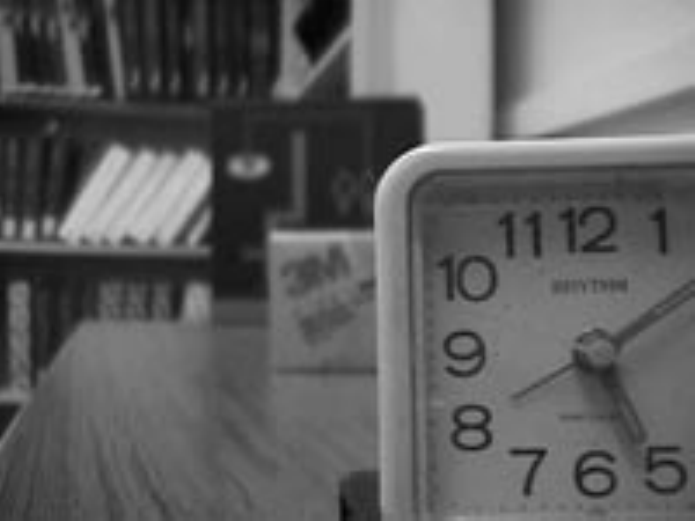}
&\includegraphics[width =1\linewidth]{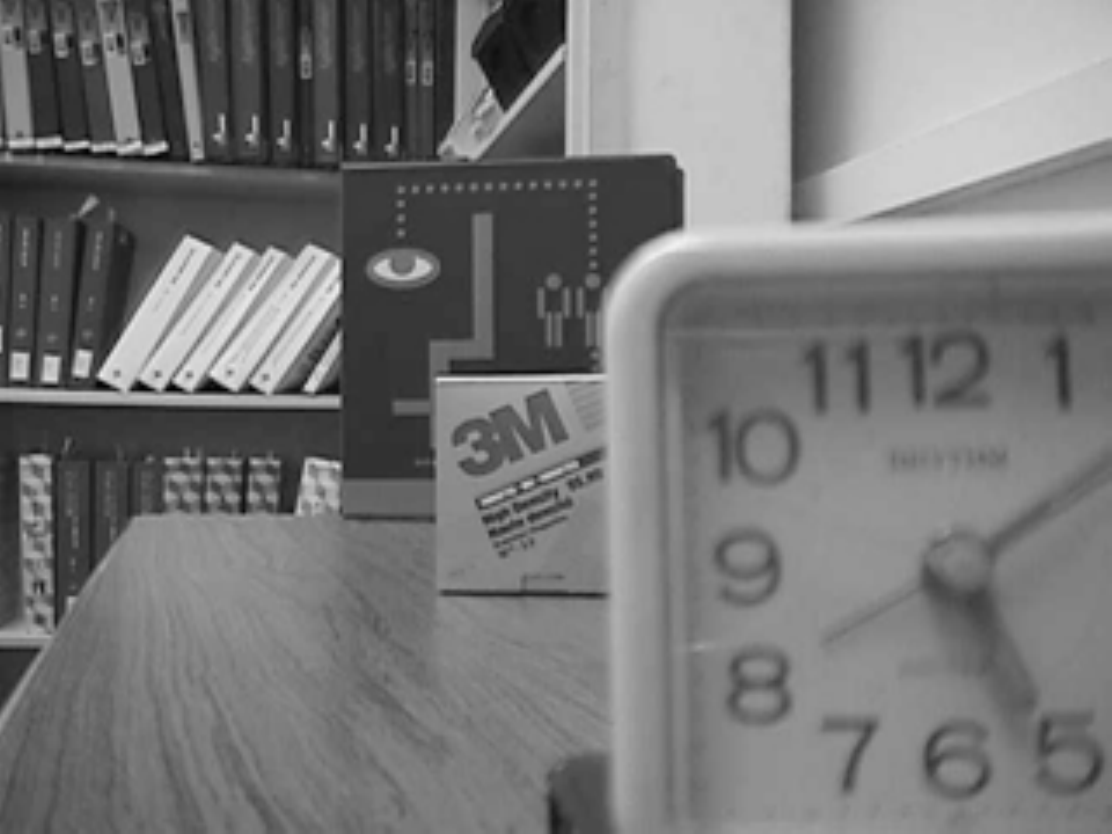}
&\includegraphics[width =1\linewidth]{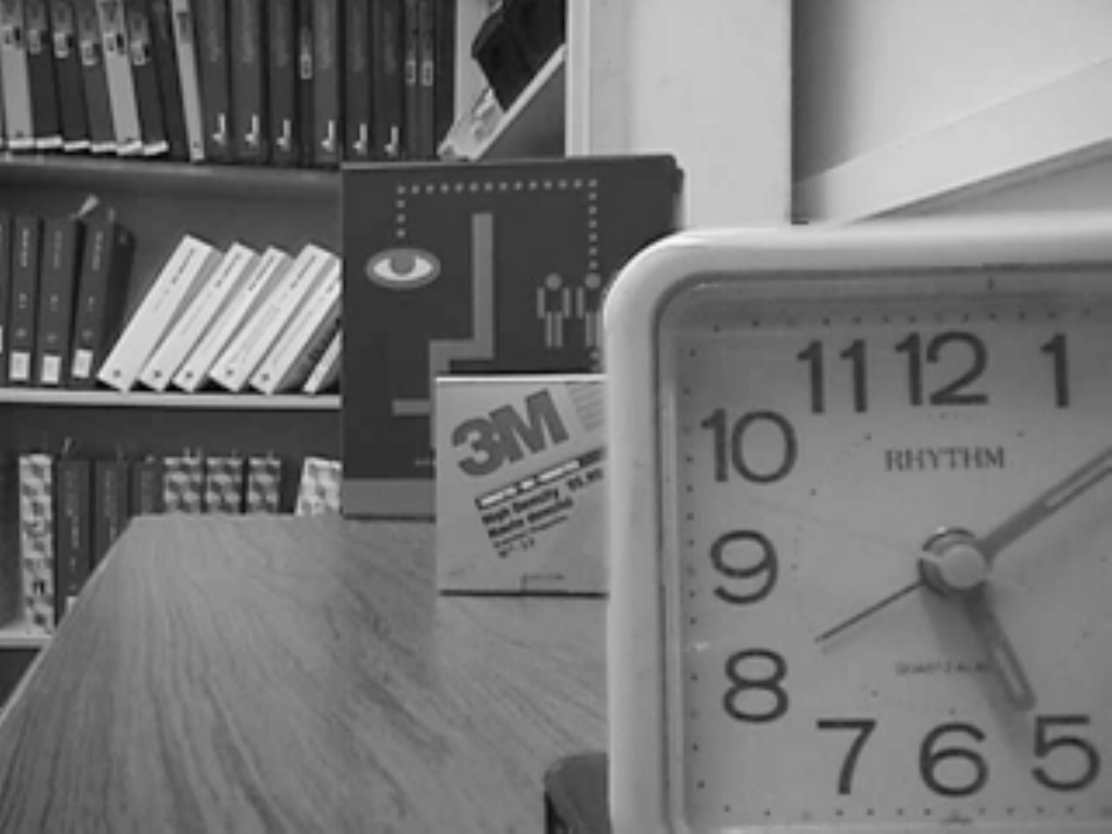}
\\
&\includegraphics[width =1\linewidth]{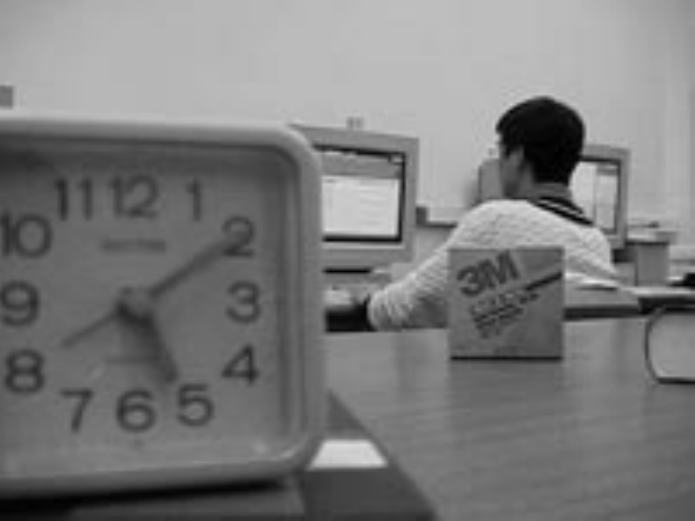}
&\includegraphics[width =1\linewidth]{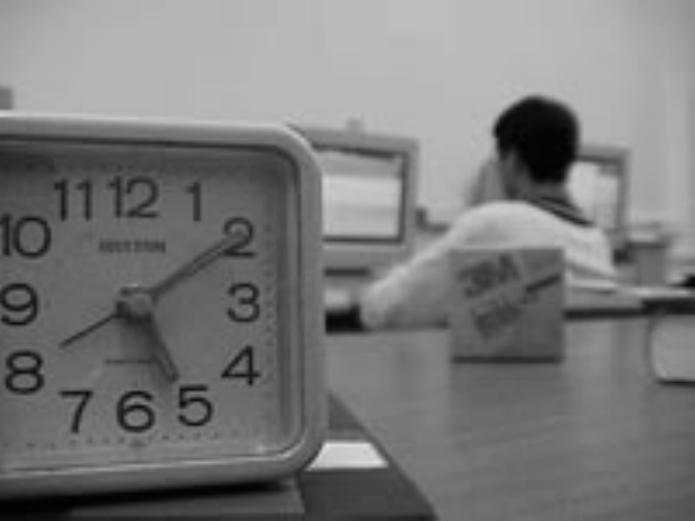}
&\includegraphics[width =1\linewidth]{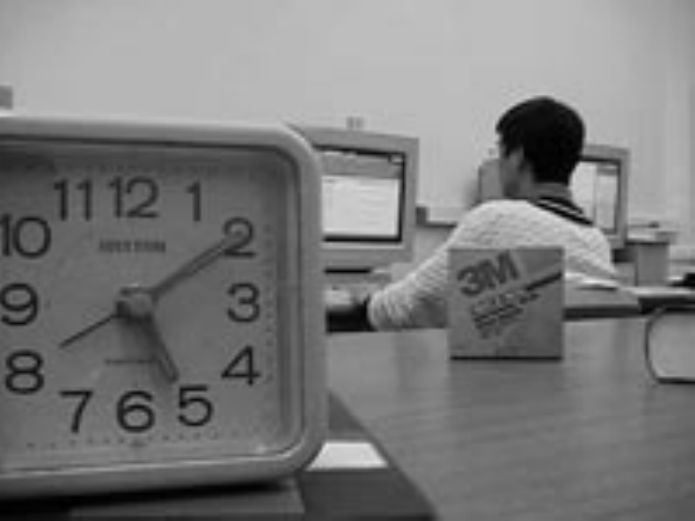}
\\
&\includegraphics[width =1\linewidth]{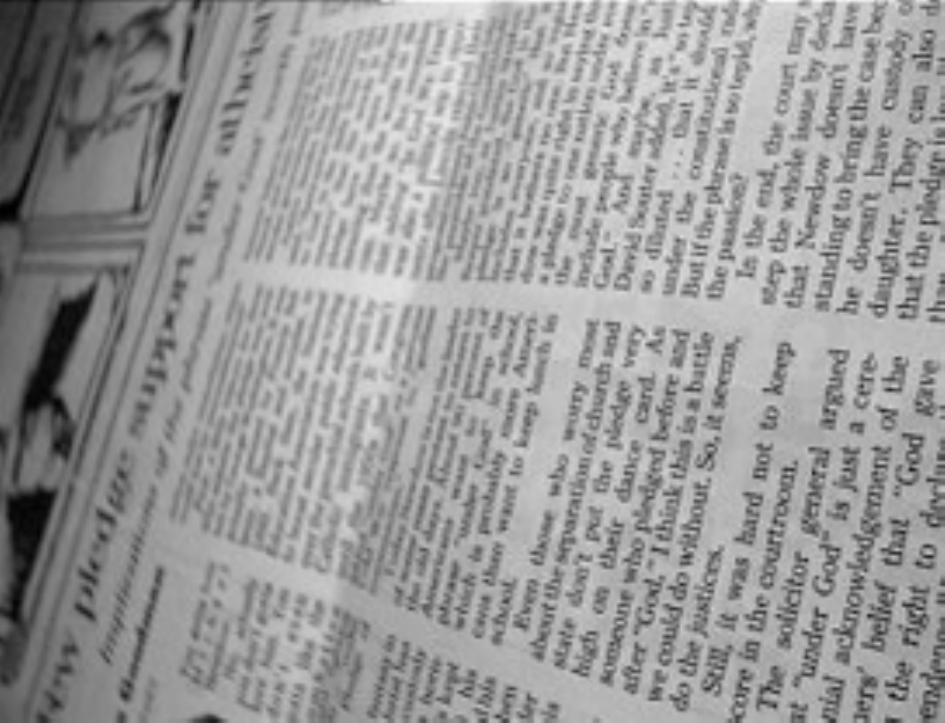}
&\includegraphics[width =1\linewidth]{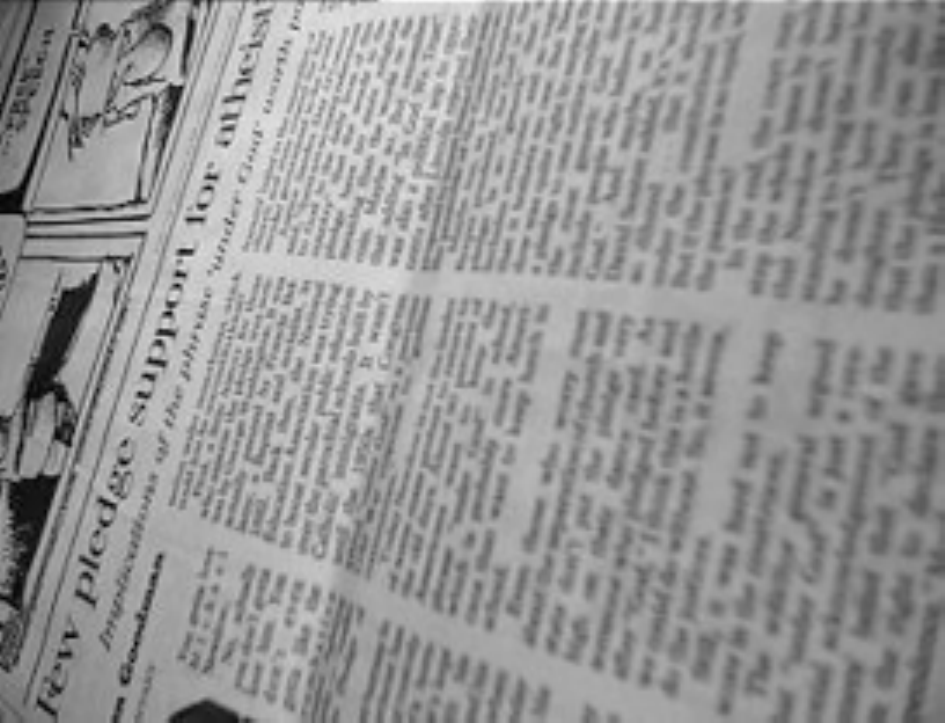}
&\includegraphics[width =1\linewidth]{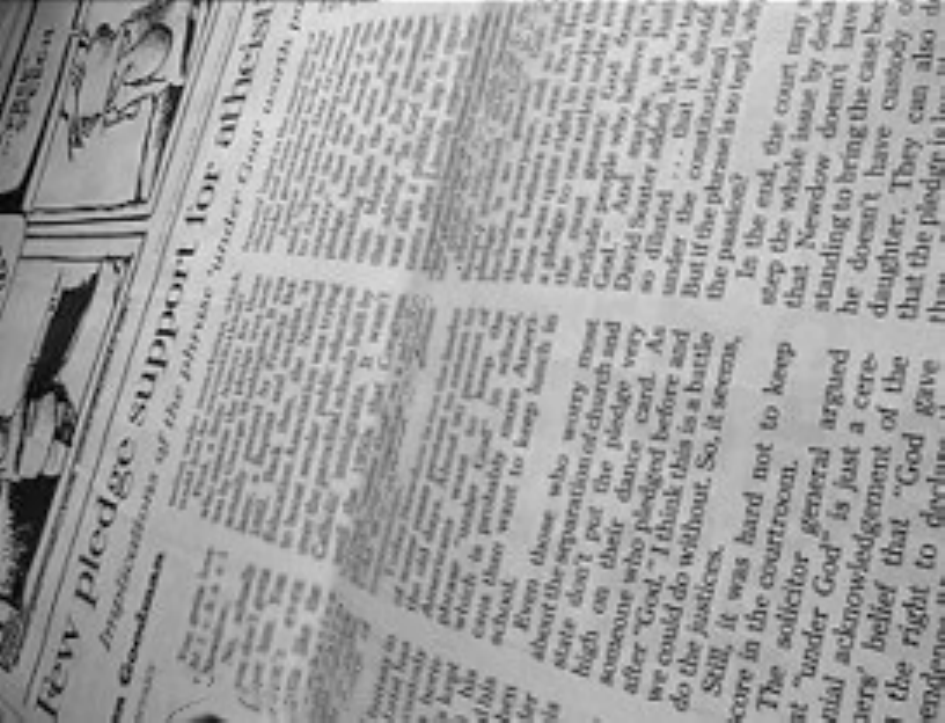}
\\
&\includegraphics[width =1\linewidth]{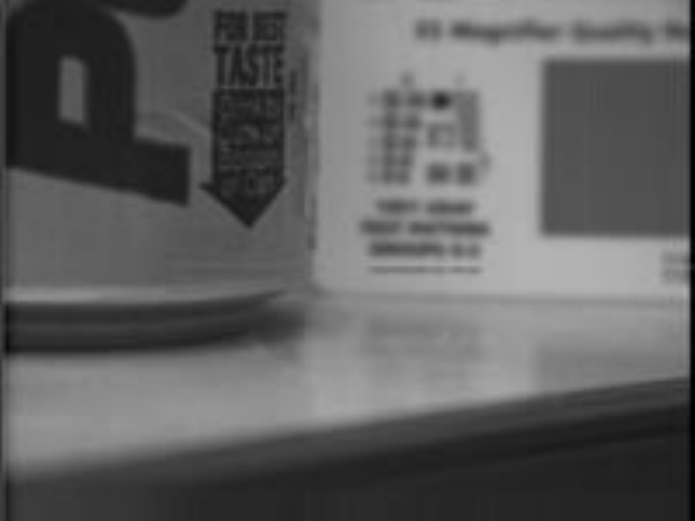}
&\includegraphics[width =1\linewidth]{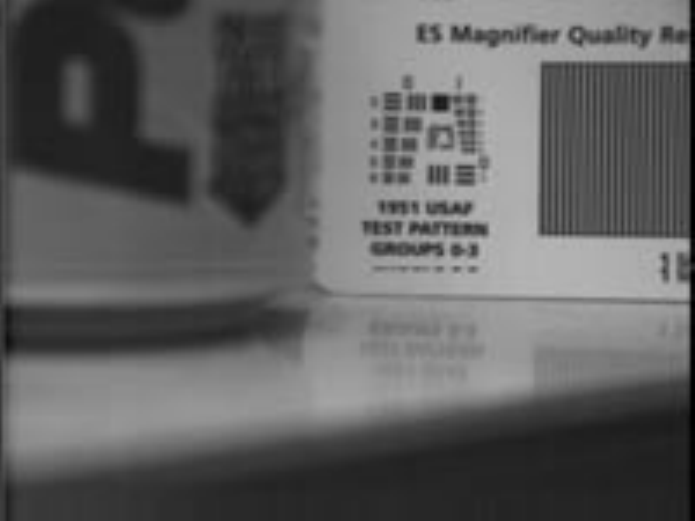}
&\includegraphics[width =1\linewidth]{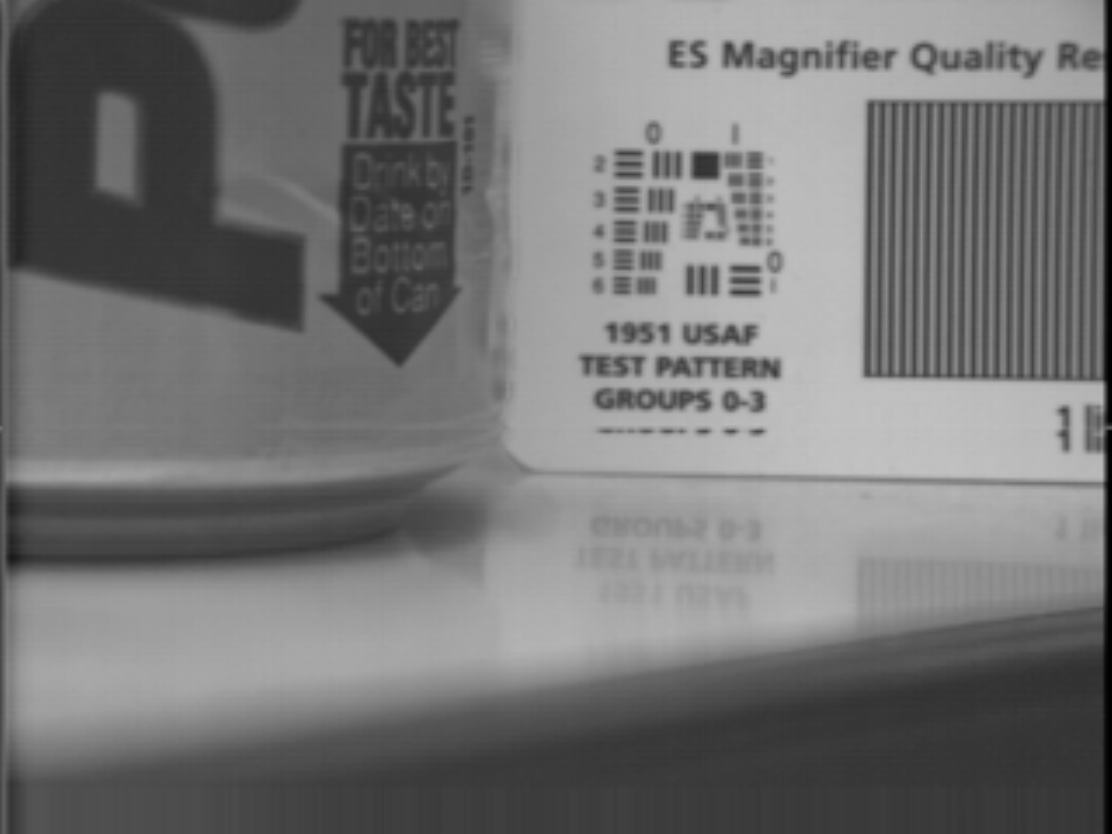}
\\
&$\textbf{I}_1$
&\vspace{0mm}$\textbf{I}_2$
&\vspace{0mm}$\textbf{F}$
 \\
\end{tabular}
\caption{{\small
The multi-focus source images $\textbf{I}_1, \textbf{I}_2$ and the fusion results $\textbf{F}$ by the proposed BS-QEBIF method. These images are `clock', `disk', `lab', `paper' and `pepsi' from top to bottom.
}}
\label{fig:fusionResults}
\vspace{-2mm}
\end{figure}

\subsection{Experimental configurations}
\label{sec:Config}
\subsubsection{Multi-focus source images}
$25$ pairs of source images were utilized for the evaluation.
The first $5$ pairs are commonly used images pairs, `clock', `pepsi', `lab', `paper' and `disk', as shown in Fig. \ref{fig:fusionResults}. 
The remaining ones are from the Lytro multi-focus dataset \cite{Lytro}. 
These image pairs vary in multiple aspects, which provides a good representation of various situations encountered in practice.

\subsubsection{Objective evaluation metrics}
In this work, three widely used evaluation metrics, $Q_{G}$ \cite{Qabf}, $Q_{NMI}$ \cite{NMI} and NCIE \cite{NCIE}, were adopted.
Default parameters provided by respective metrics were used and bigger score indicates better performance. 
These metrics evaluate the fusion results in multiple respects and provide a solid evaluation.


\begin{enumerate}
\item $Q_{G}$ \cite{Qabf} measures the success of edge information transferred from the multi-focus source images to the fusion results. 
Take $\cal{I}$$ = \{ \textbf{I}_1, \textbf{I}_2\}$ as an example, $Q_{G}$ can be defined as,
\begin{align}
\hspace{-6mm}Q_{G} = \frac{\sum_{x=1}^{N_1}\sum_{y=1}^{N_2}(Q^{\textbf{I}_1 \textbf{F}}(x,y)\tau^{\textbf{I}_1}(x,y)+Q^{{\textbf{I}_2} \textbf{F}}(x,y)\tau^{\textbf{I}_2}(x,y))}{\sum_{x=1}^{N_1}\sum_{y=1}^{N_2}(\tau^{\textbf{I}_1}(x,y)+\tau^{\textbf{I}_2}(x,y))}
\end{align}
where $N_1$ and $N_2$ represents the number of coordinates in spatial domain.
$Q^{\textbf{I}_1 \textbf{F}}$ is the sum of  $Q^{\textbf{I}_1 \textbf{F}}_g$ and $Q^{\textbf{I}_1 \textbf{F}}_o$ of $\textbf{I}_1$, where $Q^{\textbf{I}_1 \textbf{F}}_g$ and $Q^{\textbf{I}_1 \textbf{F}}_o$ represent element-wise product of edge strength and orientation preservation value respectively. 
$Q^{\textbf{I}_2 \textbf{F}}$ is defined similarly for $\textbf{I}_2$.
$\tau^{\textbf{I}_1}$ and $\tau^{\textbf{I}_2}$ represents weights for $\textbf{I}_1$ and $\textbf{I}_2$ respectively.

\item Normalized mutual information $Q_{NMI}$ \cite{NMI} measures the success of mutual information transferred from the multi-focus source images to the fusion results based on the information theory. 
The normalized mutual information is more stable than traditional mutual information metric \cite{MI}. $Q_{NMI}$ can be defined as,  
\begin{align}
Q_{NMI}= 2[\frac{MI( \textbf{I}_1, \textbf{F})}{H({\textbf{I}_1})+H(\textbf{F})}+\frac{MI( \textbf{I}_2, \textbf{F})}{H({\textbf{I}_2})+H(\textbf{F})}]
\end{align}

where $H(\textbf{I}_1), H(\textbf{I}_2)$ and $H(\textbf{F})$ represent the marginal entropy of source image $\textbf{I}_1,  \textbf{I}_2$ and $\textbf{F}$ respectively.
$MI( \textbf{I}_1, \textbf{F}) = H(\textbf{I}_1)+H(\textbf{F})-H(\textbf{I}_1,\textbf{F})$ measures the mutual information between $\textbf{I}_1$ and $\textbf{F}$.
$H(\textbf{I}_1,\textbf{F})$ is the joint entropy between $\textbf{I}_1$ and $\textbf{F}$.
$MI( \textbf{I}_2, \textbf{F})$ can be calculated similarly.

\item Nonlinear Correlation Information Entropy NCIE \cite{NCIE} measures the nonlinear correlation of source images $\cal{I}$ and the fused image $\textbf{F}$.
\begin{align}
NCIE = 1- H_{R} = 1+ \sum_{k = 1}^{K} \frac{\lambda_k^{R}}{K} log_{256}{ \frac{\lambda_k^{R}}{K}}
\end{align}
where $H_{R}$ represents the nonlinear joint entropy.
$K = N+1$ is the number of concerned variables including the source images $\cal{I}$ and the fusion result $\textbf{F}$. 
$R$ is the nonlinear correlation matrix of the concerned variables, including $\cal{I}$ and $\textbf{F}$.
$\lambda_k^{R}$ is the $k^{th}$ eigenvalue of $R$.
\end{enumerate}
\begin{table*}
\begin{center}
\caption{Performance of different fusion methods on $25$ pairs of multi-focus images.}
\label{table:sub}
    \hspace{-5.5mm}
\begin{tabular}
   {|C{25mm}|C{15mm}|C{15mm}|C{15mm}||C{25mm}|C{15mm}|C{15mm}|C{15mm}|}\hline
\multicolumn{4}{|c||}{{clock}}&\multicolumn{4}{c|}{disk}\\\hline
 Method& $Q_{G}$\cite{Qabf}&$Q_{NMI}$ \cite{NMI}&NCIE \cite{NCIE} & Method&$Q_{G}$ \cite{Qabf}&$Q_{NMI}$ \cite{NMI}&NCIE \cite{NCIE} \\ \hline
AVE       & 0.6163          & 1.0058          & 0.8283          & AVE       & 0.5018          & 0.8139          & 0.8194          \\
LAP \cite{44}         & 0.7129          & 1.0030          & 0.8301          & LAP \cite{44}         & 0.6769          & 0.8649          & 0.8242          \\
FSD \cite{45}       & 0.6808          & 0.8480          & 0.8235          & FSD \cite{45}       & 0.6511          & 0.7411          & 0.8190          \\
GRP \cite{46}       & 0.6828          & 0.8493          & 0.8236          & GRP \cite{46}       & 0.6577          & 0.7441          & 0.8191          \\
RAP  \cite{47}       & 0.5999          & 0.9975          & 0.8295          & RAP  \cite{47}       & 0.5974          & 0.8875          & 0.8247          \\
NSCT-SR \cite{MST-SR}   & 0.7302          & 1.0623          & 0.8327          &  NSCT-SR \cite{MST-SR}   & 0.7061          & 0.9149          & 0.8263          \\
NSCT  \cite{22}       & 0.7282          & 1.0135          & 0.8306          & NSCT  \cite{22}       & 0.7063          & 0.9172          & 0.8264          \\

SR \cite{26}        & 0.7310          & 1.1098          & 0.8351          & 
SR \cite{26}        & 0.7159          & 1.0527          & 0.8338          \\
 GF  \cite{GF}         & 0.7434          & 1.1123          & 0.8357          &  GF  \cite{GF}         & \textbf{0.7224}          & 1.0539          & 0.8338          \\
MWGF \cite{28} \textit{(m)} & 0.7396          & 1.1405          & 0.8364          & MWGF \cite{28} \textit{(m)} & 0.7135          & 1.0298          & 0.8327          \\
MWGF \cite{28} \textit{(r)} & 0.7430          & 1.1982          & 0.8396          & MWGF \cite{28} \textit{(r)} & 0.7146          & 1.0681          & 0.8346          \\
IM \cite{37}        & 0.7461          & 1.2119          & \textbf{0.8431} & IM \cite{37}         & 0.7208          & 1.1254          & 0.8382          \\
CBF \cite{38}       & 0.7385          & 1.0969          & 0.8344          & CBF \cite{38}       & 0.7104          & 0.9447          & 0.8277          \\
QEBIF \cite{QEBIF}     & 0.7460          & 1.1675          & 0.8381          & QEBIF \cite{QEBIF}     & 0.7190          & 1.0748          & 0.8353          \\ \hline
Ours      & \textbf{0.7518} & \textbf{1.2407} & 0.8425          & Ours      & {0.7223} & \textbf{1.1834} & \textbf{0.8423} \\
\hline\hline
\multicolumn{4}{|c||}{{lab}}&\multicolumn{4}{c|}{paper}\\\hline
 Method& $Q_{G}$\cite{Qabf}&$Q_{NMI}$ \cite{NMI}&NCIE \cite{NCIE} & Method&$Q_{G}$ \cite{Qabf}&$Q_{NMI}$ \cite{NMI}&NCIE \cite{NCIE} \\ \hline
AVE       & 0.5484          & 0.8854          & 0.8214          & AVE       & 0.1424          & 0.2882          & 0.8033          \\
LAP \cite{44}           & 0.7309          & 1.0469          & 0.8329          & LAP \cite{44}          & 0.4768          & 0.2995          & 0.8045          \\
FSD \cite{45}       & 0.7123          & 0.8436          & 0.8243          & FSD \cite{45}       & 0.4514          & 0.2934          & 0.8044          \\
GRP \cite{46}       & 0.7166          & 0.8468          & 0.8244          & GRP \cite{46}       & 0.4595          & 0.2932          & 0.8044          \\
RAP  \cite{47}       & 0.6652          & 1.0508          & 0.8329          & RAP  \cite{47}       & 0.2053          & 0.2970          & 0.8043          \\
 NSCT-SR \cite{MST-SR}   & 0.7501          & 1.0947          & 0.8352          &  NSCT-SR \cite{MST-SR}   & 0.5749          & 0.3109          & 0.8047          \\
NSCT  \cite{22}       & 0.7500          & 1.1055          & 0.8355          & NSCT  \cite{22}       & 0.5778          & 0.3131          & 0.8047          \\

SR \cite{26}        & 0.7590          & 1.1973          & 0.8405          & 
SR \cite{26}        & 0.6457          & 0.7813          & 0.8182          \\
 GF  \cite{GF}         & 0.7639          & 1.1929          & 0.8400          &  GF  \cite{GF}         & 0.6481          & 0.5859          & 0.8109          \\
MWGF \cite{28} \textit{(m)}& 0.7539          & 1.1913          & 0.8400          & MWGF \cite{28} \textit{(m)} & 0.6337          & 0.6559          & 0.8130          \\
MWGF \cite{28} \textit{(r)} & 0.7565          & 1.2263          & 0.8419          &MWGF \cite{28} \textit{(r)} & 0.6567          & 0.7061          & 0.8147          \\
IM \cite{37}         & 0.7627          & 1.2764          & 0.8448          & IM \cite{37}         & 0.6555          & 0.8130          & 0.8191          \\
CBF \cite{38}       & 0.7598          & 1.1195          & 0.8363          & CBF \cite{38}       & 0.5301          & 0.3524          & 0.8053          \\
QEBIF \cite{QEBIF}     & 0.7610          & 1.2264          & 0.8420          & QEBIF \cite{QEBIF}     & 0.6422          & 0.7113          & 0.8150          \\ \hline
Ours      & \textbf{0.7660} & \textbf{1.2895} & \textbf{0.8457} & Ours      & \textbf{0.6601} & \textbf{0.8565} & \textbf{0.8212}
\\
\hline
\hline
\multicolumn{4}{|c||}{{pepsi}}&\multicolumn{4}{c|}{Averaging result of 20 pairs in Lytro dataset \cite{Lytro}}\\\hline
 Method& $Q_{G}$\cite{Qabf}&$Q_{NMI}$ \cite{NMI}&NCIE \cite{NCIE} & Method&$Q_{G}$ \cite{Qabf}&$Q_{NMI}$ \cite{NMI}&NCIE \cite{NCIE} \\ \hline
AVE       & 0.4833          & 0.9082          & 0.8220          & AVE       & 0.3594          & 0.7172          & 0.8166          \\
LAP \cite{44}            & 0.7313          & 1.0126          & 0.8310          & LAP \cite{44}            & 0.7336          & 0.9012          & 0.8280          \\
FSD \cite{45}       & 0.6858          & 0.8417          & 0.8240          & FSD \cite{45}       & 0.6949          & 0.7360          & 0.8205          \\
GRP \cite{46}       & 0.6909          & 0.8436          & 0.8240          & GRP \cite{46}       & 0.6989          & 0.7379          & 0.8206          \\
RAP  \cite{47}       & 0.5835          & 1.0228          & 0.8314          & RAP  \cite{47}       & 0.5613          & 0.8911          & 0.8272          \\
 NSCT-SR \cite{MST-SR}   & 0.7441          & 1.0774          & 0.8340          &  NSCT-SR \cite{MST-SR}   & 0.7511          & 0.9874          & 0.8326          \\
NSCT  \cite{22}       & 0.7449          & 1.0703          & 0.8337          & NSCT  \cite{22}       & 0.7502          & 0.9592          & 0.8311          \\

SR \cite{26}        & 0.7557          & 1.1251          & 0.8367          & 
SR \cite{26}       & 0.7581          & 1.1096          & 0.8402          \\
 GF  \cite{GF}         & 0.7672          & 1.1762          & 0.8392          &  GF  \cite{GF}         & \textbf{0.7622} & 1.0993          & 0.8396          \\
MWGF \cite{28} \textit{(m)} & 0.7581          & 1.1829          & 0.8396          & MWGF \cite{28} \textit{(m)} & 0.7479          & 1.1020          & 0.8402          \\
MWGF \cite{28} \textit{(r)} & 0.7585          & 1.2213          & 0.8419          & MWGF \cite{28} \textit{(r)} & 0.7497          & 1.1264          & 0.8418          \\
IM \cite{37}         & 0.7712          & 1.2834          & 0.8457          & IM \cite{37}         & 0.7575          & 1.1465          & 0.8428          \\
CBF \cite{38}      & 0.7616          & 1.1147          & 0.8358          & CBF \cite{38}       & 0.7570          & 1.0312          & 0.8352          \\
QEBIF \cite{QEBIF}     & 0.7695          & 1.1955          & 0.8407          & QEBIF \cite{QEBIF}     & 0.7550          & 1.1025          & 0.8396          \\ \hline
Ours      & \textbf{0.7748} & \textbf{1.2955} & \textbf{0.8476} & Ours      & 0.7578          & \textbf{1.1762} & \textbf{0.8446}
\\
\hline
    \noalign{\smallskip}
    \end{tabular}
\end{center}
\vspace{-6mm}
\end{table*}

\subsubsection{Image fusion methods for comparison}
 13 traditional and state-of-the-art multi-focus image fusion methods were utilized for comparison, including AVE, LAP \cite{44}, FSD \cite{45}, GRP \cite{46}, RAP \cite{47}, NSCT-SR \cite{MST-SR}, NSCT \cite{22}, SR \cite{26}, GF \cite{GF}, MWGF \cite{28}, IM \cite{37}, CBF \cite{38} and QEBIF \cite{QEBIF}. 
Code of implementing these methods were obtained online or directly from the authors, and their default parameters are used to make a fair comparison.
Especially, MWGF \cite{28} provides two sets of parameters for two cases. 
In the implementation, we provide the results under both assumptions and adopt \textit{(m)} and \textit{(r)} for mis-registered and well registered cases respectively.
Since the proposed method targets on fusing gray images, all color images were converted into gray ones before the fusion process in all methods.

\begin{figure*}[t!]
\begin{tabular}
{@{\hspace{-10mm}}c@{\hspace{+3mm}}c@{\hspace{+3mm}}c@{\hspace{+3mm}}c@{\hspace{+3mm}}c@{\hspace{+3mm}}c}
&\includegraphics[width =0.2\linewidth]{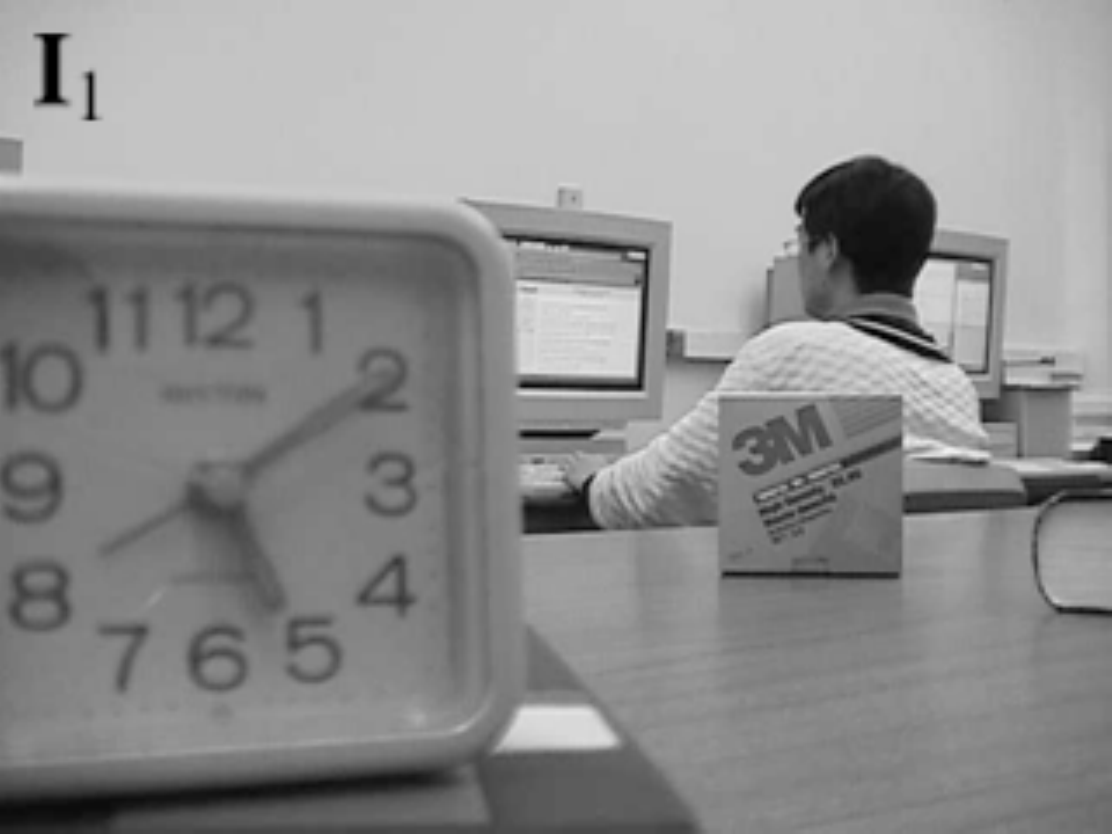}
&\includegraphics[width =0.2\linewidth]{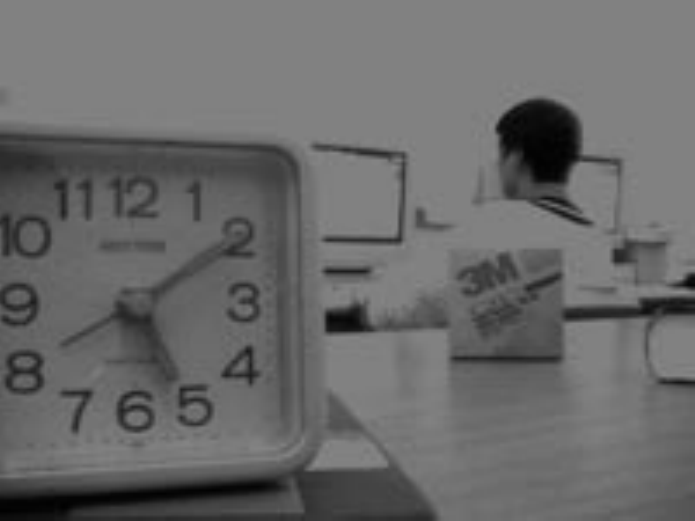}
&\includegraphics[width =0.2\linewidth]{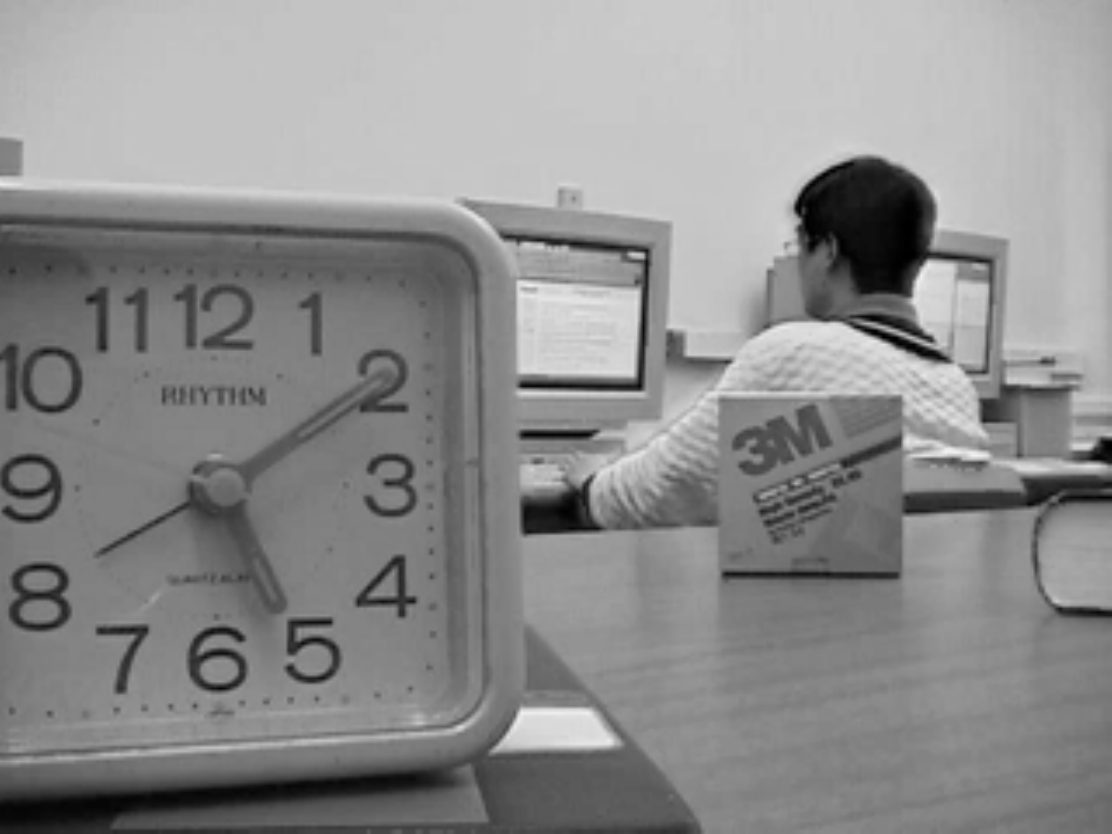}
&\includegraphics[width =0.2\linewidth]{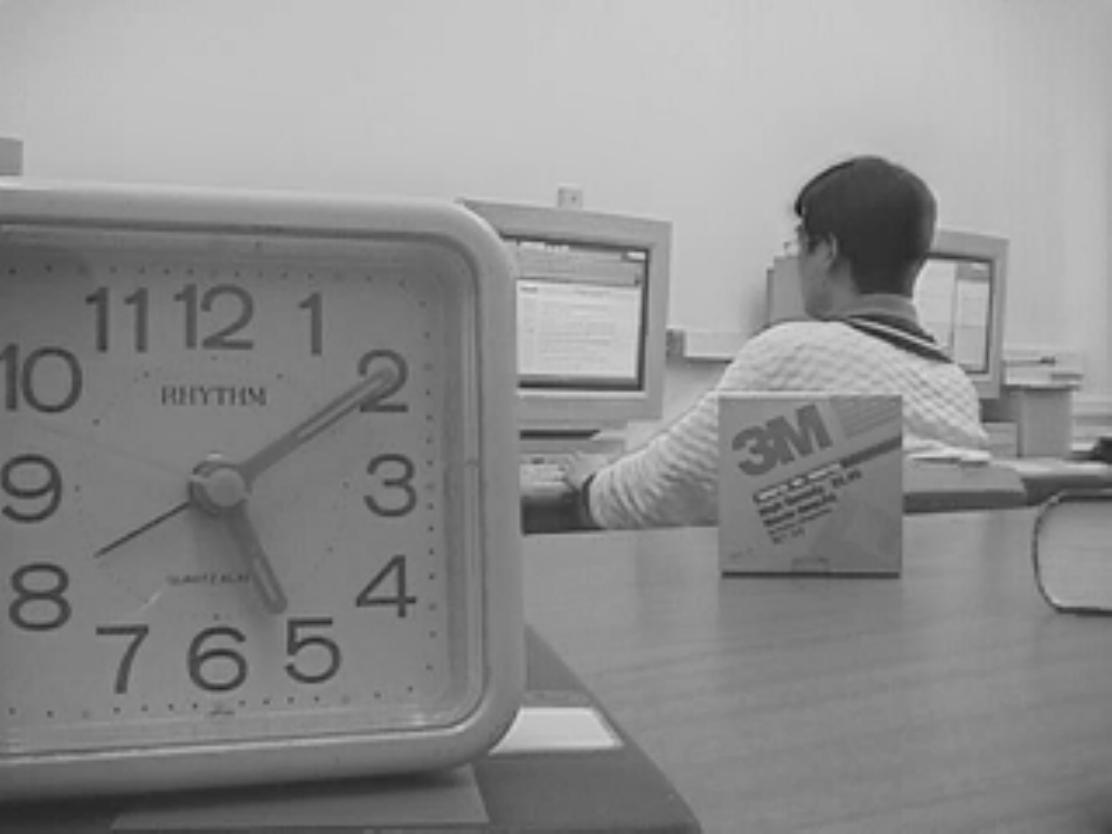}
&\includegraphics[width =0.2\linewidth]{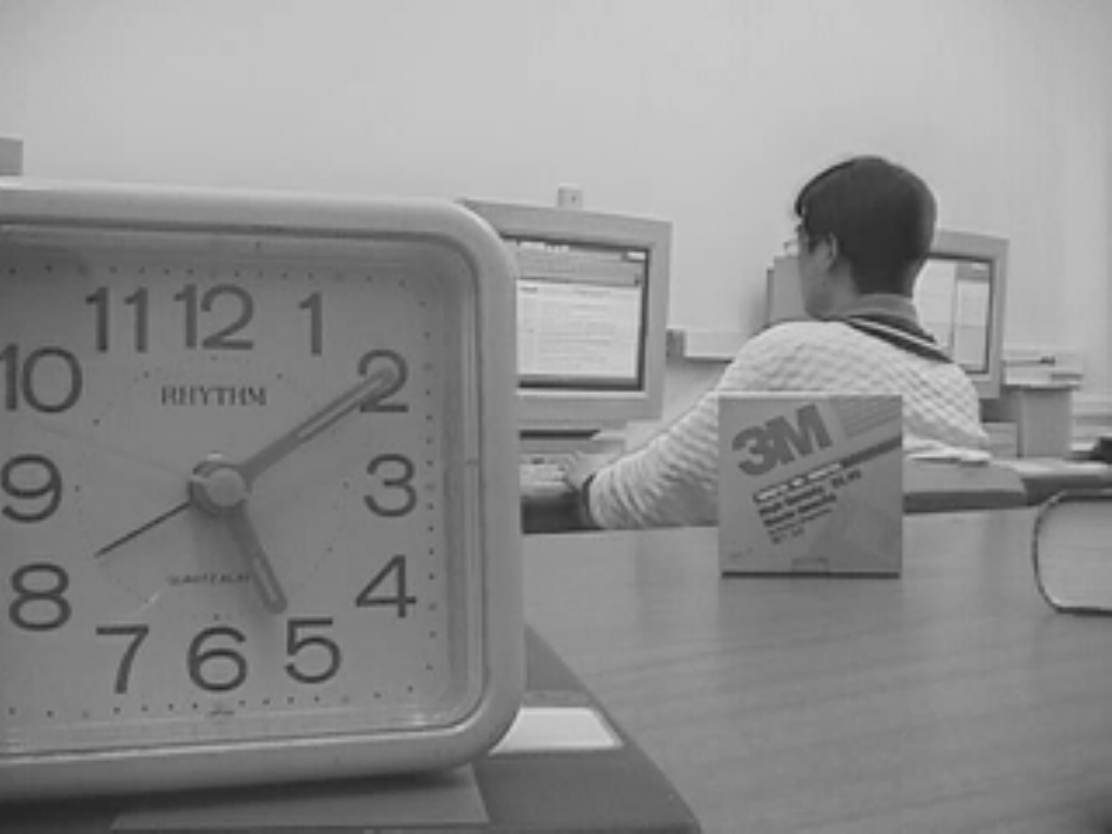}
\\
&\includegraphics[width =0.2\linewidth]{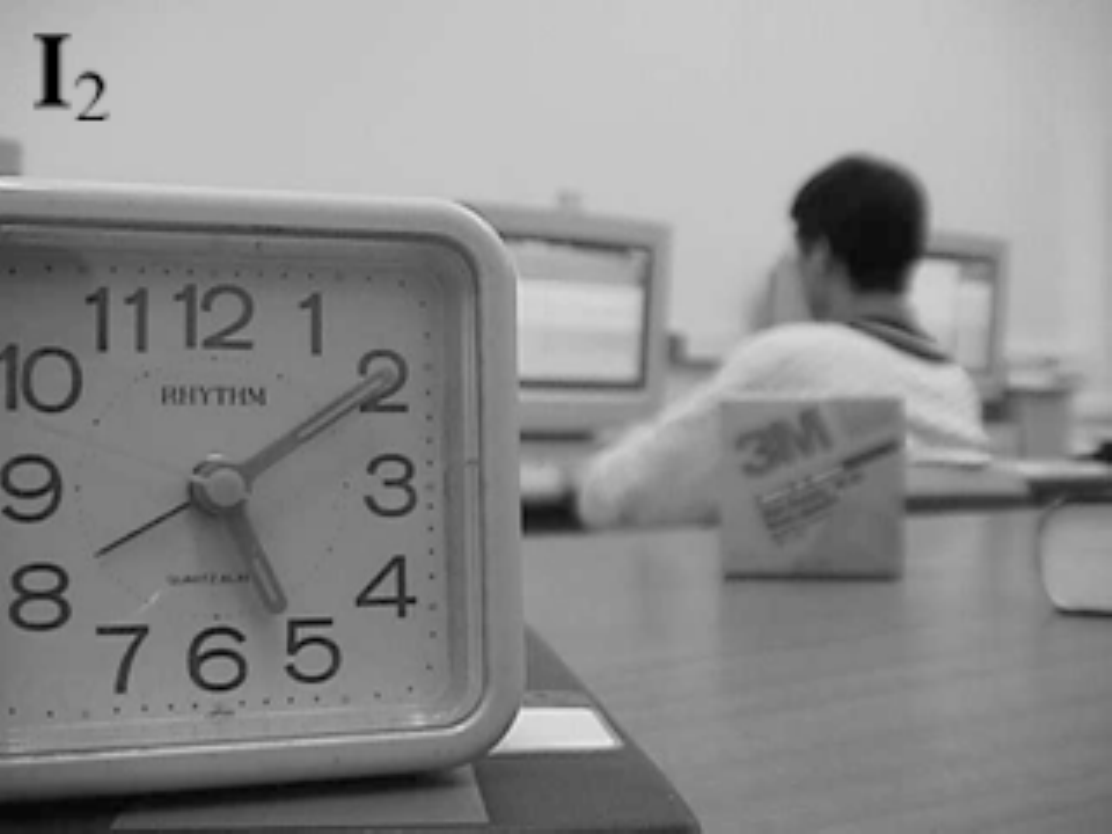}
&\includegraphics[width =0.2\linewidth]{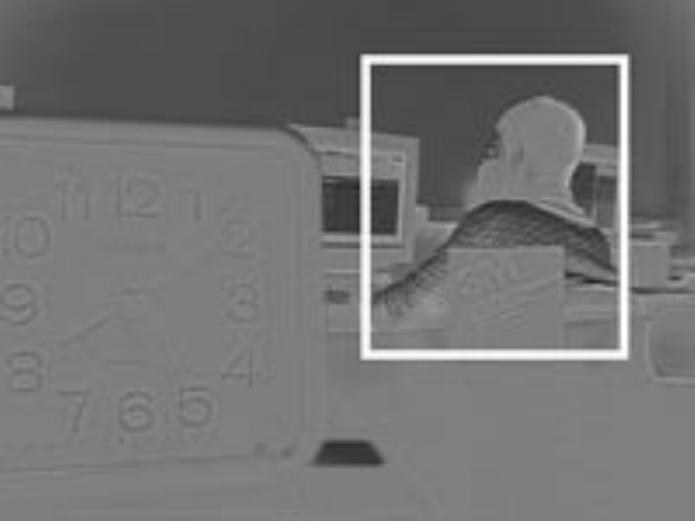}
&\includegraphics[width =0.2\linewidth]{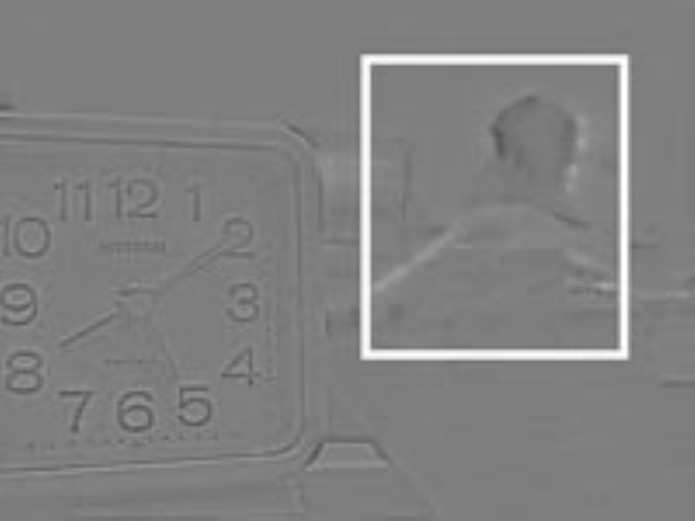}
&\includegraphics[width =0.2\linewidth]{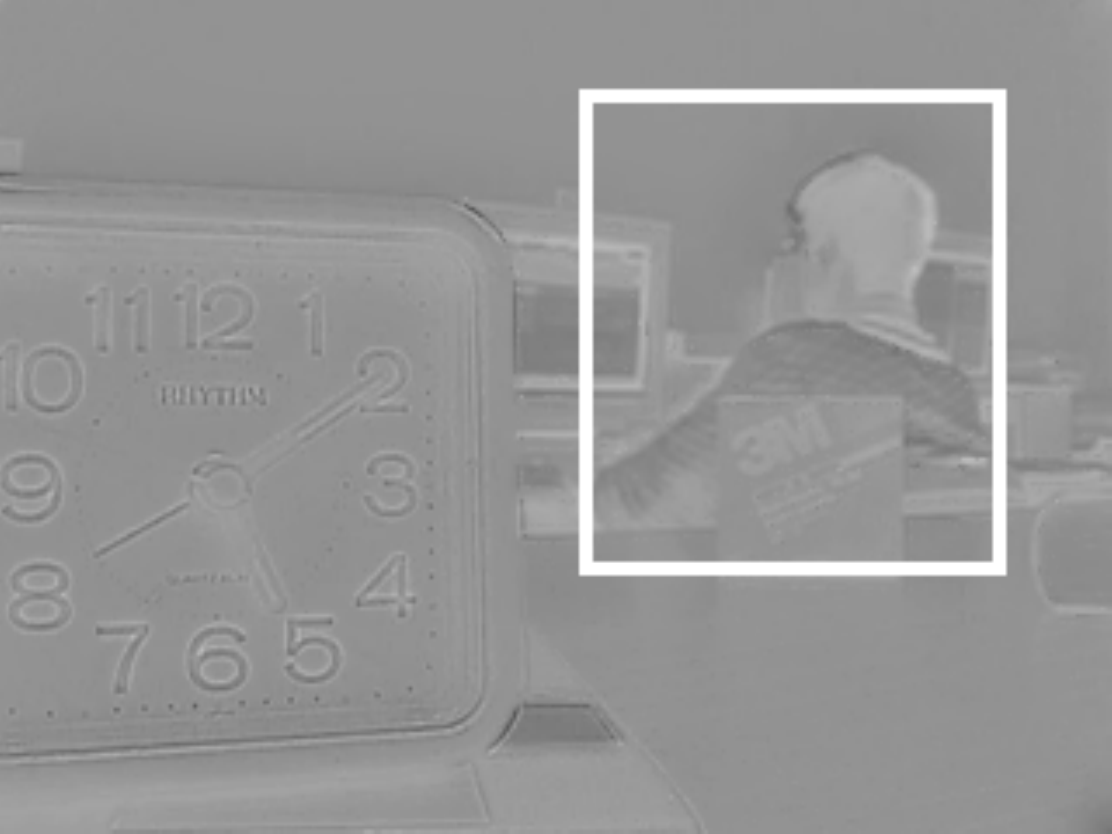}
&\includegraphics[width =0.2\linewidth]{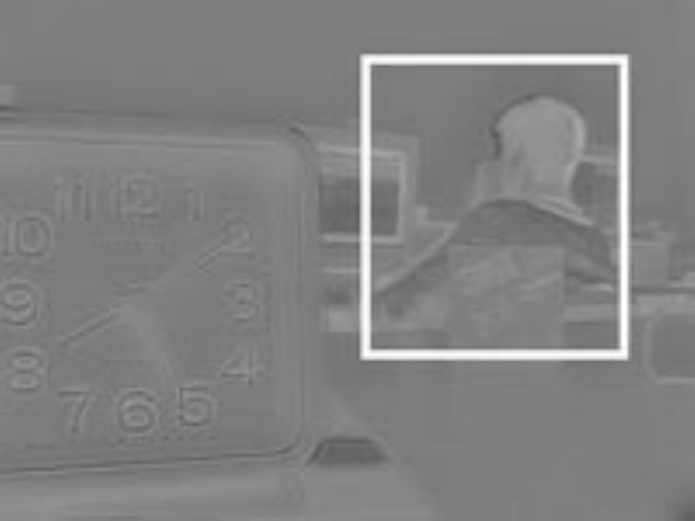}
\\
& (a) Multi-focus images& (b) AVE & (c) LAP \cite{44} & (d) FSD \cite{45} & (e) GRP \cite{46} \\
&\includegraphics[width =0.2\linewidth]{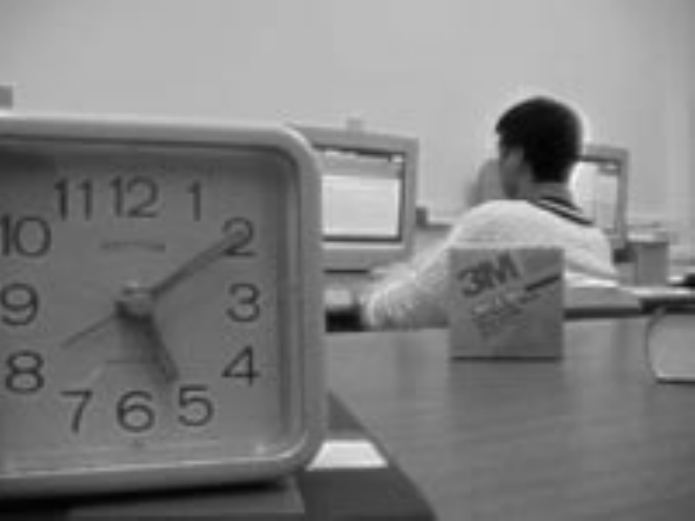}
&\includegraphics[width =0.2\linewidth]{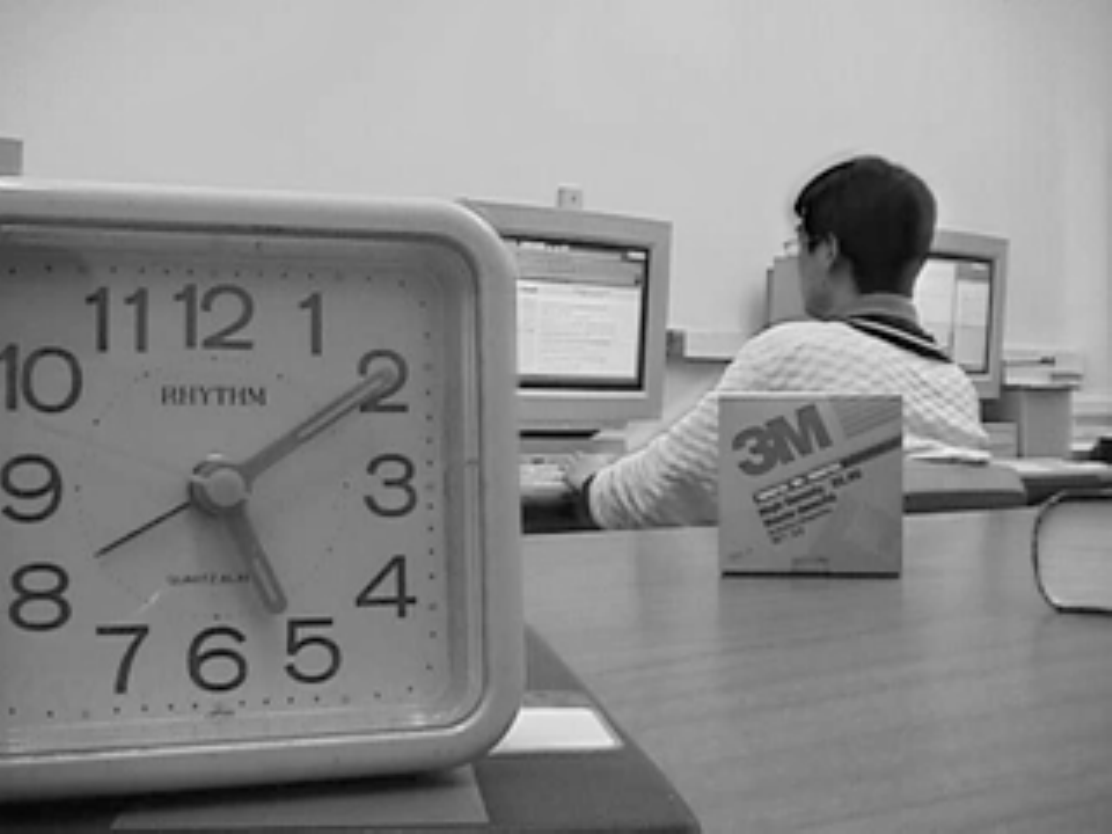}
&\includegraphics[width =0.2\linewidth]{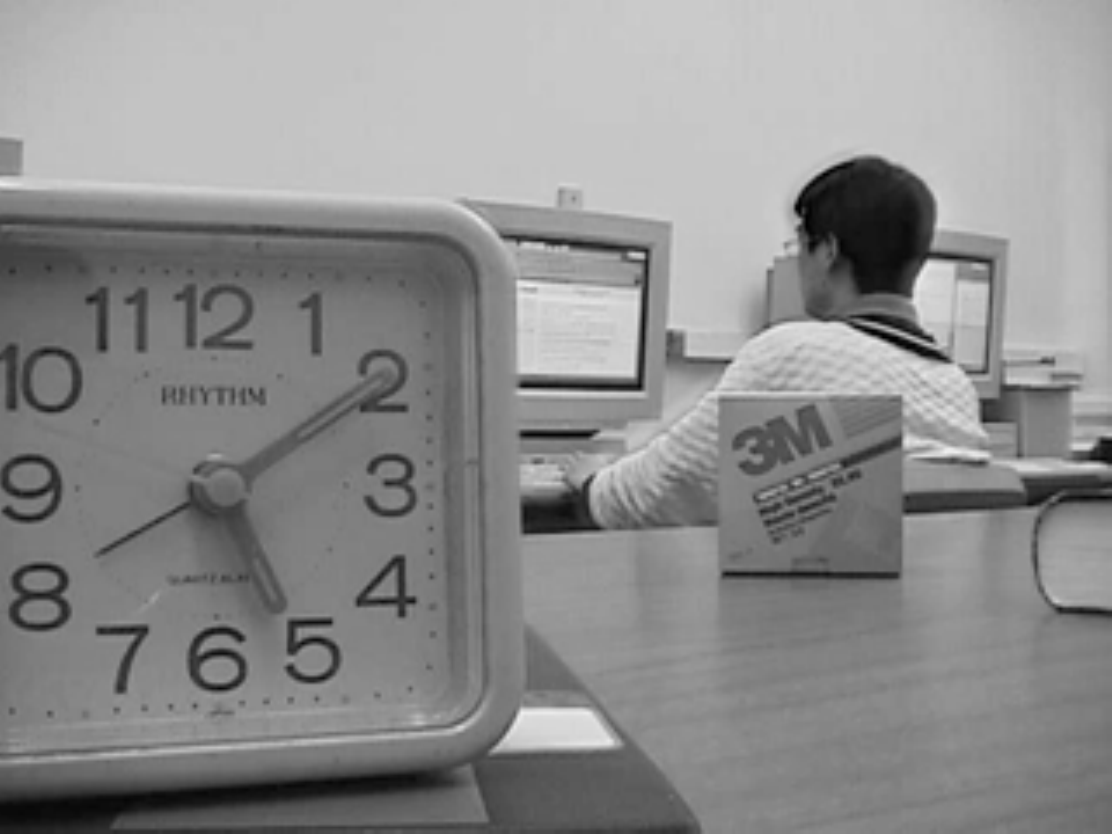}
&\includegraphics[width =0.2\linewidth]{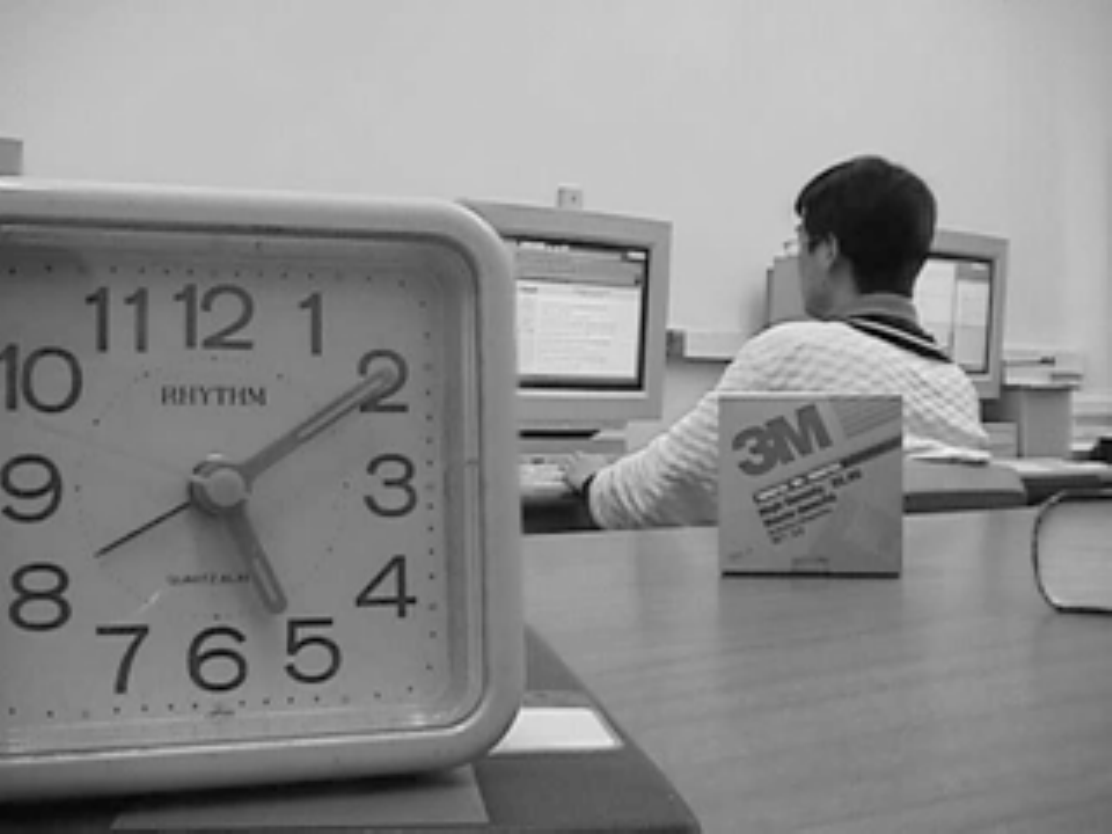}
&\includegraphics[width =0.2\linewidth]{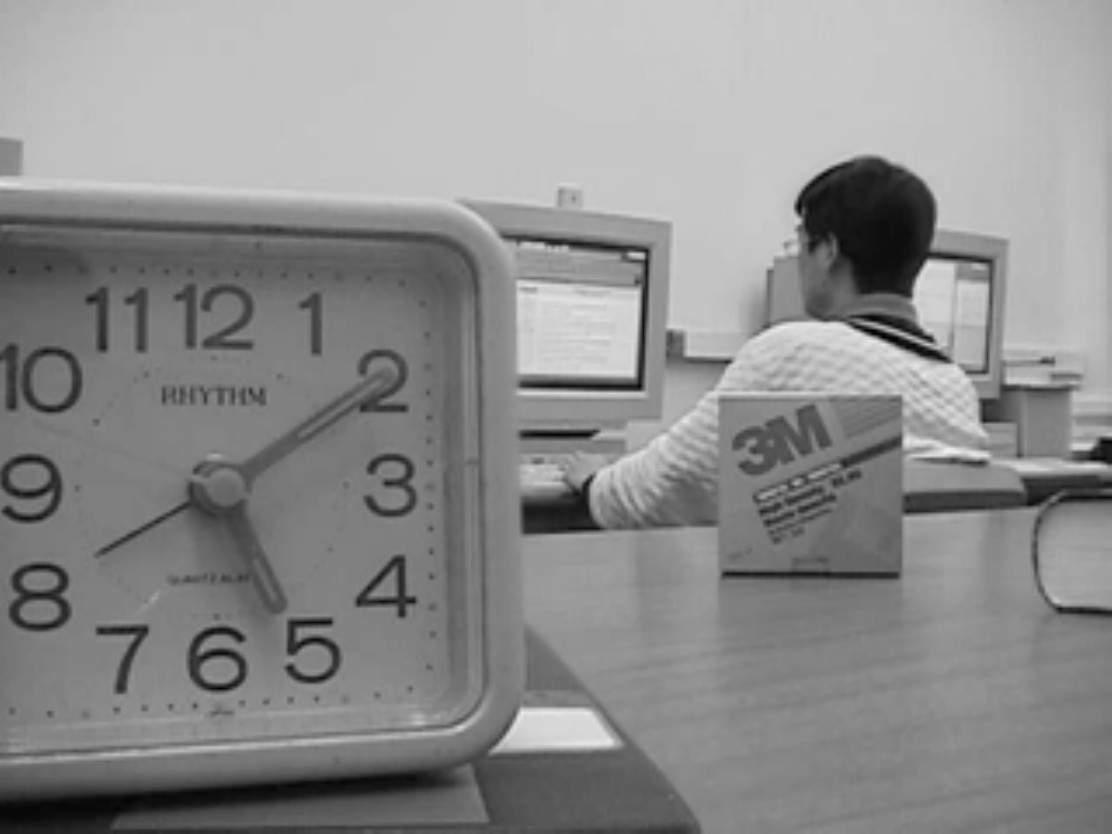}
\\
&\includegraphics[width =0.2\linewidth]{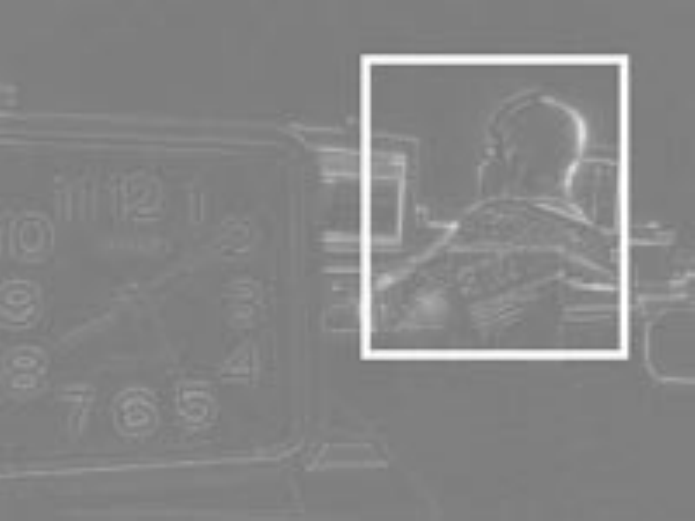}
&\includegraphics[width =0.2\linewidth]{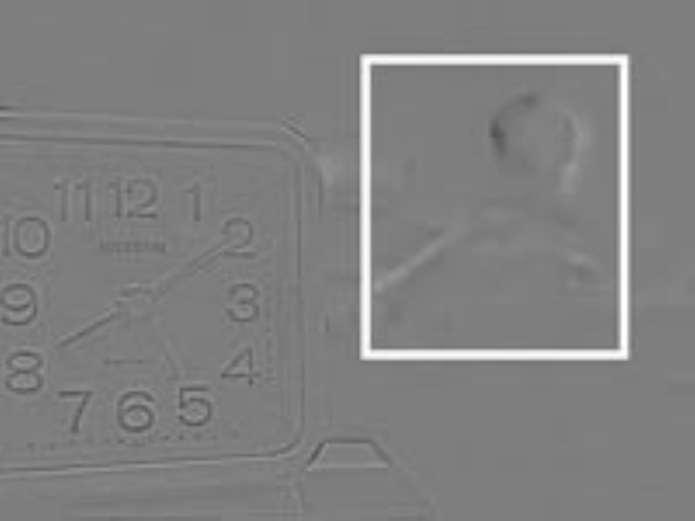}
&\includegraphics[width =0.2\linewidth]{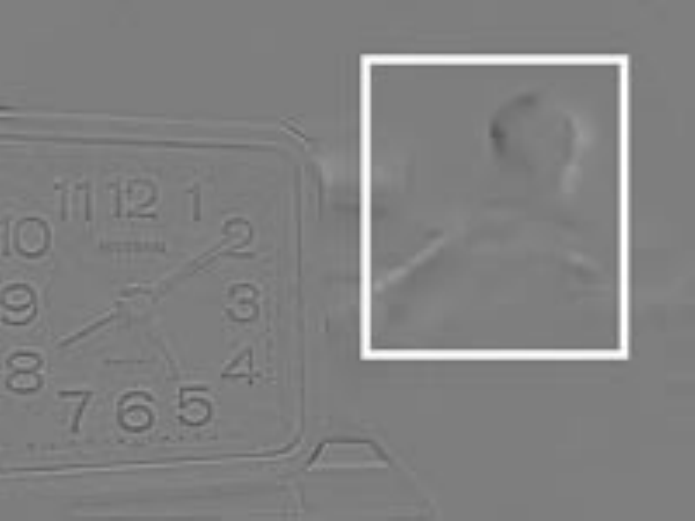}
&\includegraphics[width =0.2\linewidth]{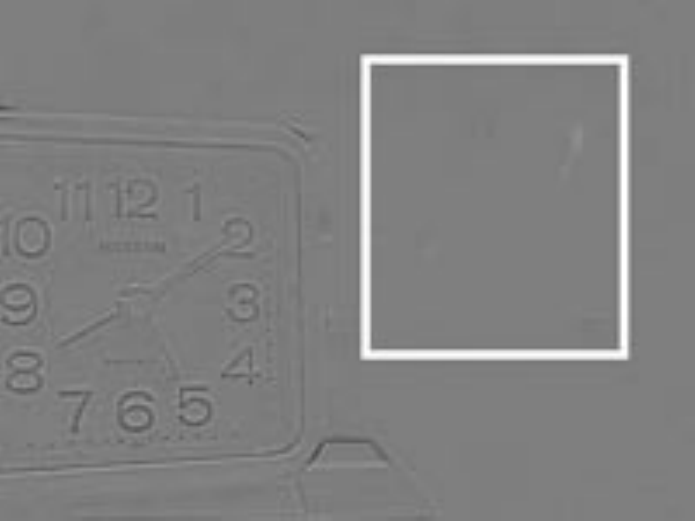}
&\includegraphics[width =0.2\linewidth]{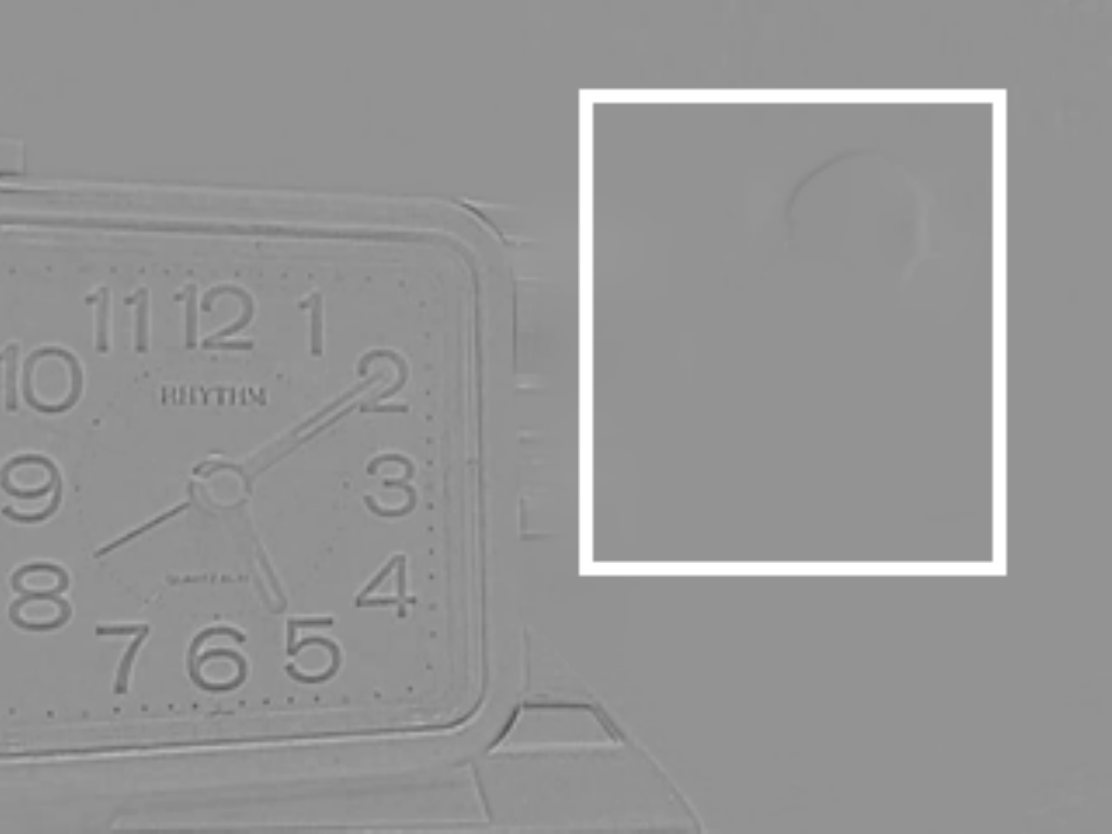}
\\
& (f) RAP \cite{47} & (g) NSCT-SR \cite{MST-SR} & (h) NSCT \cite{22} & (i) SR \cite{26} & (j) GF \cite{GF}\\
&\includegraphics[width =0.2\linewidth]{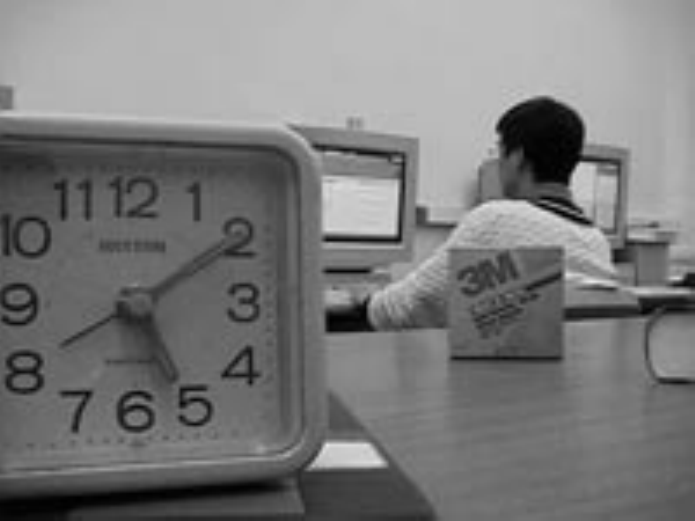}
&\includegraphics[width =0.2\linewidth]{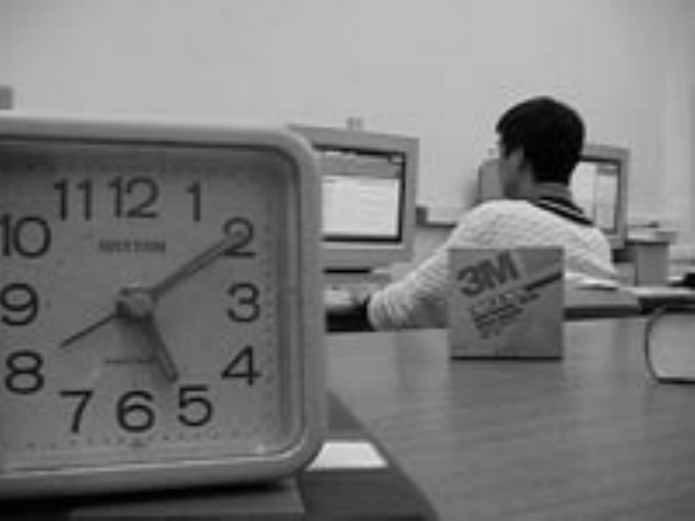}
&\includegraphics[width =0.2\linewidth]{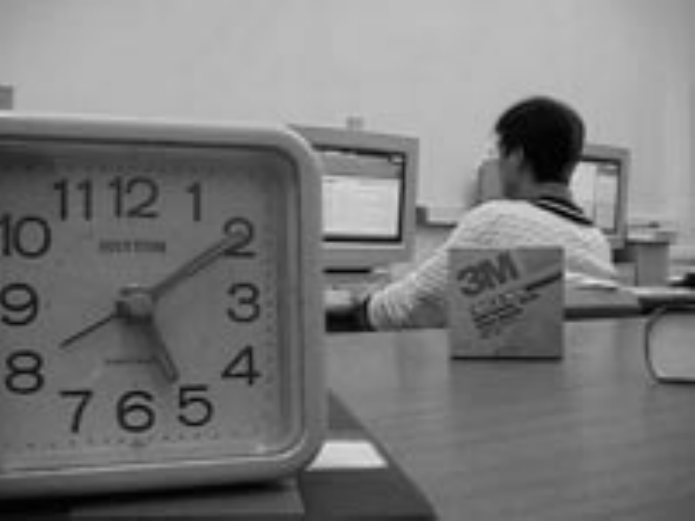}
&\includegraphics[width =0.2\linewidth]{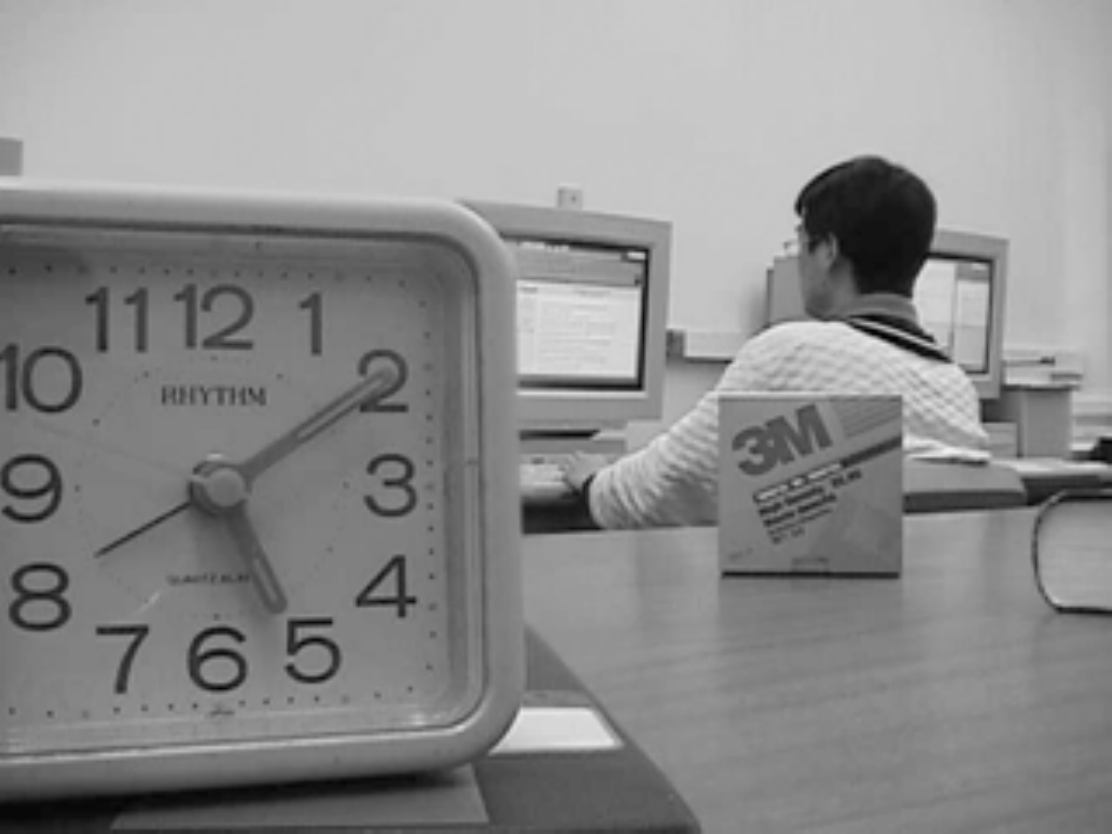}
&\includegraphics[width =0.2\linewidth]{./Figure/lab400}\\
&\includegraphics[width =0.2\linewidth]{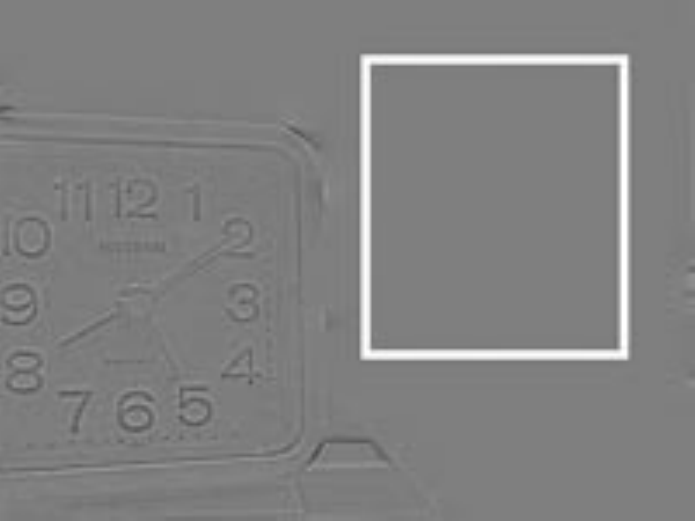}
&\includegraphics[width =0.2\linewidth]{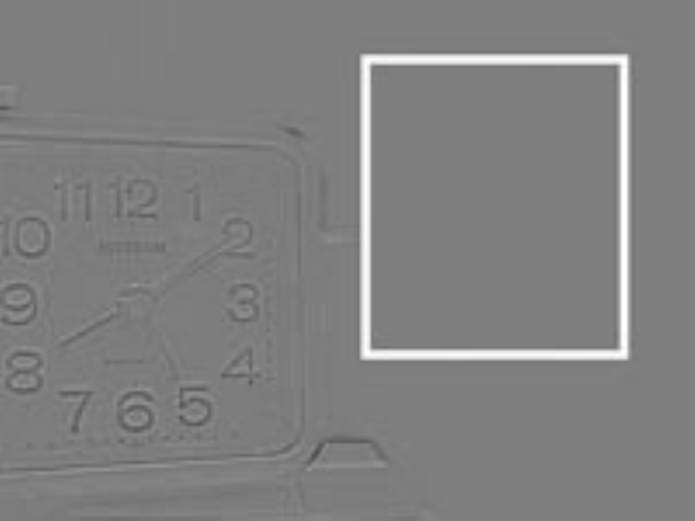}
&\includegraphics[width =0.2\linewidth]{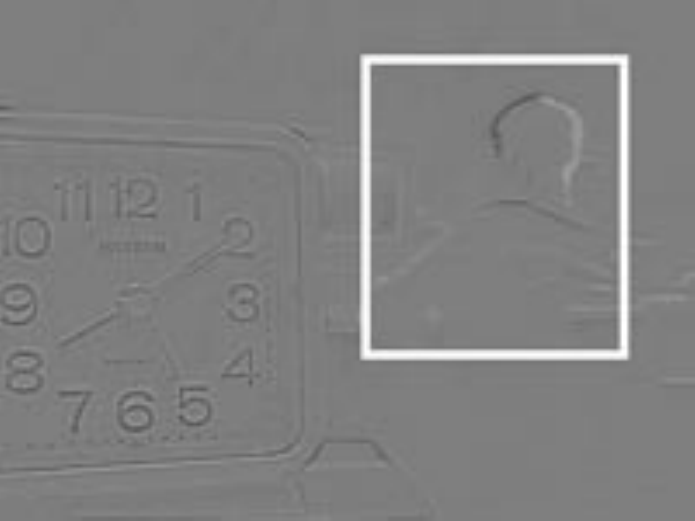}
&\includegraphics[width =0.2\linewidth]{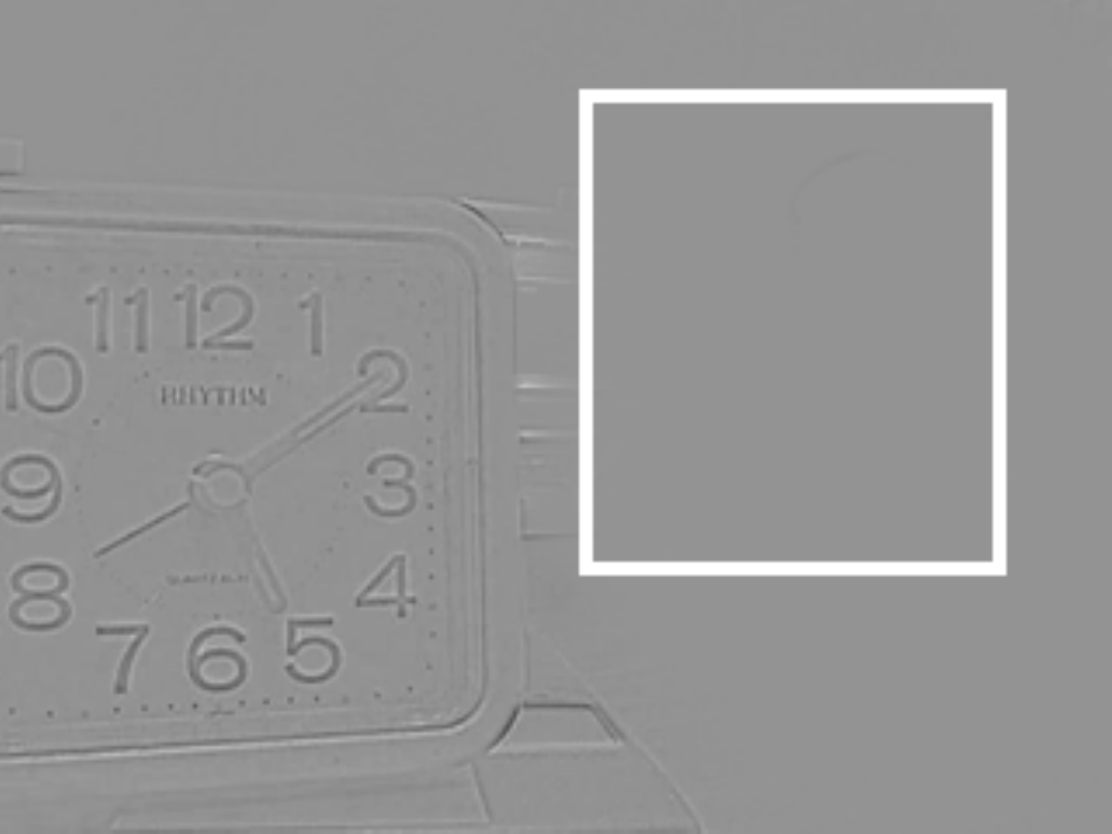}
&\includegraphics[width =0.2\linewidth]{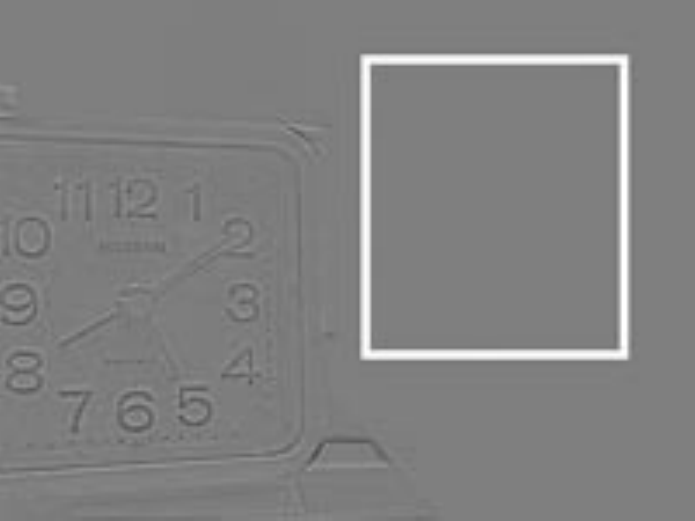}\\
 & (k) MWGF \cite{28} & (m) IM \cite{37} & (s) CBF \cite{38} & (n) QEBIF \cite{QEBIF} & (o) Ours \\
\end{tabular}
\caption{{
Multi-focus source images and the fusion results of the proposed method and some state-of-the-art methods on `disk'.
Images in the first line are the fusion result, while those in the second line are the normalized differences $\textbf{D}_{norm}$ between the fusion result and input $\textbf{I}_1$.  
}}
\label{fig:lab}
\vspace{-4mm}
\end{figure*} 
\begin{figure*}[t!]
\begin{tabular}
{@{\hspace{-10mm}}c@{\hspace{+3mm}}c@{\hspace{+3mm}}c@{\hspace{+3mm}}c@{\hspace{+3mm}}c@{\hspace{+3mm}}c}
&\includegraphics[width =0.2\linewidth]{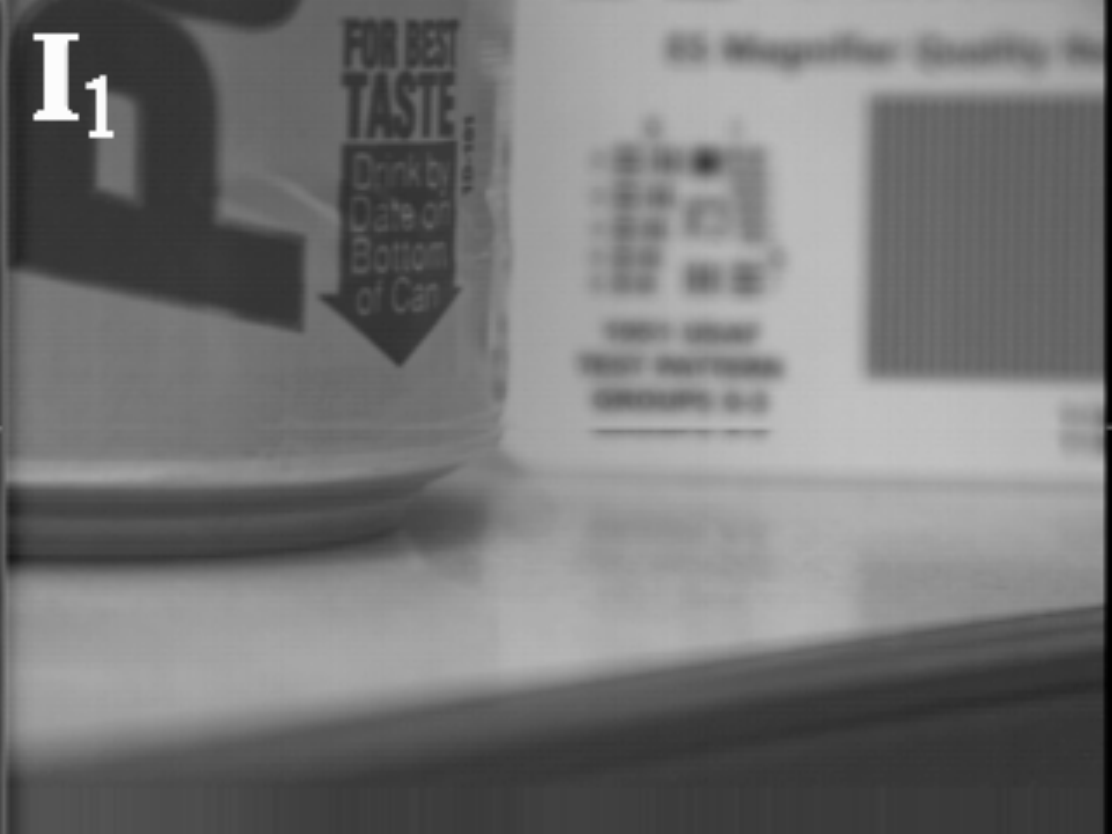}
&\includegraphics[width =0.2\linewidth]{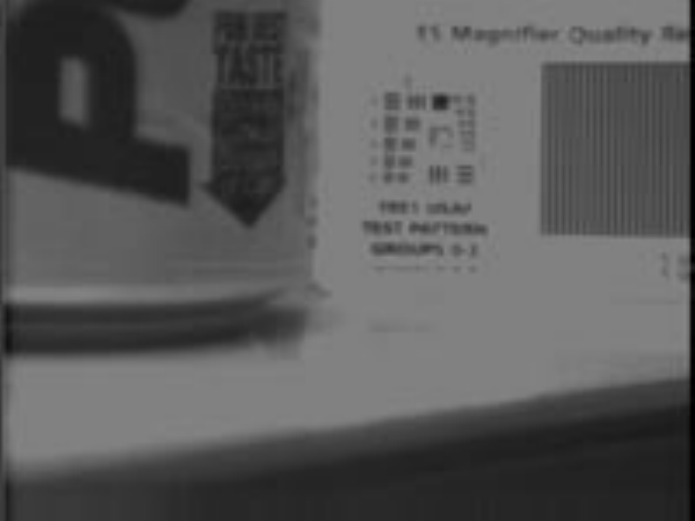}
&\includegraphics[width =0.2\linewidth]{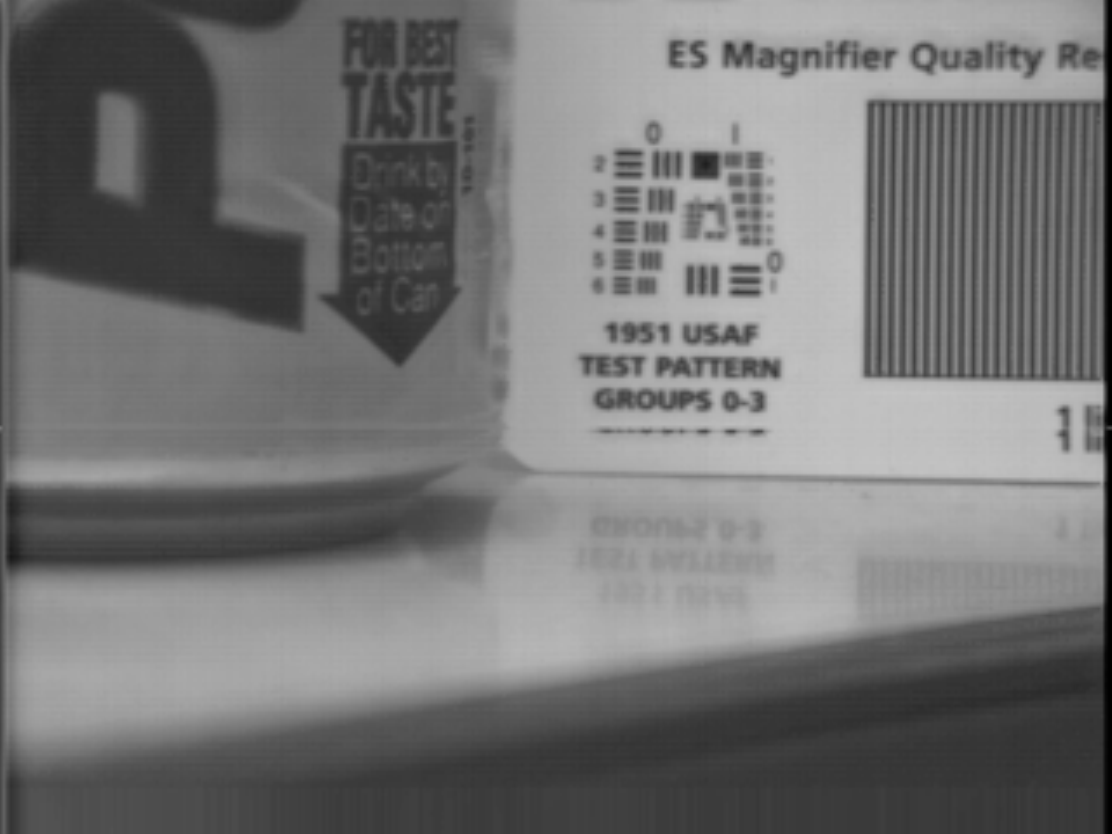}
&\includegraphics[width =0.2\linewidth]{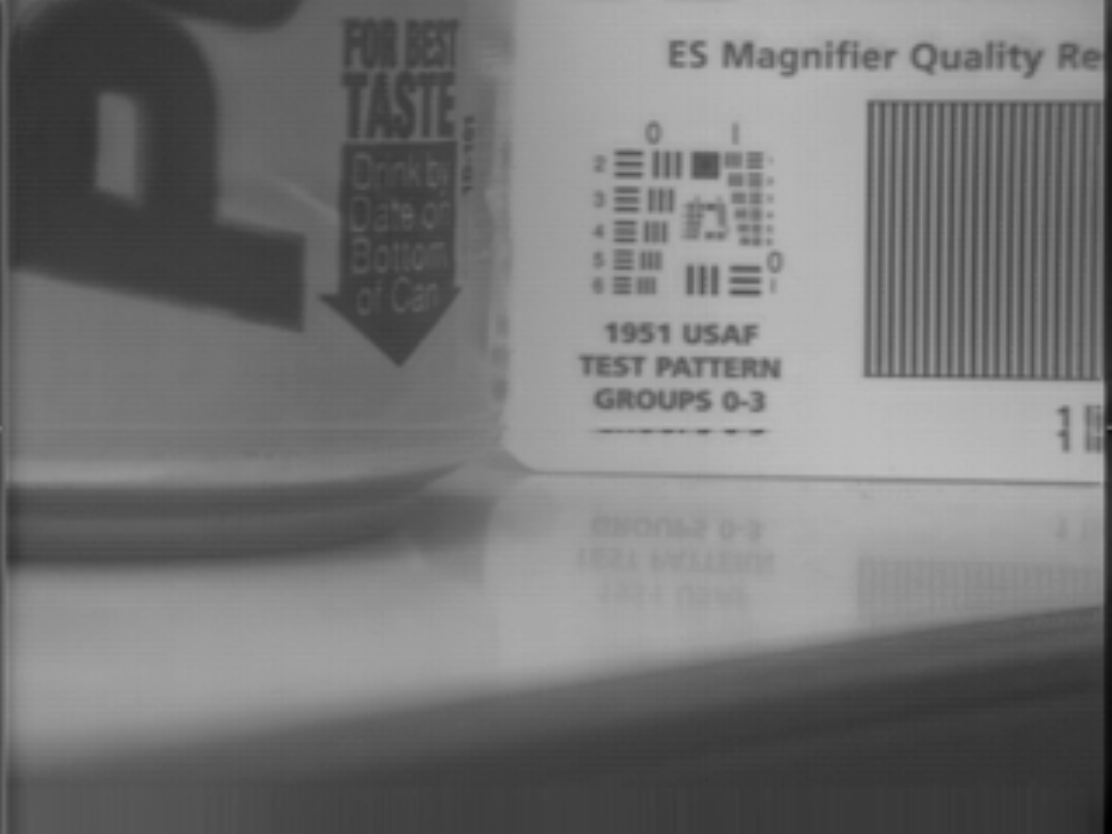}
&\includegraphics[width =0.2\linewidth]{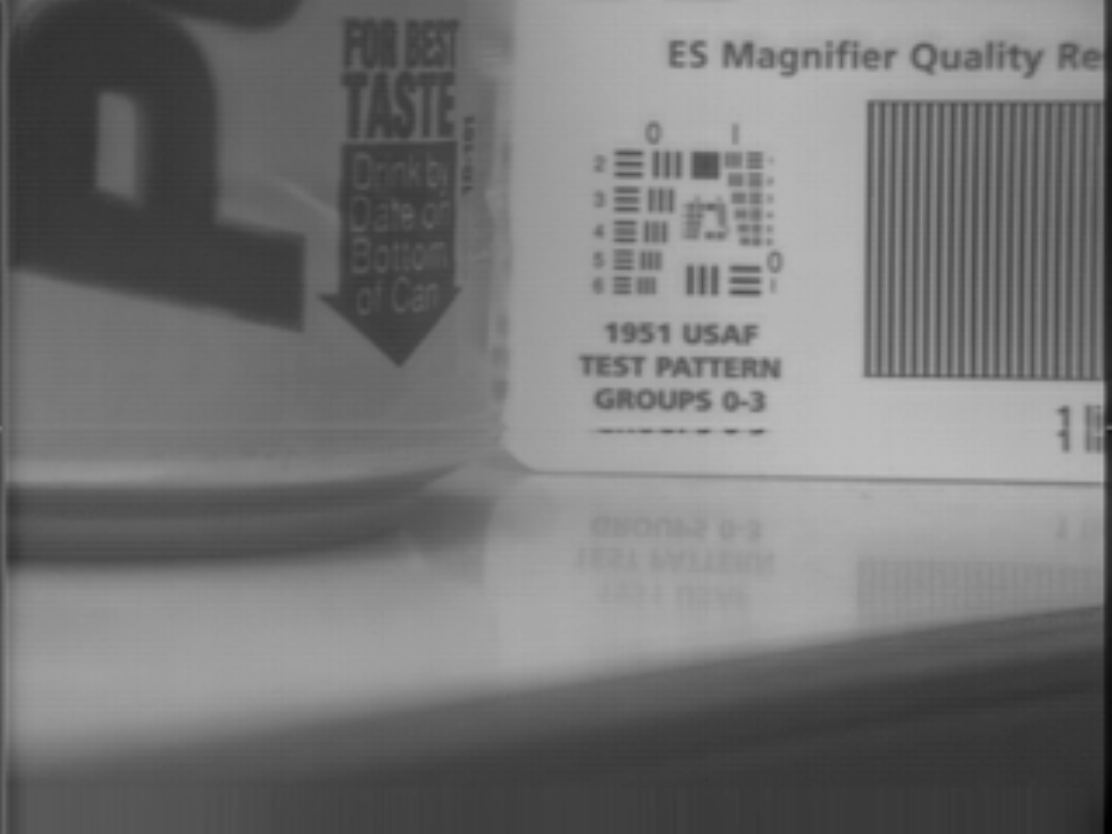}\\
&\includegraphics[width =0.2\linewidth]{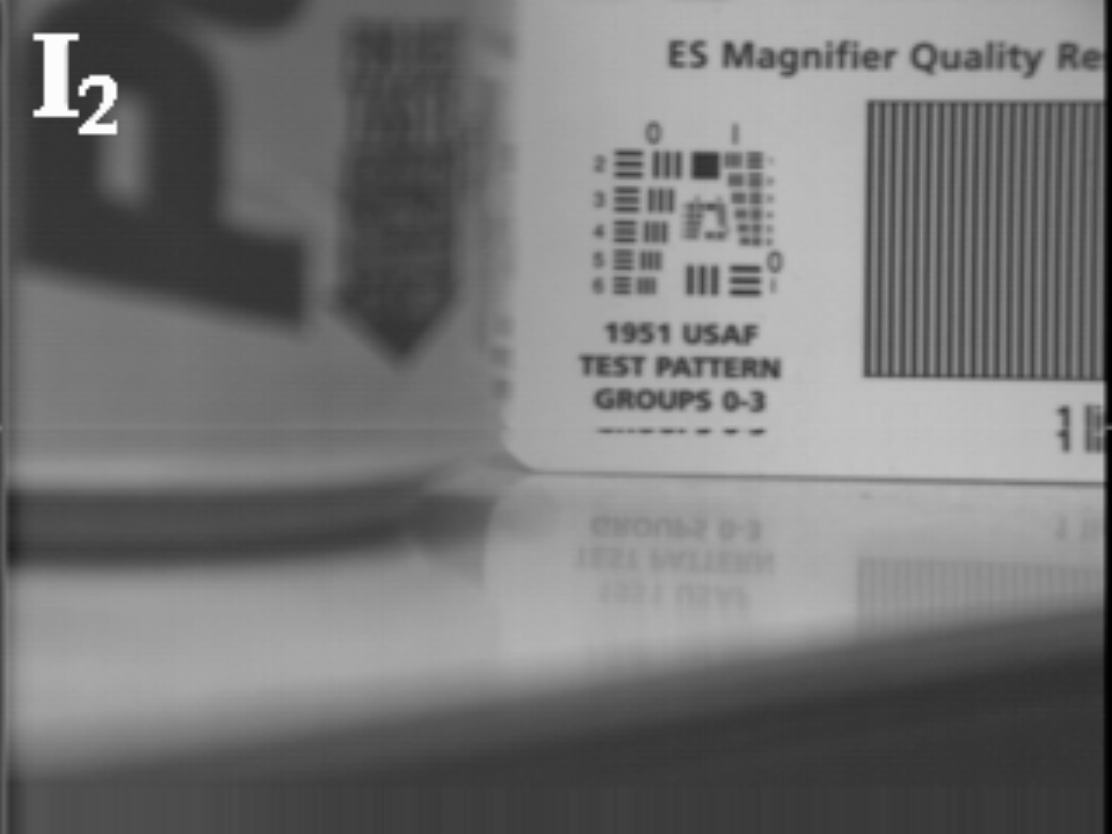}
&\includegraphics[width =0.2\linewidth]{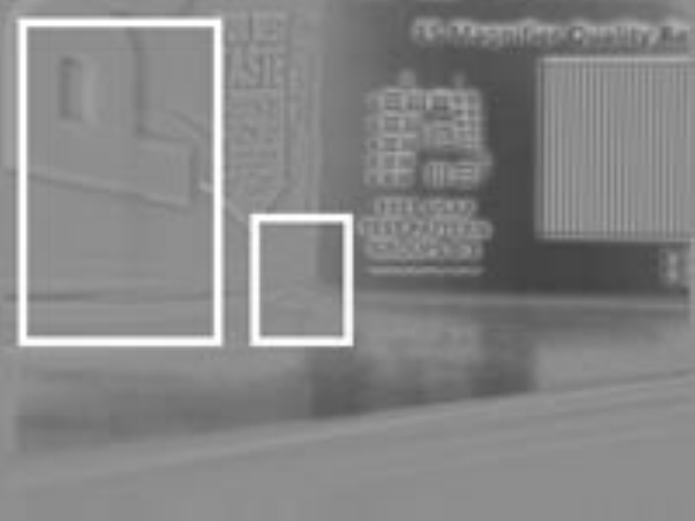}
&\includegraphics[width =0.2\linewidth]{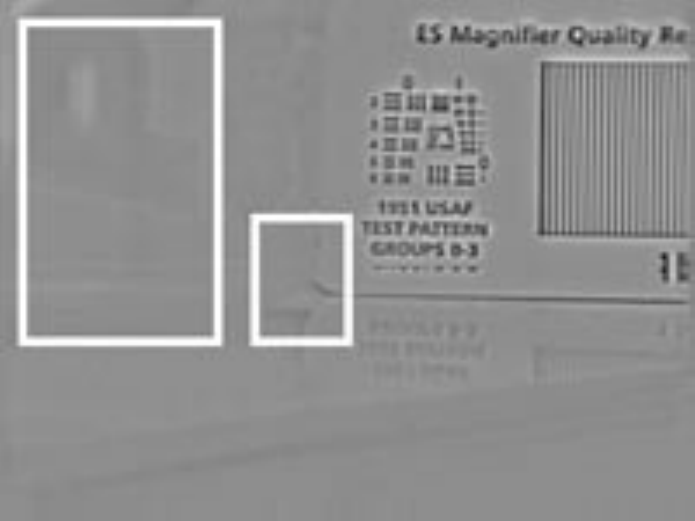}
&\includegraphics[width =0.2\linewidth]{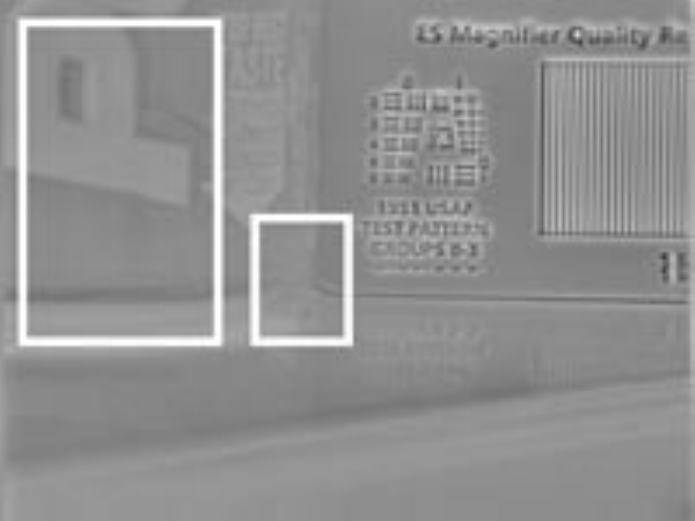}
&\includegraphics[width =0.2\linewidth]{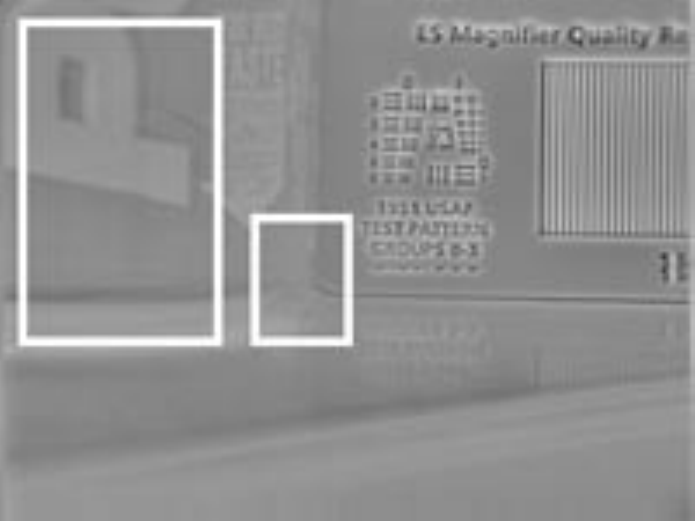}
\\
& (a) Multi-focus images& (b) AVE & (c) LAP \cite{44} & (d) FSD \cite{45} & (e) GRP \cite{46} \\
&\includegraphics[width =0.2\linewidth]{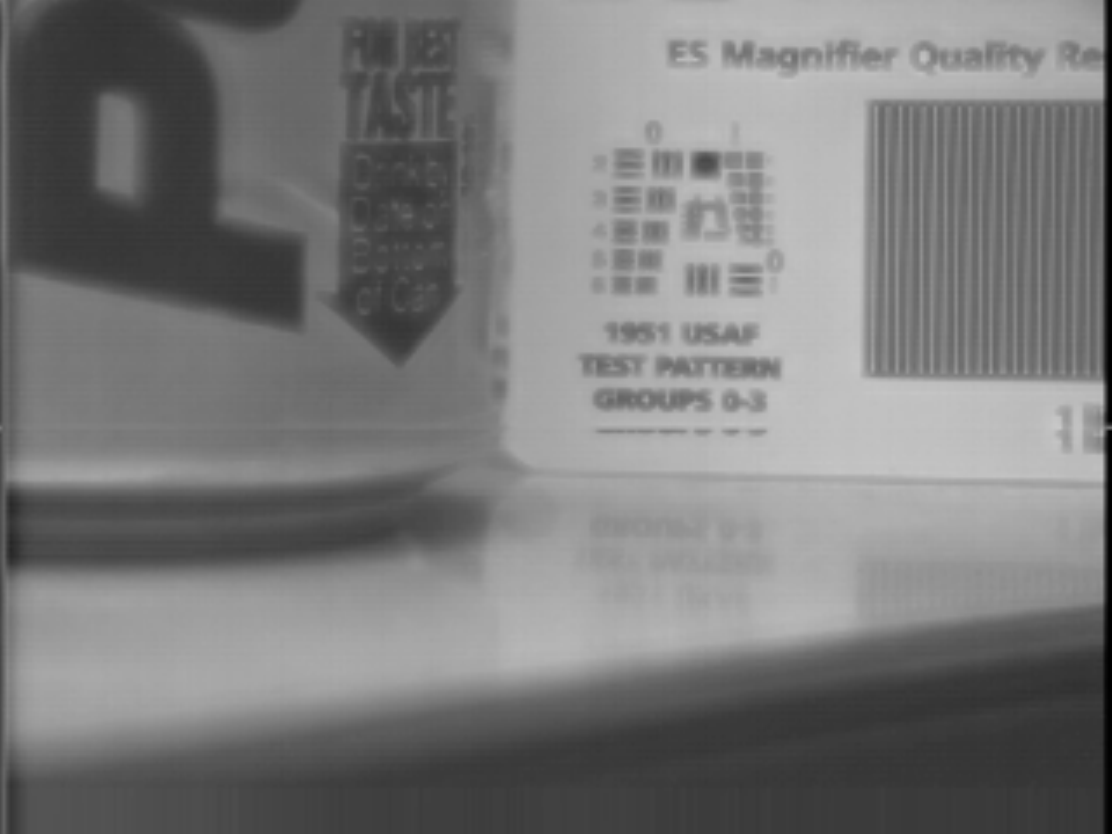}
&\includegraphics[width =0.2\linewidth]{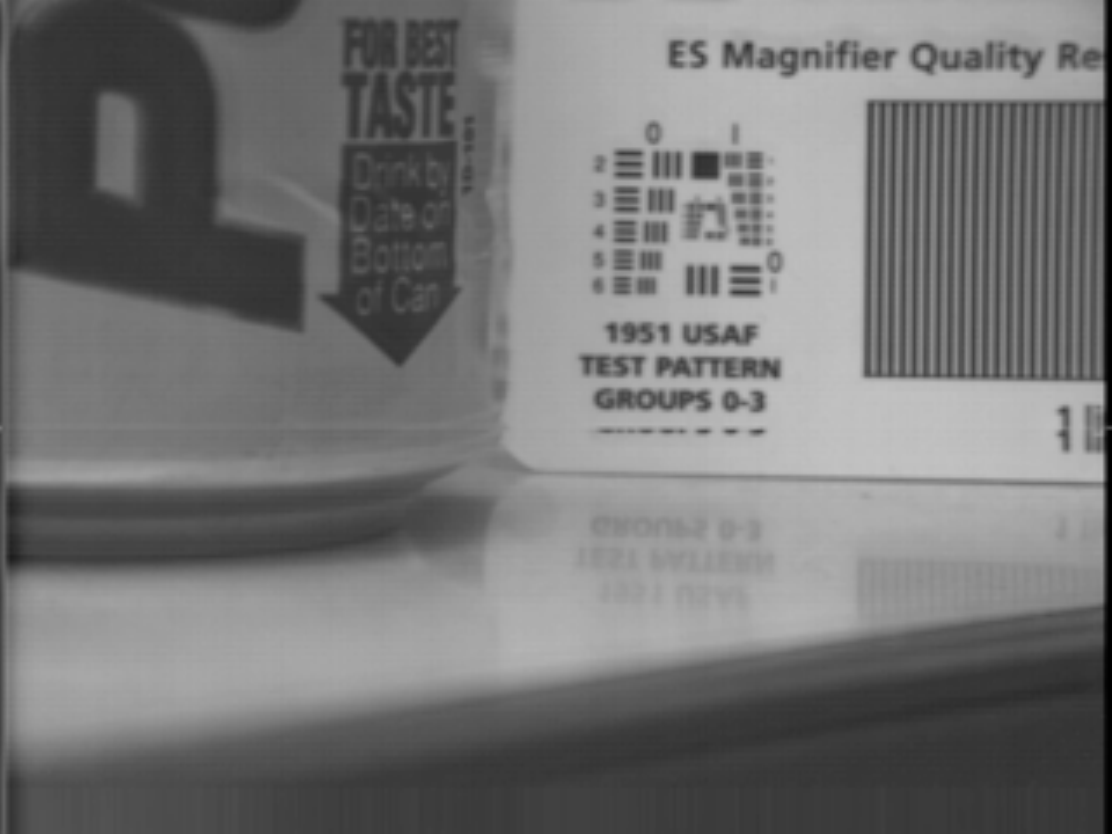}
&\includegraphics[width =0.2\linewidth]{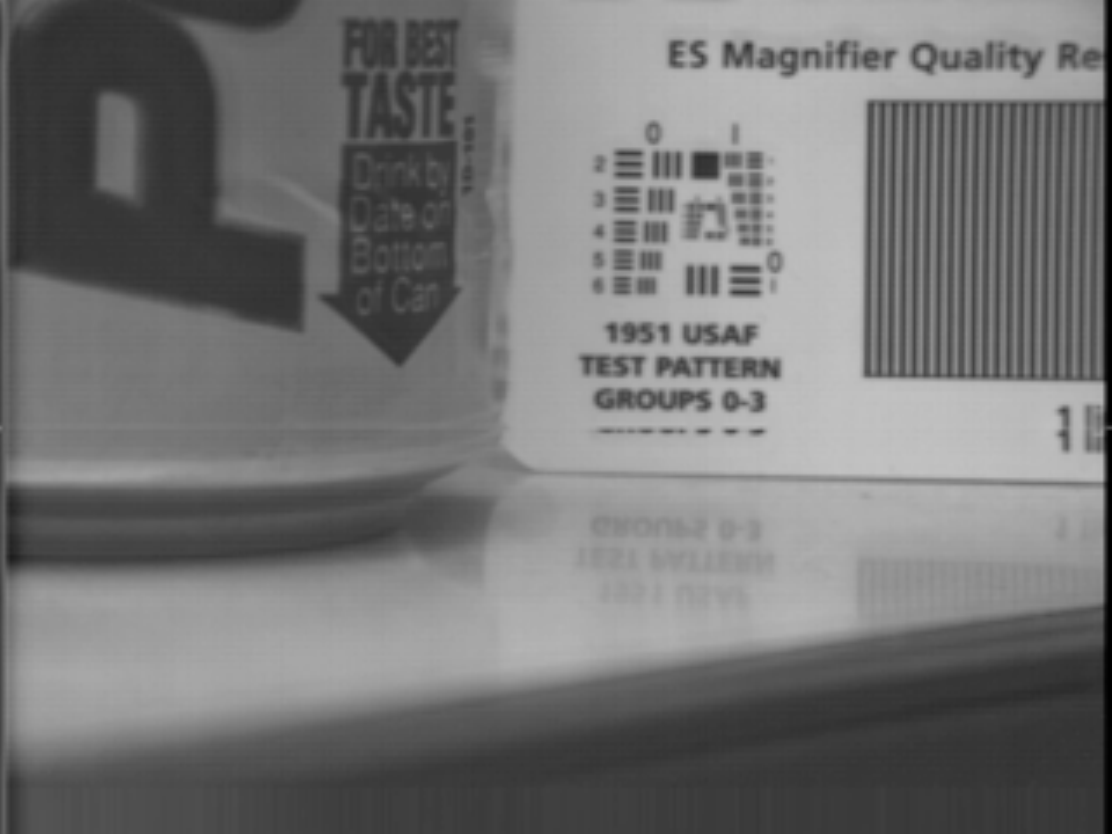}
&\includegraphics[width =0.2\linewidth]{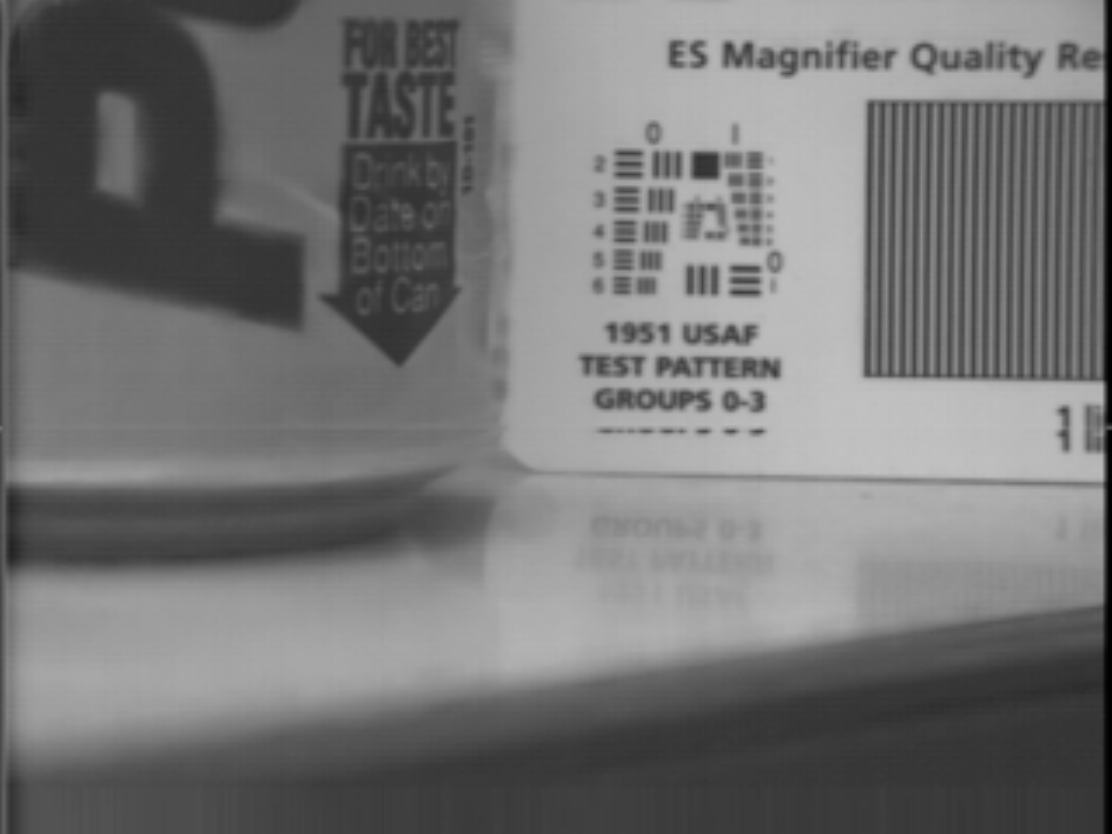}
&\includegraphics[width =0.2\linewidth]{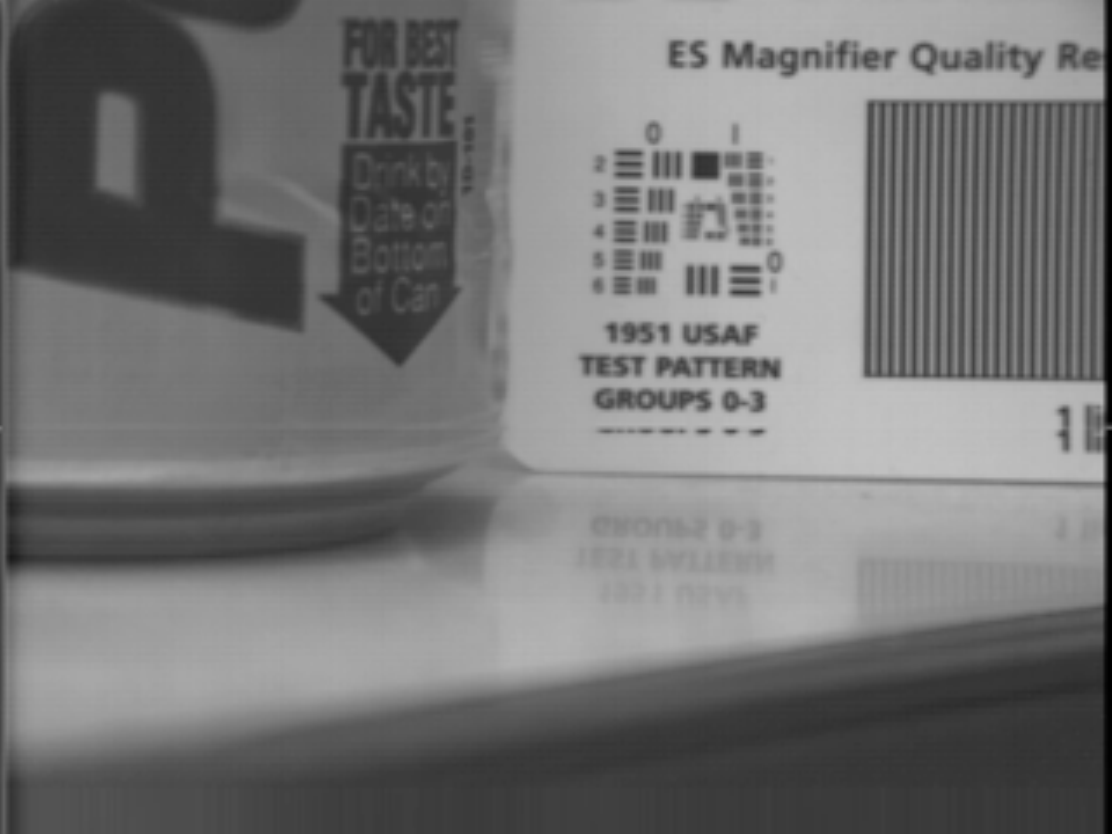}
\\

&\includegraphics[width =0.2\linewidth]{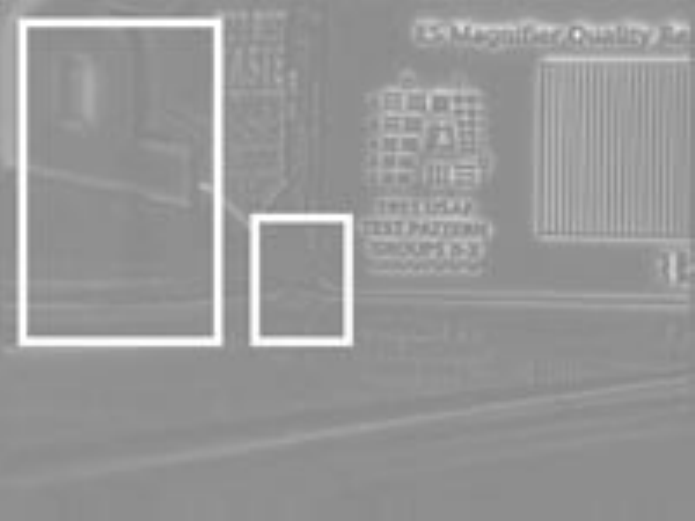}
&\includegraphics[width =0.2\linewidth]{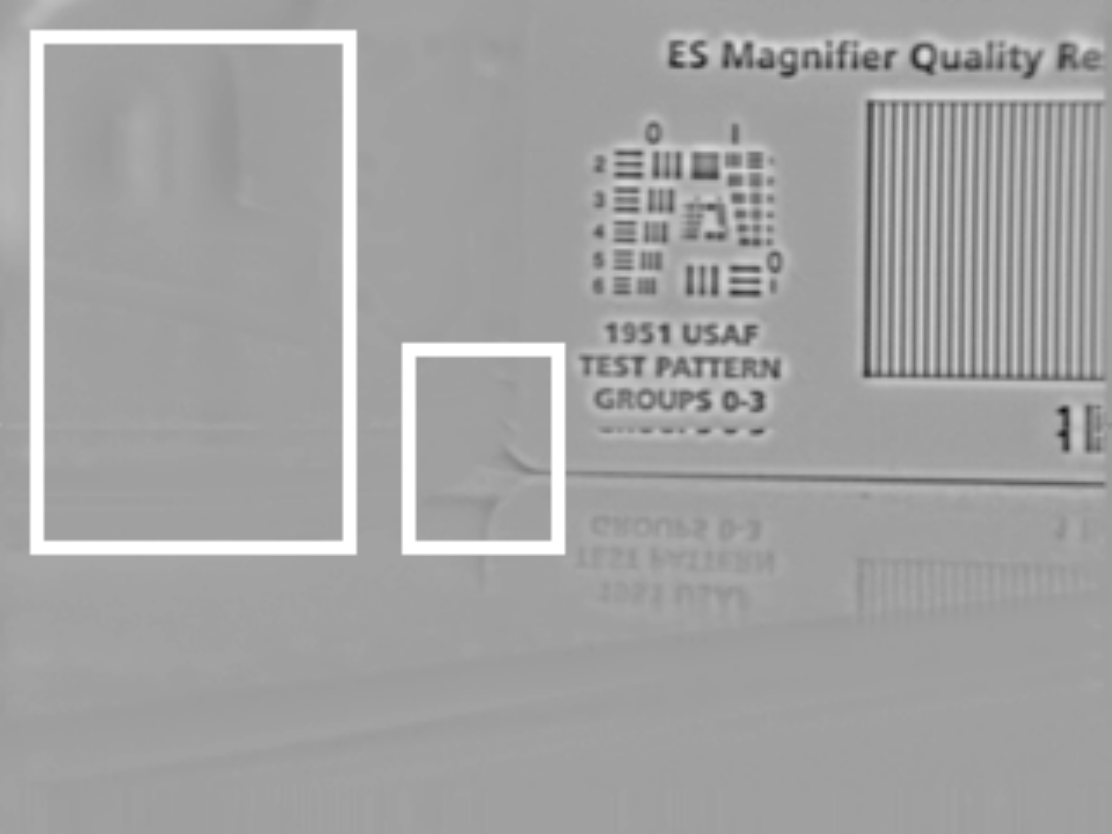}
&\includegraphics[width =0.2\linewidth]{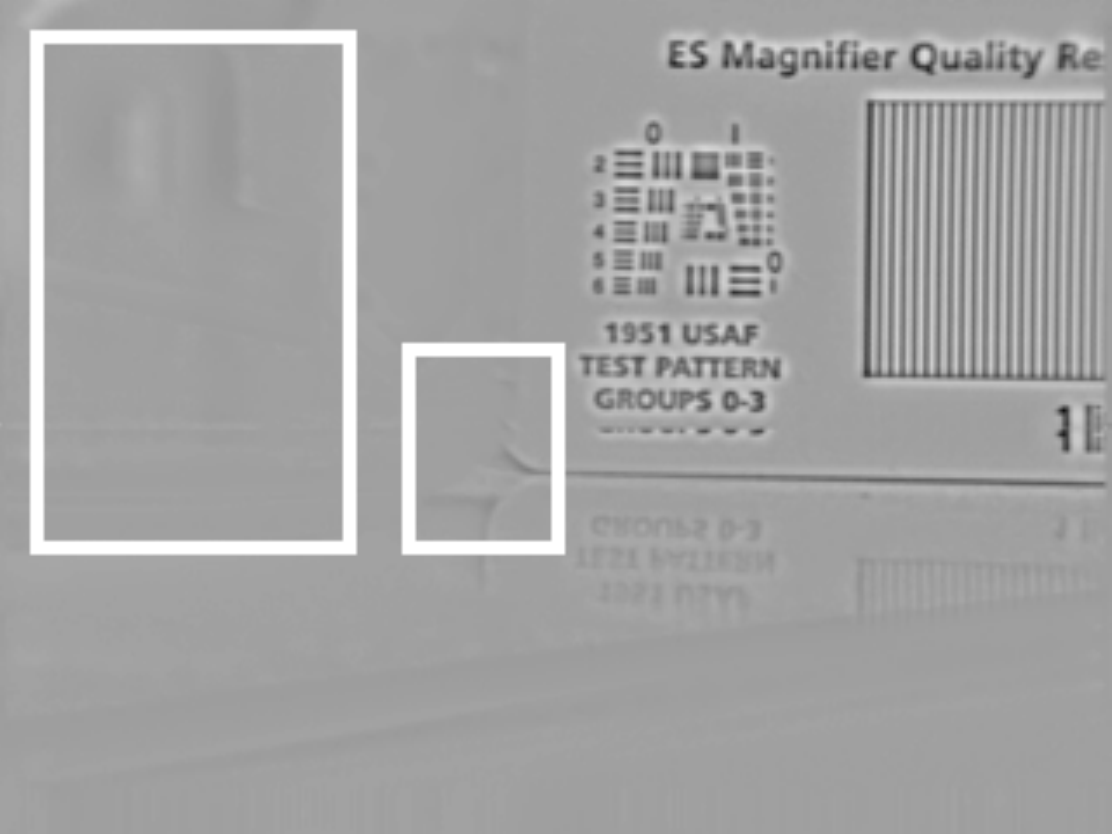}
&\includegraphics[width =0.2\linewidth]{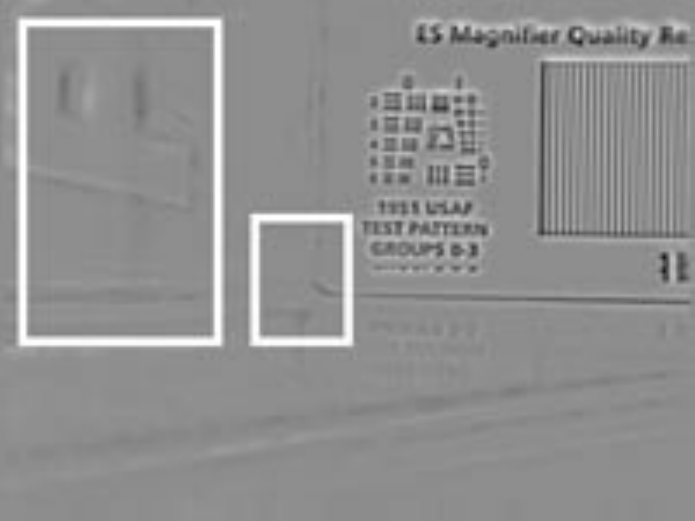}
&\includegraphics[width =0.2\linewidth]{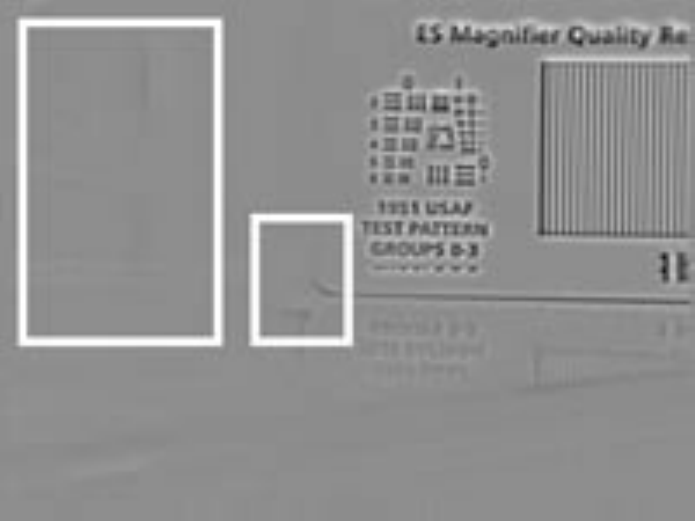}
\\
& (f) RAP \cite{47} & (g) NSCT-SR \cite{MST-SR} & (h) NSCT \cite{22} & (i) SR \cite{26} & (j) GF \cite{GF}\\
&\includegraphics[width =0.2\linewidth]{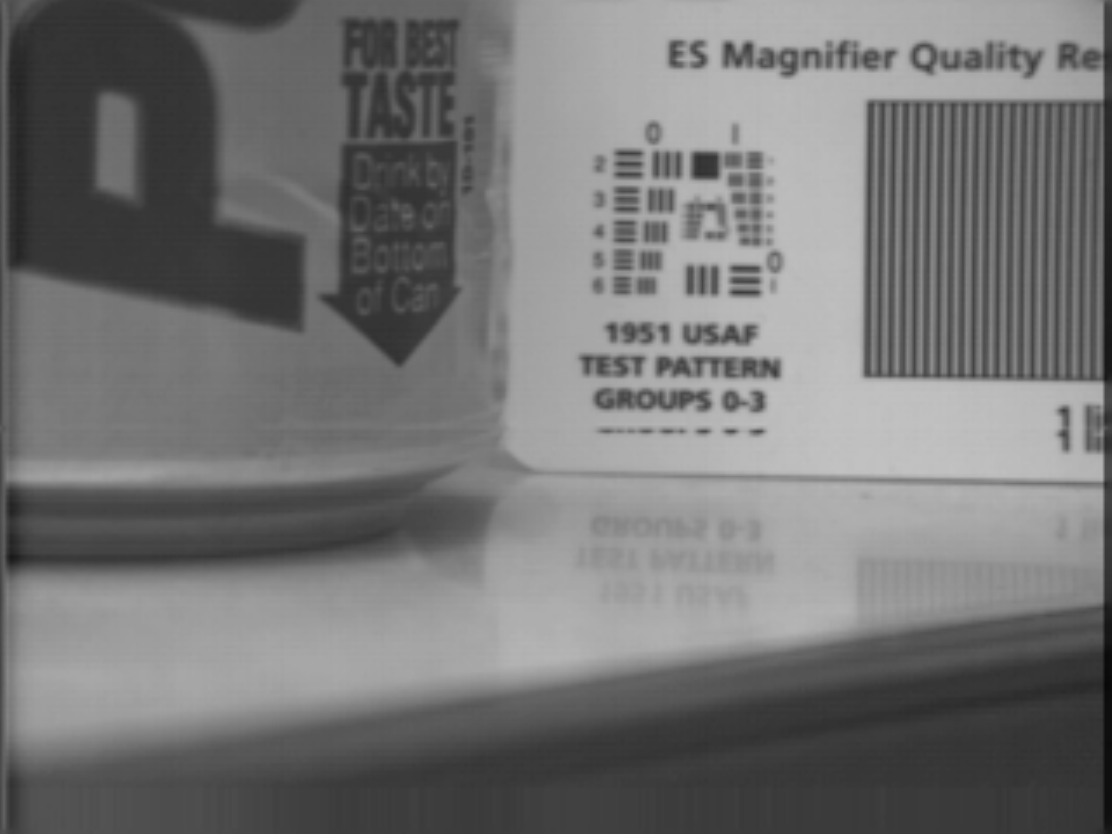}
&\includegraphics[width =0.2\linewidth]{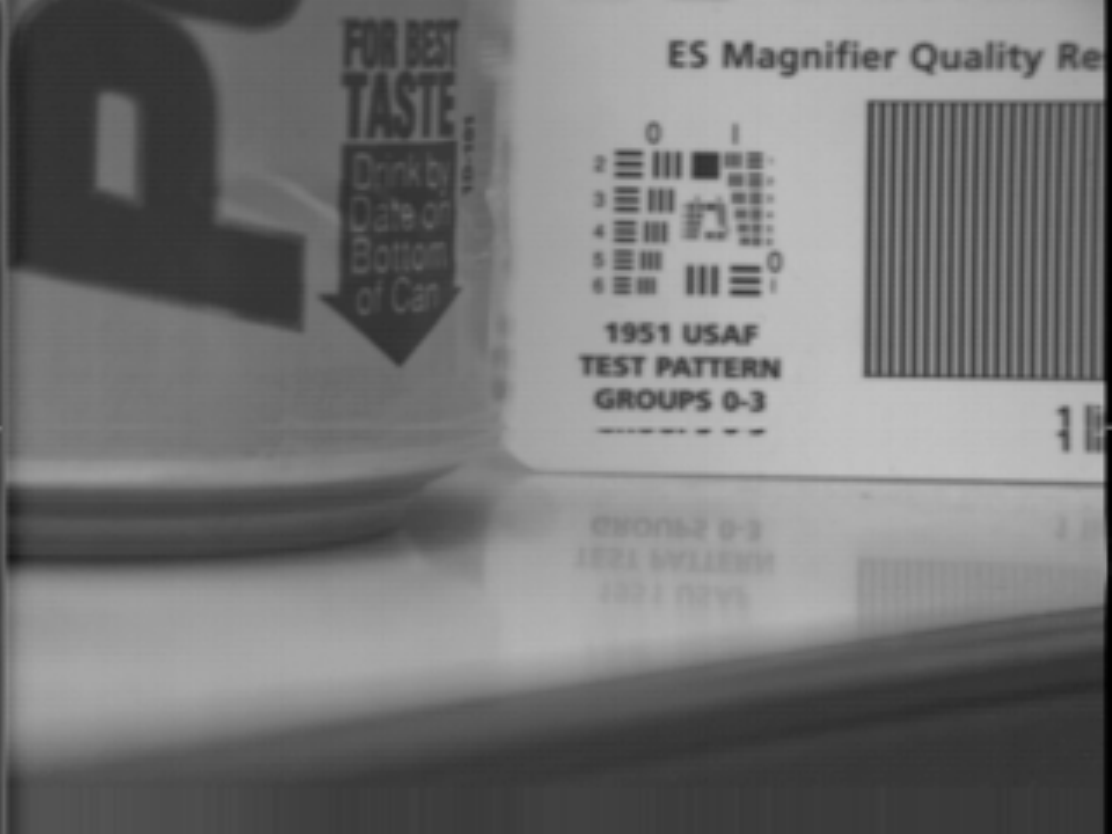}
&\includegraphics[width =0.2\linewidth]{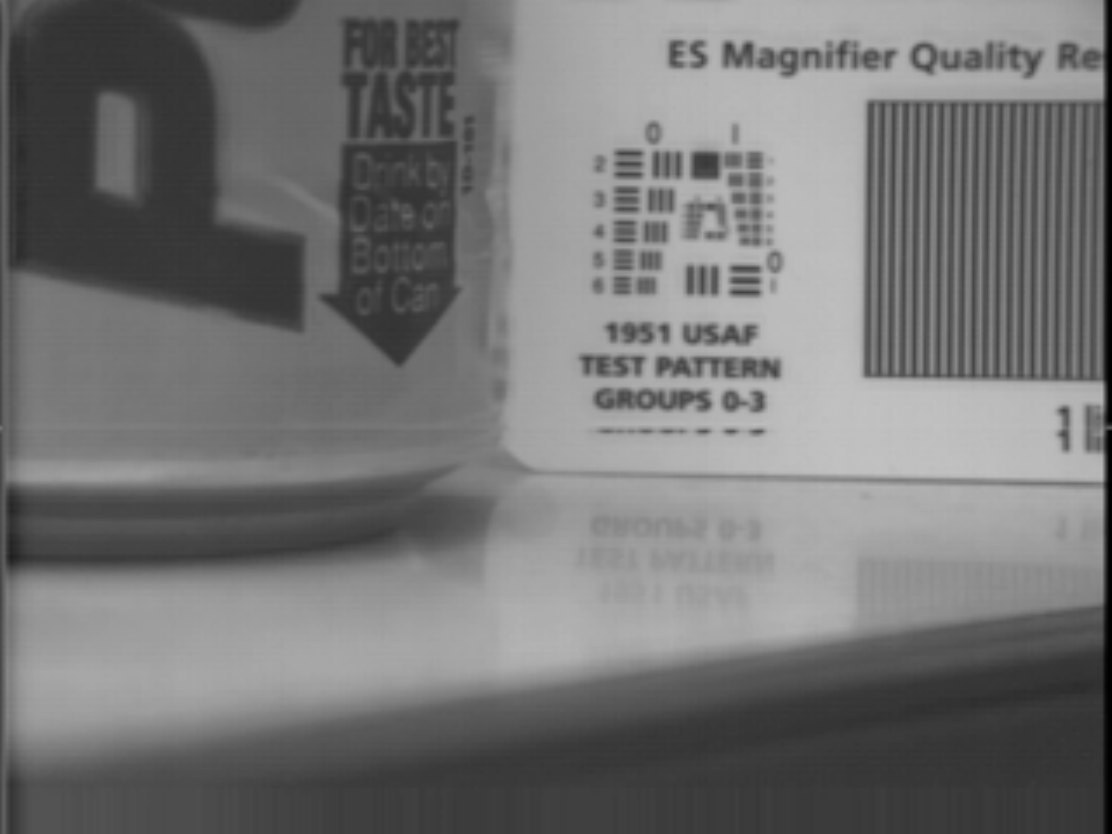}
&\includegraphics[width =0.2\linewidth]{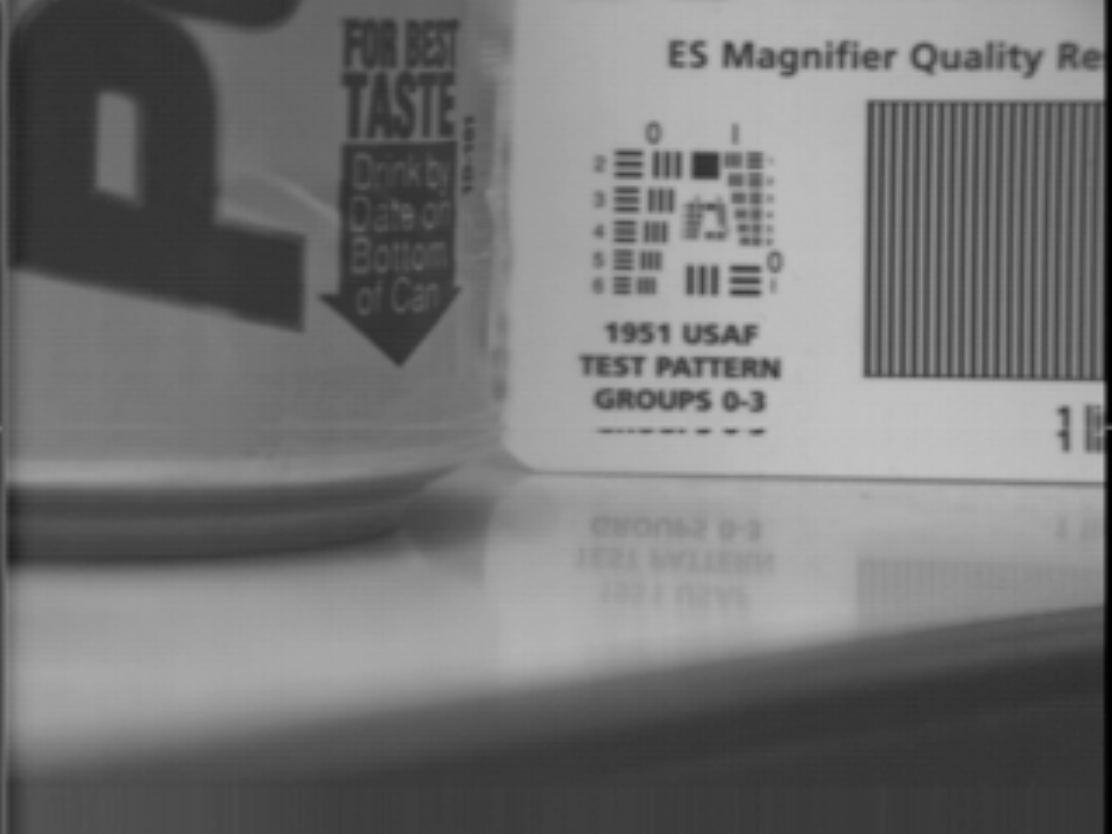}
&\includegraphics[width =0.2\linewidth]{./Figure/pepsi400}\\
&\includegraphics[width =0.2\linewidth]{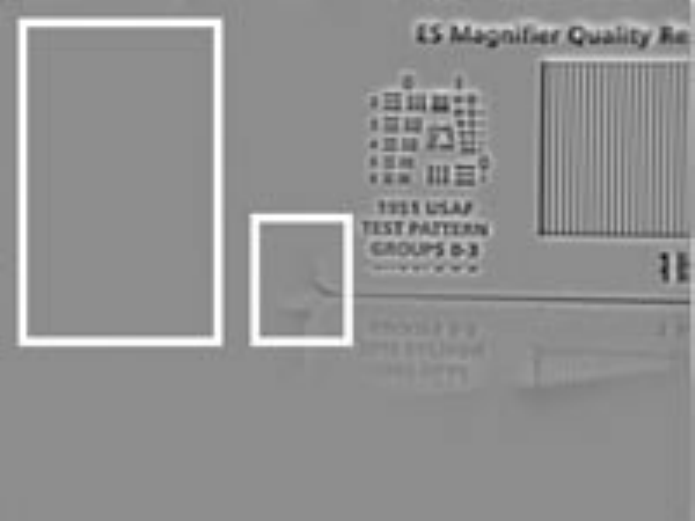}
&\includegraphics[width =0.2\linewidth]{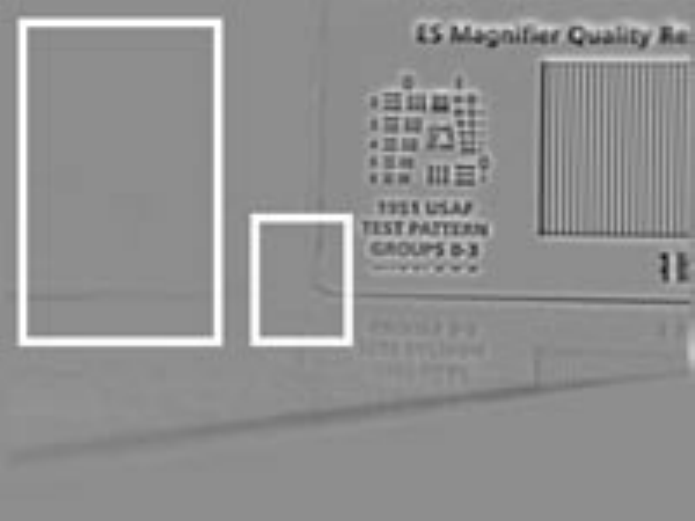}
&\includegraphics[width =0.2\linewidth]{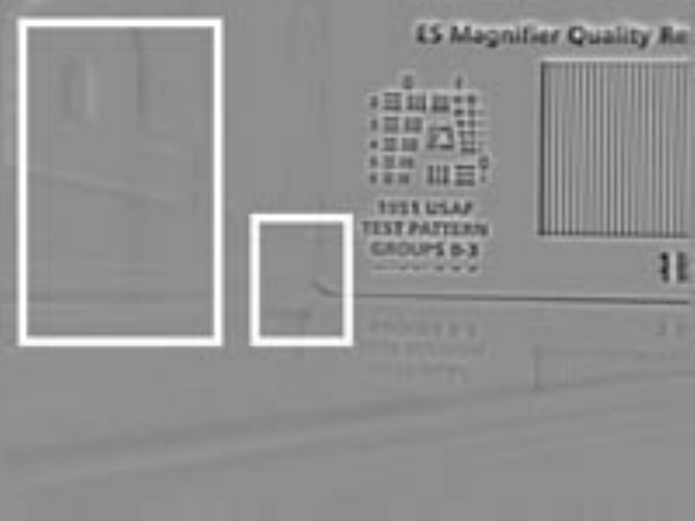}
&\includegraphics[width =0.2\linewidth]{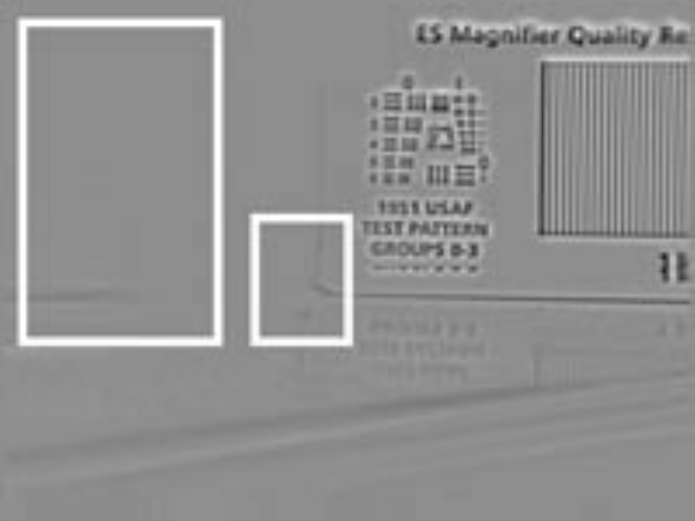}
&\includegraphics[width =0.2\linewidth]{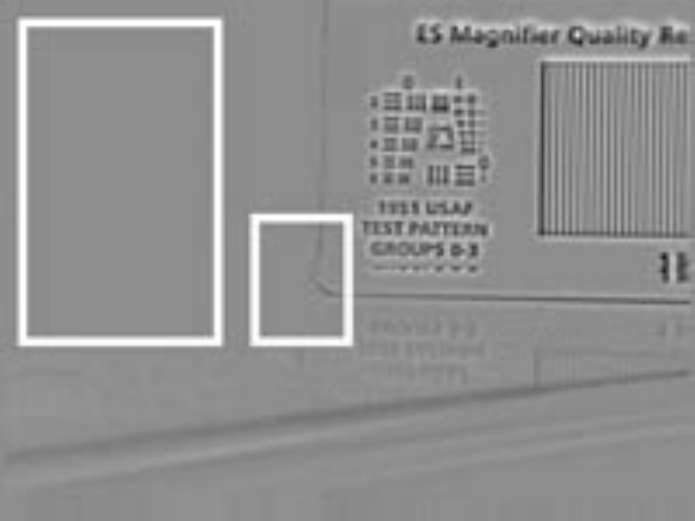}\\
& (k) MWGF \cite{28}& (m) IM \cite{37} & (s) CBF \cite{38} & (n) QEBIF \cite{QEBIF} & (o) Ours \\
\end{tabular}
\caption{{
Multi-focus source images and the fusion results of the proposed method and some state-of-the-art methods on `pepsi'.
Images in the first line are the fusion result, while those in the second line are the normalized differences $\textbf{D}_{norm}$ between the fusion result and input $\textbf{I}_1$.  
}}
\label{fig:pepsi}
\vspace{-4mm}
\end{figure*}

\subsection{Fusion Results}
\label{sec:fusionResults}
The proposed method was evaluated on all $25$ pairs of source images, objectively and subjectively.

\subsubsection{Objective evaluation}
TABLE \ref{table:sub} summarized $Q_{G}$ \cite{Qabf}, $Q_{NMI}$ \cite{NMI} and NCIE \cite{NCIE} of the proposed BS-QEBIF method and the $13$ other image fusion methods.
The results on each source image pair are grouped together with the name of the source images ahead.
Results of the first five pairs were summarized individually. 
The averaging results of all $20$ pairs in Lytro dataset \cite{Lytro} were summarized in the last block in the table.
The best performing methods are shown in bold. 

The proposed method achieves the best performance in almost all situations. 
The consistent satisfactory performance on $25$ pairs of source images prove the stability of the proposed method.
Moreover, the good performance of the proposed method in terms of $Q_{G}$ \cite{Qabf} and $Q_{NMI}$ \cite{NMI} indicate its strong ability to maintain sharpness information and mutual information from the source images.
Besides, the fusion result of the proposed method is highly correlated with the source images which can be concluded from NCIE \cite{NCIE}.
Nevertheless, the proposed method BS-QEBIF with the fast bilateral solver works better than QEBIF.
It shows that the fast bilateral solver is more suitable than the guided filter in the proposed pipeline.
%


\subsubsection{Subjective evaluation}
To further evaluate the performance of the proposed method, 
subjective evaluation were conducted by visualizing the fusion results as shown in Fig \ref{fig:fusionResults}, \ref{fig:lab} and \ref{fig:pepsi}.
Fig. \ref{fig:fusionResults} illustrates the fusion result of the proposed method on the first 5 pairs of source images.
The fusion results can well preserve the details in the source images.

Fig. \ref{fig:fusionResults} and Fig. \ref{fig:lab} analyze the results in two major challenges in image fusion task, mis-registration and fusion at focus-level-changed boundary regions. The fusion results of the proposed method are compared with those of 13 other fusion methods.
Specifically, the source images and the generated fusion results are shown in (a) and (b-o) respectively.
For MWGF \cite{28} in TABLE \ref{table:sub}, the result of \textit{(r)} performs better than \textit{(m)}.
Thus MWGF \cite{28} \textit{(r)} is adopted for visualization in image (k). 
For each method, the fused image $\textbf{F}$ (first line) and its normalized difference with $I_1$, $\textbf{D}_{norm}$ (second line), are shown.
$\textbf{D}_{norm}$ is calculated by $\textbf{D}_{norm} = \frac{\textbf{D}-\textbf{D}_{min}}{\textbf{D}_{max}-\textbf{D}_{min}}$, where $\textbf{D} = \textbf{F}-\textbf{I}_1$ is the difference map. $\textbf{D}_{max}$ and $\textbf{D}_{min}$ denote the maximum and minimum values among all $14$ difference maps, which 
ensures the fairness of the comparison.

\begin{figure*}[t!]
\begin{center}
\includegraphics[width=1.00\linewidth]{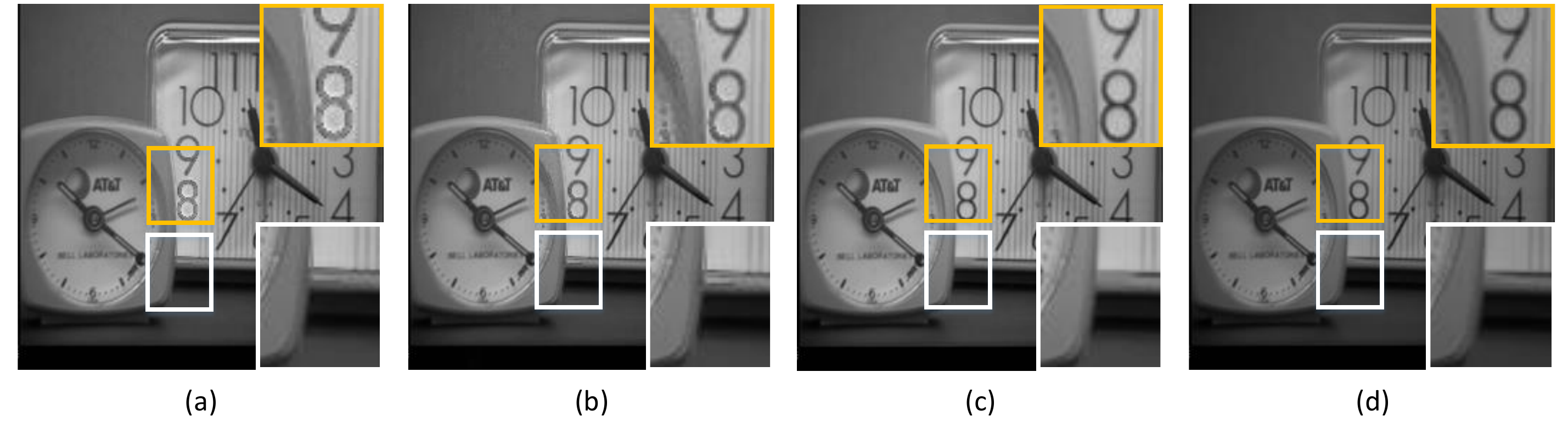}
\end{center}
\caption{{\small
Examples of fusion results under different settings.
These images are (a) `Baseline' (b) `Baseline+CM' (c) `Baseline+Norm' (d) `Ours'.}}
\label{fig:component}
\end{figure*}
In Fig \ref{fig:lab}, multi-focus source images `lab'  suffer from the mis-registration, especially on the man's head as shown in (a). 
Since $\textbf{I}_1$ focuses on the man's head, thus this area of $\textbf{D}_{norm}$ should be zero in the ideal case.
It can be seen that AVE, LAP \cite{44}, FSD \cite{45}, GRP \cite{46}, RAP \cite{47}, NSCT-SR \cite{MST-SR}, NSCT \cite{22} and CBF \cite{38} failed to handle the mis-registered well, because big residual errors can be observed in the blocks in their normalized difference maps.
SR \cite{26}, GF \cite{GF} and QEBIF \cite{QEBIF} works better, but few residual errors still exist in the white block.
The proposed method, MWGF \cite{28} and IM \cite{37} can handle thus mis-registered problems well since no residual errors exist in the block.
This example indicates the proposed method can handle source images with mis-registration well and fusion result does not suffer from the blocking, ringing or blurring artifacts, which may occur in other methods.

Similar results can be observed in Fig. \ref{fig:pepsi}.
The left block concentrates on measuring the performance of mis-registration.
In this case, only the proposed method and MWGF \cite{28} work well and do not suffer from any residual errors.
The right blocks concentrate on measuring the accuracy in the focus-level-changed boundary region. 
The left and right parts within the block are focused on $\textbf{I}_1$ and $\textbf{I}_2$ respectively, Therefore, the left part in $\textbf{D}_{norm}$ should be zero in the ideal case.
However, only the proposed method and IM \cite{37} work well and accurately fuse in the focus-level-changed boundary regions.
For all other methods, clear horizontal lines can be observed in the left part of $\textbf{D}_{norm}$.
Considering these two blocks, the proposed method works the best on `pepsi'.

From these examples, it can be seen that the proposed method is effective and works well subjectively.
The fusion result of the proposed method does not suffer from any residual errors with source images mis-registered. 
Besides, it can deal well with the focus-level-changed boundary regions, avoid blocking, ringing and blurring artifacts.

\subsection{Component analysis}
\label{Sec:EX_CM}
%
%
In this section, the effectiveness of the proposed confidence map (CM) and normalization (Norm) are evaluated objectively and subjectively.
Objectively, they were evaluated on all 25 pairs of source image in term of $Q_G$ \cite{Qabf}, $Q_{NMI}$ \cite{NMI} and NCIE \cite{NCIE}.
The averaging results on these 25 pairs were summarized in TABLE \ref{table:Component}, 
where `baseline' corresponds to the fusion results using $\cal{W}'$ with values in the confidence map were set to identical; 
`baseline+CM' corresponds to the fusion results using $\cal{W}'$ with the proposed confidence map; 
`baseline+Norm' corresponds to the fusion result while adding normalization on the `baseline'.
`Ours' represents the proposed method which utilizes the confidence map and normalization.
 
The proposed confidence map (CM) and normalization (Norm) are proved to be effective by comparing the results in `Baseline+CM' and `Baseline+Norm' to `Baseline' respectively.
Performance gains are achieved in both cases.
`Ours' combines them together and achieved the best results indicating `CM' and `Norm' cooperate well with each other.
Besides, `CM' and `Norm' are effective to transfer mutual information from source images to the fusion results concluded from $29.6\%$ performance gain in $Q_{NMI}$ \cite{NMI}.
%
Subjectively, fusion results of `clock' under different settings are shown in Fig. \ref{fig:component}.
The fusion results of (a) `Baseline',  (b) `Baseline+CM', (c) `Baseline+Norm' and (d) `Ours' are shown. 
Without normalization, image (a) and (b) suffers from the residual errors on the focus-level-changed boundary region as shown in the upper block.
Benefiting from the normalization process, image (c) is much smoother and appears better.
However, (c) still cannot handle focus-level-changed boundary regions accuratly as we can see on the bottom block.
The confidence map effectively improve the fusion results as we can see from the bottom block in (d), where the fusion errors have be effectively eliminated.
To sum up, CM and Norm are both effective.
Especially, the proposed confidence map achieved the desired goal (discussed in `Boundary regions' of Fig. \ref{fig:CM}) to improve focus-level-changed boundary regions even it does not achieved an apparent performance improvement during objective evaluation.

\begin{table}
\begin{center}
\caption{Component analysis}
\label{table:Component}
\begin{tabular}
   {|C{25mm}|C{15mm}|C{15mm}|C{15mm}|}\hline
 Method& $Q_{G}$\cite{Qabf}&$Q_{NMI}$ \cite{NMI}&NCIE \cite{NCIE}\\ \hline
Baseline         & 0.6992 & 0.9074 & 0.8284 \\
Baseline+CM      & 0.7118 & 0.9650 & 0.8314 \\
Baseline+Norm    & 0.7487 & 1.1713 & 0.8435 \\ \hline
Ours & \textbf{0.7532} & \textbf{1.1756} & \textbf{0.8436}\\
\hline
    \noalign{\smallskip}
    \end{tabular}
\end{center}
\end{table}


\section{Conclusions}
In this work, a quality estimation based multi-focus image fusion method with the fast bilateral solver (BS-QEBIF) is proposed.
The visual quality is estimated to help measure the focus levels since the visual quality of an image is highly correlated with its focus level.
{In addition, the confidence map is proposed to measure the reliability of different local regions}.
The fusion results of BS-QEBIF can well maintain the details in the multi-focus images and do not suffer from the ringing or blocking artifacts.

\ifCLASSOPTIONcaptionsoff
  \newpage
\fi

\bibliographystyle{IEEEtran}
\bibliography{IEEEabrv,egbib}
\end{document}